\documentclass[%
reprint,
superscriptaddress,
 amsmath,amssymb,
 aps,
 prb,
]{revtex4-2}

\usepackage{graphicx}
\usepackage{dcolumn}
\usepackage{bm}
\usepackage{hyperref}
\usepackage{cleveref} 
\usepackage[draft]{changes}
\usepackage[version=4]{mhchem}
\usepackage{lineno}
\usepackage{booktabs} 
\usepackage{multirow}
\usepackage{threeparttable}

\begin{document}

\title{Accurate Modeling of Gate Leakage Currents in SiC Power MOSFETs}

\author{Ang Feng}
\email{ang.feng@onsemi.com}
 \affiliation{Institute for Microelectronics, Vienna University of Technology (TU Wien), Gußhausstraße 27-29/E360, A-1040 Vienna, Austria}
 \affiliation{Reliability Physics Group, Onsemi, Schaliënhoevedreef 20B, 2800 Mechelen, Belgium}
\author{Alexander Karl}
 \affiliation{Institute for Microelectronics, Vienna University of Technology (TU Wien), Gußhausstraße 27-29/E360, A-1040 Vienna, Austria}
 \author{Dominic Waldh\"or}
 \affiliation{Institute for Microelectronics, Vienna University of Technology (TU Wien), Gußhausstraße 27-29/E360, A-1040 Vienna, Austria}
\author{Marina Avramenko}
  \affiliation{Reliability Physics Group, Onsemi, Schaliënhoevedreef 20B, 2800 Mechelen, Belgium}
\author{Peter Moens}
 \affiliation{Reliability Physics Group, Onsemi, Schaliënhoevedreef 20B, 2800 Mechelen, Belgium}
\author{Tibor Grasser}%
  \email{grasser@iue.tuwien.ac.at}
 \affiliation{Institute for Microelectronics, Vienna University of Technology (TU Wien), Gußhausstraße 27-29/E360, A-1040 Vienna, Austria}

\date{\today}

\begin{abstract}
Silicon carbide (SiC) metal-oxide-semiconductor field-effect-transistors (MOSFETs) enable high-voltage and high-temperature power conversion. Compared to Si devices, they suffer from pronounced gate leakage due to the reduced electron tunneling barrier at the interface between SiC and amorphous silicon dioxide (a‑SiO$_2$). We develop a self-consistent, physics-based simulation framework that couples electrostatics, quantum tunneling, carrier transport, impact ionization, and charge trapping for both electrons and holes. The model quantitatively reproduces measured gate-current-voltage characteristics of SiC MOS capacitors over a wide temperature (80-573 K) range and a wide bias range without empirical fitting. Simulations reveal that conduction electrons in a-SiO$_2$ can trigger impact ionization, which generates electron-hole pairs, and leads to capture of holes in the oxide bulk, thereby enhancing gate leakage current. The framework captures these coupled processes across multiple orders of magnitude in time and field, providing predictive capability for oxide reliability. Although demonstrated for SiC devices, the methodology also applies to Si technologies that uses the same gate dielectric.
\end{abstract}

\maketitle


\section{Introduction}
SiC MOSFETs have emerged as the leading power-switching devices for high-voltage, high-temperature applications due to the wide band gap, high critical field, and superior thermal conductivity of 4H-SiC \cite{langpoklakpamReviewSiliconCarbide2022a,buffoloReviewOutlookGaN2024}. Yet SiC MOS devices face distinct reliability challenges. SiC crystals exhibit elevated densities of bulk and near-interface defects \cite{SchleichIEDM19}, while the wide band gap extends the energy range over which traps can exchange charge with a surrounding insulator \cite{Waltl_DefectSpectroscopy}. The charge trapping manifests as mobility degradation, bias temperature instability (BTI), hysteresis of the electrical response, and time-dependent dielectric breakdown, etc \cite{grasserStochasticChargeTrapping2012}. Compounding these effects is the lower electron tunneling barrier at the SiC/a-SiO$_2$ interface compared with the n$^+$-Si/a-SiO$_2$ interface, which leads to significantly higher gate-oxide leakage currents under comparable fields \cite{yangAdvancesChallenges4H2024}. Gate leakage itself constitutes a reliability concern, but more critically, it acts as a precursor and catalyst that accelerates charge trapping and stress-induced degradation \cite{moensIntrinsicExtrinsicReliability2024,nissancohenHighFieldCurrent1983,Schroder98,transportMC_ArnoldPRB1994}.

The gate leakage current is a critical indicator of oxide integrity because it is highly sensitive to changes in tunneling potential \cite{FowlerNordheim1928}. In practice, J$_\text{g}$-V$_\text{g}$ sweeps are widely employed to screen for extrinsic defects and to probe intrinsic failure pathways under electrical stress \cite{shiGateOxideReliability2024}. However, experiments alone cannot disentangle the coupled roles of quantum tunneling, carrier transport, impact ionization, and charge trapping over the full span of electric fields (up to 10 MV/cm), temperatures (80-600 K), and timescales (nanoseconds-months). On the simulation side, existing models either treat these phenomena in isolation or rely on empirical 'laws' \cite{StathisReliabgateinsulatorCMOS}, which cannot be applied to this complicated situation.

\begin{figure}
\centerline{\includegraphics[width=\linewidth]{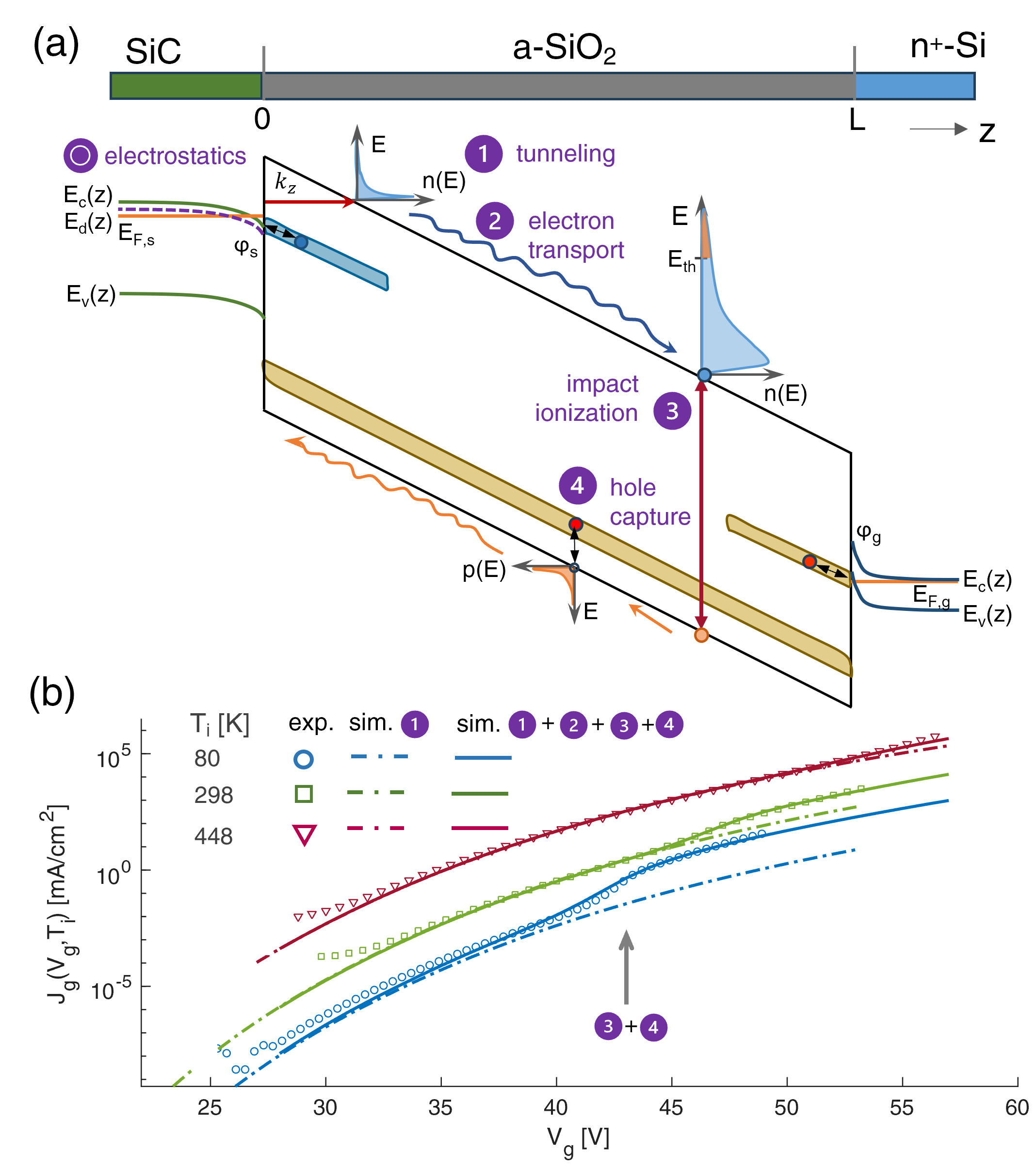}}
\caption{Device-physics framework and experimental validation. (a) Self-consistent one-dimensional MOSC (1D MOSC) model: electrostatics set the potential and band profiles across SiC, a-SiO$_2$, and n$^+$-Si; quantum tunneling injects electrons into the oxide; drift transport carries hot electrons, which undergo impact ionization to generate holes; these holes are captured by defect states in the oxide. Conduction-band edge $E_\text{c}(z)$, valence-band edge $E_\text{v}(z)$, and Fermi levels $E_\textsc{f}$ are indicated in each region. (b) Gate leakage current density $J_\text{g}$ versus gate bias $V_\text{g}$ at 80, 298, and 448 K for the MOSC (oxide thickness 54 nm). Experimental data (points) are compared with simulations that include only tunneling (dashed) or both transport and charge capture (solid). }
\label{fig_summary}
\end{figure}

To address this gap, we have developed a self-consistent, physics-based simulation framework that captures all key processes controlling gate leakage and charge trapping in MOS structures. As illustrated in Fig. \ref{fig_summary}, our model integrates (i) electrostatics, (ii) quantum tunneling from the SiC into the oxide, (iii) one-dimensional Boltzmann transport including hot-carrier dynamics, (iv) impact ionization, and (v) nonradiative multiphonon charge trapping for both electrons and holes. Implemented within the open-source Comphy platform \cite{Comphy_Rzepa2018,Comphyv3_Waldoer2023}, the framework quantitatively reproduces measured J$_\text{g}$-V$_\text{g}$ characteristics of SiC MOS capacitors (MOSC) from 80 K to 573 K across a broad bias range without any empirical fitting, while simultaneously tracking flat-band voltage shifts due to hole capture. 

This predictive tool closes a crucial gap in reliability-oriented simulation for wide band gap power devices. The methodology extends directly to advanced Si CMOS because it shares the same a-SiO$_2$ gate dielectric and poly-Si gate stacks. It enables quantitative lifetime projections, optimized accelerated stress tests, and informed design of more robust gate oxides across semiconductor platforms.

The remainder of this manuscript is organized as follows. Section II introduces the self-consistent simulation framework, detailing models for electrostatics, quantum tunneling, carrier transport, impact ionization, and charge trapping within the Comphy simulator. Section III presents experimental validation, and compares measured and simulated J$_\text{g}$-V$_\text{g}$ data across temperature and bias, elucidating how coupled tunneling, ionization, and trapping drive gate leakage currents and flat-band shifts.  Section IV concludes the study by summarizing the most important findings. 

\section{Simulation framework}
\label{sec:simnexp_details}
A n-SiC/SiO$_2$/n$^+$-Si MOSC was used for both the simulation and experimental measurement of the  gate leakage current under a sweep of gate bias (1 V/s) at temperatures from 80 to 573 K. This device was fabricated  according to the process detailed in Ref. \cite{moensChargetoBreakdownQBDApproach2020a} and has a donor concentration of $N_\text{d}=1.0\times10^{16} \text{ cm}^{-3}$,  a polycrystalline n$^{+}$-Si gate, and a 54-nm thermally-grown a-SiO$_2$ which was exposed to a NO post oxidation annealling. 

The detailed physical models required to capture all described effects have been implemented in the open-source simulator Comphy \cite{Comphy_Rzepa2018,Comphyv3_Waldoer2023}. For simplicity, the elementary charge has been set to unity in this work.
\subsection{Gate leakage current in MOSC}
\begin{figure}
\centerline{\includegraphics[width=0.96\linewidth]{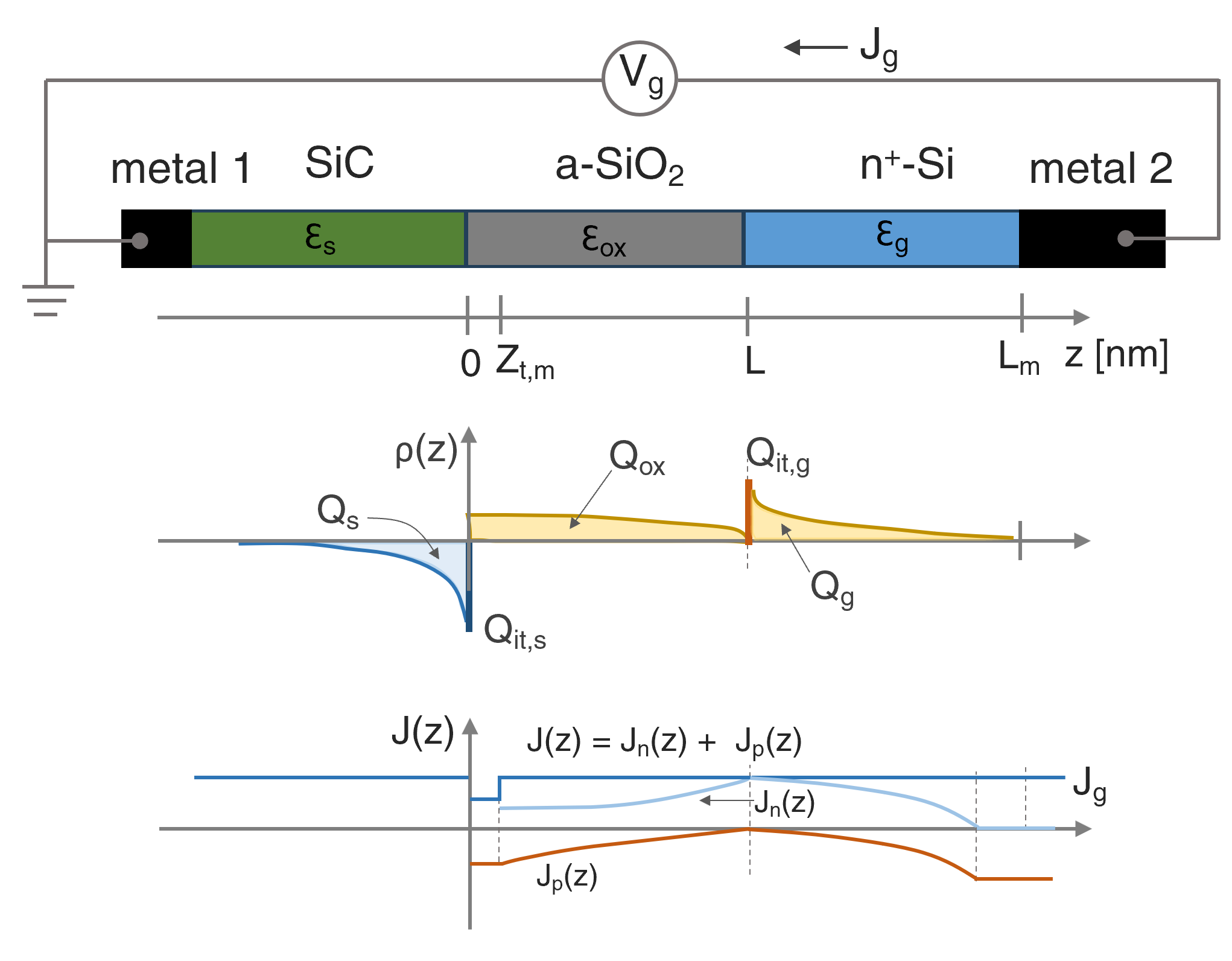}}
\caption{The 1D MOSC model for gate leakage current simulation. Top: External circuit used for analysis, with voltage and conventional current (arrow) defined as positive. Middle: Illustrative space-charge profile $\rho(z)$ showing oxide charge trapping and possible interface charges   at the SiC/a‑SiO$_2$ and n$^+$-Si/a‑SiO$_2$ interfaces, respectively. Bottom: Illustrative current density profile $J(z)$ across the MOSC, indicating Fowler-Nordheim tunneling near $z_\text{tm}$ and amplification of electron and hole currents via impact ionization.}
\label{fig_1dmoscap}
\end{figure}

Fig. \ref{fig_1dmoscap} depicts the one-dimensional MOSC (1D MOSC) model used to analyze the gate leakage current. An a-SiO$_2$ layer of thickness $L$ separates an n-type 4H-SiC from an n$^+$-Si gate, each contacted by a metal electrode.  A time-dependent gate bias $V_\text{g}(t)$ is applied to gate electrode, and the gate leakage current $J_\text{g}(t)$ is measured in the external circuit. Upon assuming translational invariance along the lateral directions, the electrostatic potential $\varphi(z,t)$ obeys the 1D Poisson equation
\begin{equation}
\frac{\partial}{\partial z}\left[\epsilon(z)\frac{\partial\varphi(z,t)}{\partial z}\right]=-\rho(z,t)
\label{eqn_Poissoneq_general}
\end{equation}
where $\epsilon(z)$ is the local permittivity and $\rho(z,t)$ comprises free-carrier and fixed charges. Solving Eq. \ref{eqn_Poissoneq_general} in the  SiC and n$^+$-Si regions yields the surface potential $\varphi_\text{s}(t)$, $\varphi_\text{g}(t)$ and their corresponding surface charges $Q_\text{s}(t)$, $Q_\text{g}(t)$, including any interface-trapped charges $Q_\text{it,s}(t)$, $Q_\text{it,g}(t)$. Enforcing the Kirchhoff's voltage law \cite{tsividisOperationModelingMOS2011} across the stack gives
\begin{subequations}
\begin{align}
\varphi_\text{s}(t)  -\frac{Q_\text{s}(t)+Q_\text{it,s}(t)}{C_\text{ox}} + \Delta V_\text{ox,s}(L,t) &= V_\text{g}(t) - \phi_\text{ms} +\varphi_\text{g}(t) \\ 
\varphi_\text{s}(t) + \frac{Q_\text{g}(t)+Q_\text{it,g}(t)}{C_\text{ox}}  + \Delta V_\text{ox,g}(0,t) &= V_\text{g}(t)- \phi_\text{ms} +\varphi_\text{g}(t)
\end{align}
\label{eqns_surfpot}
\end{subequations}
where $C_\text{ox}=\epsilon_\text{ox}/L$, $\phi_\text{ms}$   the metal-semiconductor work-function difference, and $\Delta V_\text{ox,s}(L,t)$ [$\Delta V_\text{ox,g}(0,t)$] the voltage drop across the oxide due to trapped charges that is projected to the  SiC/a-SiO$_2$ (n$^+$-Si/a-SiO$_2$) interface. When charge trapping in bulk oxide is significant, the potential profile must be solved self-consistently with the trap-occupancy dynamics.

According to the continuity equation, the gate leakage current $J_\text{g}(t)$ can be expressed as
\begin{equation}
J_\text{g}(t) = J_n(L^\text{-})+\frac{\partial (Q_\text{g}+Q_\text{it,g})}{\partial V_\text{g}}\frac{\mathrm{d} V_\text{g}}{\mathrm{d} t} 
\label{eqn_Jgt_ful}
\end{equation}
in which the first term is the \textit{net} electron current arriving at the n$^+$-Si/a-SiO$_2$ interface, and the second term is the displacement current with  $\partial (Q_\text{g}+Q_\text{it,g})/\partial V_\text{g}$ representing the incremental capacitance of the gate side . In typical quasi-static measurement, the displacement current can be ignored due to the long-enough integration time. In this work, both charg capture (emission) and trap creation at the interface have been neglected. Hence, one obtains
\begin{equation}
J_\text{g}(t) \approx  J_\text{n}(L^\text{-}).
\label{eqn_Jg_JnL}
\end{equation}
where  $J_\text{n}(L^\text{-})$ is the electron current arriving at the n$^+$-Si/a-SiO$_2$ interface.

Under forward bias, the $J_\text{g}(t)$ is entirely determined by $J_\text{n}(L^\text{-})$ because hole current  $J_\text{p}(L^\text{-})$ is negligible. Physically, $J_\text{n}(L^\text{-})$ arises from (i) quantum injection of electrons from the conduction band of SiC conduction band into that of a-SiO$_2$ and (ii) field-driven transport and amplification via impact-ionization in a-SiO$_2$. A brute-force quantum mechanical solution across tens of nanometers of oxide is computationally prohibitive, especially when charge trapping must also be included in the oxide at every disctetized time step. Therefore, we use a ``tunneling + transport" approach \cite{tunnelingptransp_FilipJVSTB1999, Fischetti_shortcutpaperJAP1995}, which is supported by experimental transport studies \cite{ballistictransportaSiO2_DiMariaPRL1986, BallisticTrans_FischettiPhysRevB1987}. First, the energy distribution function (EDF) of electrons that just tunnel into the CB of a-SiO$_2$ will be calculated by a simplified tunneling model (see sec. \ref{sec:tunneling}). Second, the evolution of EDFs and amplification of the electron current is obtained by solving a simplified 1D Boltzmann transport equation (BTE) where impact ionization is also incorporated (see Sec. \ref{sec:transport}). The trapped charges in a-SiO$_2$ influence the tunneling and transport by altering the potential profile $\varphi(z)$. Within this framework, the electron current and the electron concentration are, respectively,
\begin{subequations}
\begin{align}
J_\text{n}(z,t) &= n_\text{ox}(z,t)v_\text{d}(z,t)\\
n_\text{ox}(z,t) &= n_\textsc{fn}(z_\text{t,m})\times\exp\left(\int_{z_\text{t,m}}^z\gamma_\text{ox}(x)\mathrm{d}x\right)
\end{align}
\label{eqn_elcurrent}
\end{subequations}
in which $v_\text{d}(z)$ is the average drift velocity, $\gamma_\text{ox}(z)$  the electron-initiated impact ionization coefficient in a-SiO$_2$ and $n_\textsc{fn}(z_\text{tm})$  the concentration of electrons that is injected into the CB of a-SiO$_2$ by Fowler-Norheim tunneling at the maximum classical turning point $z_\text{tm}$. The hole initiated impact ionization in a-SiO$_2$ has been neglected because of the high relaxation rate of (hot) holes \cite{holeIIquartz_KunikiyoJAP941096} and the low hole mobility in a-SiO$_2$ \cite{holemobiaSiO2_HughesAPL264362}. Hence, the hole current and hole concentration can be obtained as
\begin{subequations}
\begin{align}
J_\text{p}(z,t) &= J_\text{n}(L^\text{-})\left[1-\exp\left(-\int_{z}^L\gamma_\text{ox}(x)\mathrm{d}x\right)\right] \\
p_\text{ox}(z,t) &= J_\text{p}(z,t)/v_\text{h}(z,t)
\end{align}
\label{eqn_holecurrent}
\end{subequations}
where $v_\text{h}(z,t)$ is the average drift velocity of holes. The subsequent capture of those holes in bulk a-SiO$_2$ reshapes the electrostatic potential and hence indirectly contributes to the gate leakage current. 

\subsection{Electrostatics}
\label{sec:electrostatics}
The potential profile $\varphi(z)$ of the 1D MOSC spans three regions, i.e., the SiC substrate, the oxide a-SiO$_2$, and the n$^+$-Si gate. The n$^+$-Si gate is treated as a heavily doped semiconductor rather than a metal. In the SiC and n$^+$-Si, the space charge space-charge density arises from both free carriers and ionized dopants. Within the oxide, mobile carriers are negligible and the charge density is dominated by trapped charges. 
\subsubsection{Electrostatics of SiC and n$^+$-Si}
Upon assuming thermal equilibrium, the electrostatic potential in each semiconductor region satisfies
\begin{equation}
\epsilon_s\frac{\mathrm{d}^2\varphi(z)}{\mathrm{d}z^2}=-\left[p(\varphi)-n(\varphi) + N_\text{d}^{+}(\varphi)-N_\text{a}^-(\varphi)\right]
\label{Poissoneq}
\end{equation}
where $n(\varphi)$ and $p(\varphi)$ denote the electron and hole concentrations, $N_\text{d}^{+}(\varphi)$ and $N_\text{a}^-(\varphi)$ are the concentrations of ionized donors and acceptors, respectively.  It is noteworthy that charged point defects in semiconductors can be modeled as donors or acceptors according to their levels in the band gap. Solving this equation requires models for both carriers statistics and dopant ionization as a function of $\varphi$.

Unpon assuming thermal equilibrium, the free-carrier concentrations follow directly from Fermi-Dirac (FD) statistics applied to the band-structure and the corresponding density of states (DOS). Denoting by $E_\parallel$ and $E_\perp$ the kinetic energies associated with motion parallel to $z$ ($k_\parallel$) and within the $k_x-k_y$ plane ($k_\perp$), we obtain
\begin{subequations}
\begin{align}
n(\varphi) &= \sum_i \iint_0^{\infty}\frac{g_{\text{c},i}(E_\parallel, E_\perp)\mathrm{d}E_\parallel \mathrm{d}E_\perp}{1+\exp[\beta(E_\parallel+E_\perp+E_{\textsc{cf},i}-\varphi)]}, \\
p(\varphi) &= \sum_j \iint_0^{\infty}\frac{g_{\text{v},j}(E_\parallel, E_\perp)\mathrm{d}E_\parallel \mathrm{d}E_\perp}{1+\exp[\beta(E_\parallel+E_\perp-E_{\textsc{fv},j} + \varphi)]},
\end{align}
\label{eqn_elhconc}
\end{subequations}
where $\beta = 1/k_\textsc{b}T, E_{\textsc{cf},i}=E_{\textsc{c},i}-E_\textsc{f}$, and $ E_{\textsc{fv},j}=E_\textsc{f}-E_{\textsc{v},j}$. The index $i$ ($j$) runs over conduction (valence) subbands, each with DOS function $g_{\text{c},i}$ ($g_{\text{v},j}$). In 4H-SiC with $k_\parallel = [001]$ and $k_\perp$ in the $\Gamma$MK plane, the conduction band comprises two Kane-type nonparabolic subbands separated by a 0.115 eV mini-gap, while the valence band is represented by three parabolic subbands, as illustrated in Fig. \ref{fig_SiCbandstruct}. Details of the band models for a-SiO$_2$ and n$^+$-Si are given in Appendix \ref{sec:App_bandstructure}.

\begin{figure}
\centerline{\includegraphics[width=\linewidth]{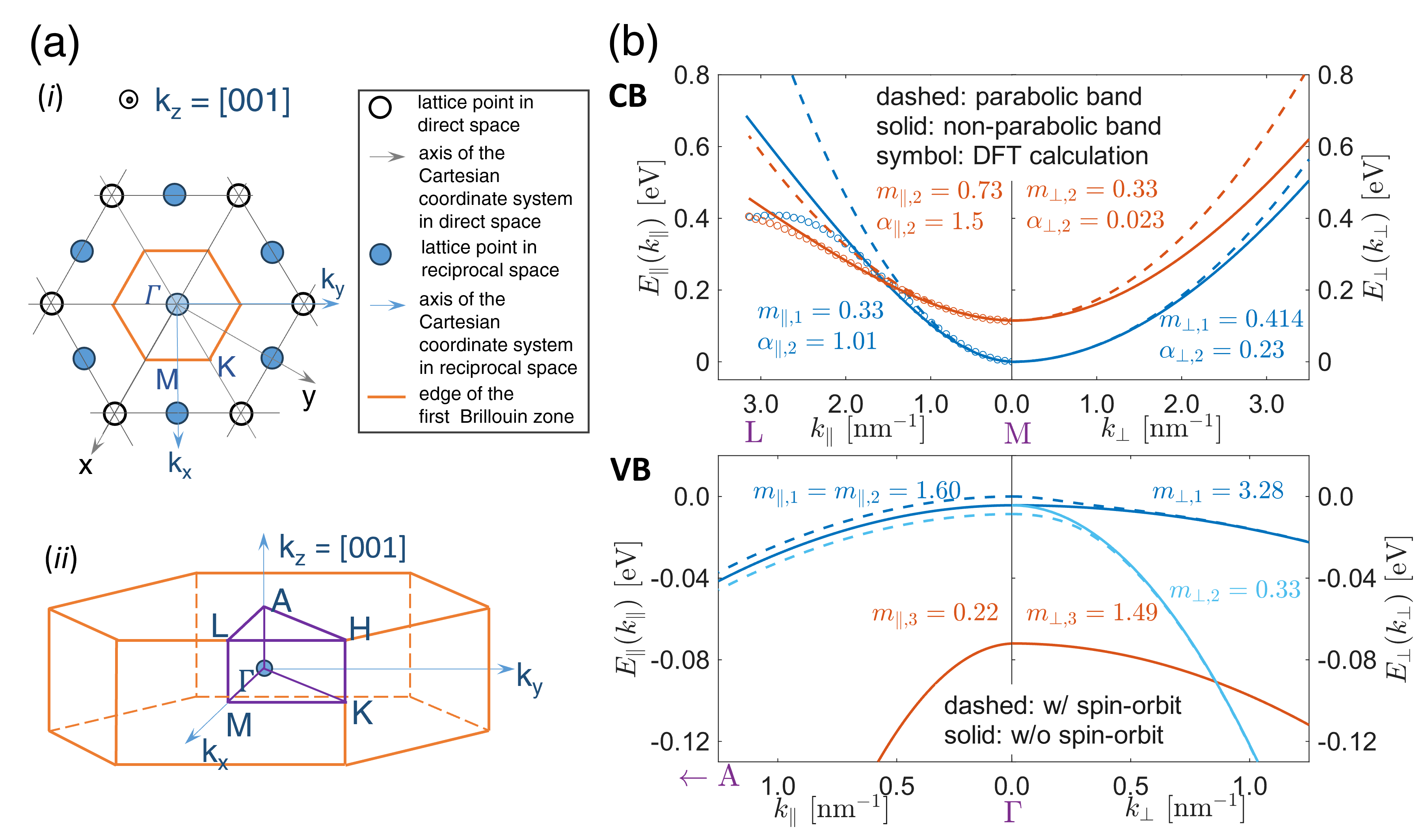}}
\caption{Band structure of 4H-SiC. (a) Crystal structure representation: (i) (0001) plane showing lattice points in direct space (open circles) and reciprocal space (filled circles), with first Brillouin zone boundary (orange hexagon) and Cartesian coordinate axes; (ii) three-dimensional view of the first Brillouin zone highlighting high-symmetry points ($\Gamma$, M, K, L, A, H) and crystallographic direction $k_z = [001]$. (b) Energy dispersion along high-symmetry directions: (top) conduction band modeled by two Kane-type non-parabolic bands ($\Delta E = 0.115$ eV) with effective masses and non-parabolicity parameters annotated, showing excellent agreement with DFT calculations (symbols) \cite{massDFTplusexp_KaczerPRB1998}; (bottom) valence band structure comprising heavy-hole , light-hole, and split-off hole bands, illustrating spin-orbit coupling [dashed (slod) lines include (exclude) spin-orbit coupling] as reported in Ref. \cite{VBofSiC_PersonLindefeltJAP1997}. Note $k_\parallel = [001]$, while $k_\perp$ is in the $\Gamma$MK plane.}
\label{fig_SiCbandstruct}
\end{figure} 

Each dopant species at crystallographic site $i$ (donor) or $j$ (acceptor) ionizes according to the local potential $\varphi$ and the site's ionization energy $E_{\text{d},i}$ or $E_{\text{a},j}$. The thermal-equilibrium ionization fractions are
\begin{subequations}
\begin{align}
\frac{N_{\text{d},i}^{+}(\varphi)}{N_{\text{d},i}} &= 1 - \frac{b(N_\text{d})}{1 + g_{\text{d}}^{-1}\exp\left\{-\beta\left[ E_{\text{d},i}(N_\text{d})-E_\textsc{cf} + \varphi\right]\right\}},\\
\frac{N_{\text{a},j}^{-}(\varphi)}{N_{\text{a},j}} &= 1 - \frac{b(N_\text{a})}{1 + g_{\text{a}}^{-1}\exp\left\{-\beta\left[ E_{\text{a},j}(N_\text{a})-E_\textsc{fv} - \varphi\right]\right\}},
\end{align}
\label{eqn_ionized_dopants}
\end{subequations}
where $g_\text{d}$ ($g_\text{a}$) is the donor (acceptor) degeneracy factor, and  $b(N)$ accounts for dopant clustering effects. Here the ionization energy $E_{\text{d},i}(N_\text{d})$ and  $E_{\text{a},j}(N_\text{a})$ may shift with local concentration via band-gap narrowing or Coulomb interactions. The Fermi level offset $E_{\textsc{cf}}$ ($E_{\textsc{fv}}$) is  defined as the separation of the Fermi level $E_\textsc{f}$ from the CB minimum $E_\textsc{c}$ (VB maximum $E_\textsc{v}$)
\begin{equation}
E_{\textsc{cf}}(T) = E_{\textsc{c}}(T)-E_\textsc{f}(T),E_{\textsc{fv}}(T) = E_\textsc{f}(T)-E_{\textsc{v}}(T).
\label{eqn_Fermilevel_offset}
\end{equation}
Both the offsets and the ionized-dopant concentrations are obtained by enforcing global charge neutrality, so their temperature dependence follows directly from the band-structure and ionization models. Detailed parameterizations of $E_{\text{d},i}$ or $E_{\text{a},j}$ and $b(N)$ are given in the Appendix \ref{sec:APP_incompleteionization}.

\subsubsection{Electrostatics in  a-SiO$_2$}
The space charge in the oxide 
\begin{equation*}
\rho_\text{ox}(z,t) = n_\text{ox}(z,t) + p_\text{ox}(z,t) + \rho_\text{ox,t}(z,t).
\end{equation*}
comprises electrons in the CB $n_\text{ox}(z)$, holes in the VB $p_\text{ox}(z)$, and charges trapped in the oxide $n_\text{ox,t}(z)$. The charges in the intervals $(0,z)$ and $(z,L)$ lead to potential shifts relative to the SiC/a-SiO$_2$ and the n$^+$Si /a-SiO$_2$ interfaces, respectively, 
\begin{subequations}
\begin{align}
\Delta V_\text{ox,s}(z,t) &= \frac{1}{\epsilon_\text{ox}}\int_0^{z}(z-x)\rho_\text{ox}(x,t)\mathrm{d}x,\\
\Delta V_\text{ox,g}(z,t) &= \frac{1}{\epsilon_\text{ox}}\int_L^{z}(z-x)\rho_\text{ox}(x,t)\mathrm{d}x
\end{align}
\label{eqn_dVox}
\end{subequations}
which enter Eq. \ref{eqns_surfpot}. We update $\rho_\text{ox}(z,t)$ in successive quasi-static time steps: $n_\text{ox}(z,t)$ and $p_\text{ox}(z,t)$ directly follow from the transport solution (see Sec. \ref{sec:transport}), while $\rho_\text{ox,t}(z,t)$ is governed by the trapping dynamics.

\begin{figure}
\centering
\includegraphics[width=0.98\linewidth]{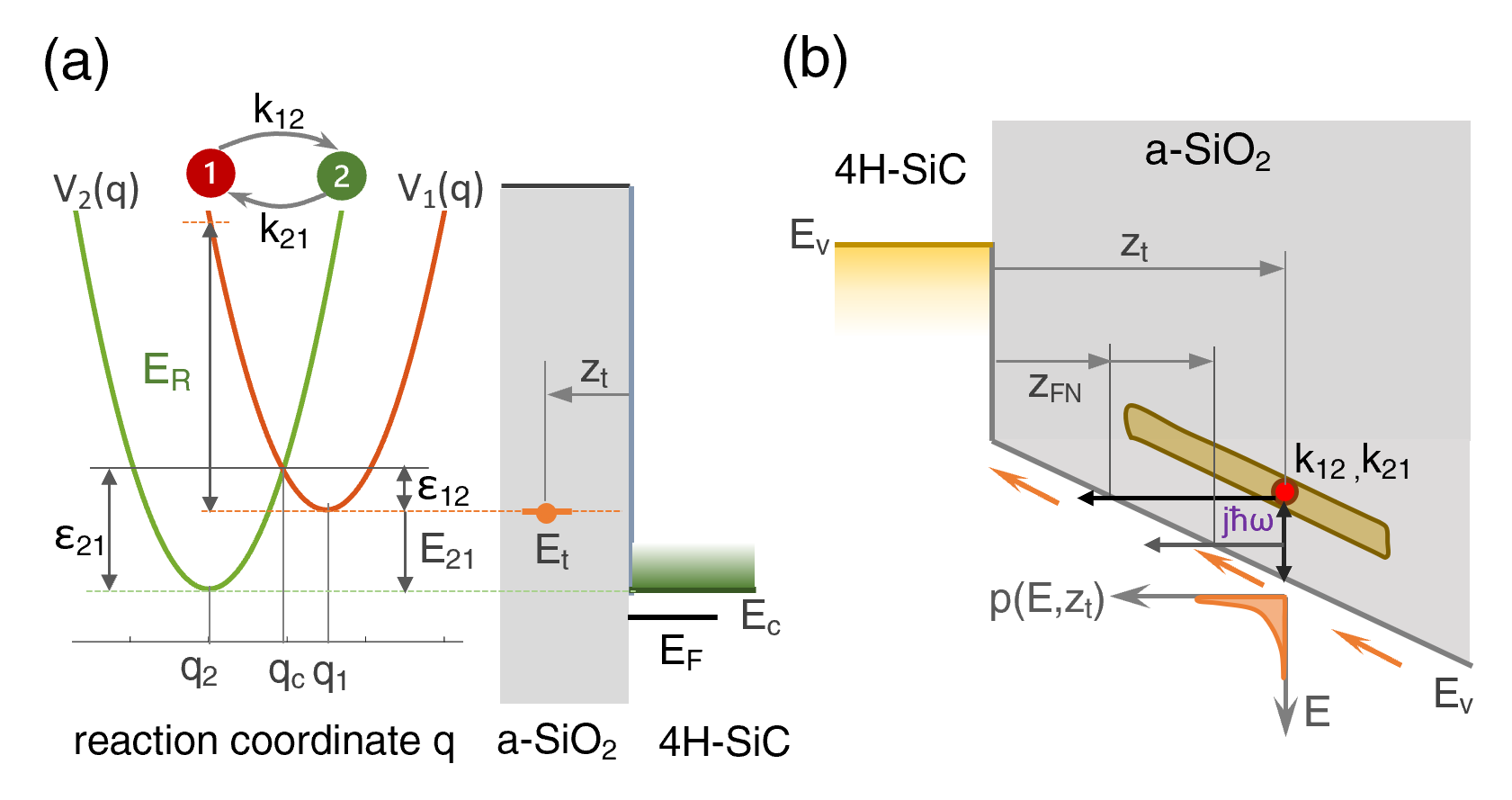}
\caption{Nonradiative multiphonon–mediated charge transitions in a-SiO$_2$. (a) Configuration–coordinate diagram for a two-state defect: parabola 1 represents the potential-energy surface when the electron is bound in the trap, parabola 2 when the electron occupies the conduction band. Their intersection defines the capture barrier  $\varepsilon_{21}$, the emission barrier $\varepsilon_{12}$, and the relaxation energy $E_\textsc{r}$. (b) Band-diagram view of a trap-band exchange in the oxide. A defect at energy $E_\text{t}$ and depth $z_\text{t}$ may undergo a capture/emission event to the oxide conduction/valence band locally at $z_\text{t}$ (multiple-trapping), and/or non-locally at the Fowler-Nordheim turning point $z_\textsc{fn}$ (phonon-assisted tunneling).}
\label{fig_chargecapture}
\end{figure}

The concentration of trapped oxide charges is obtained by integrating the product of the local DOS $g_\text{t}(E_\text{t},z,\text{P})$ and the corresponding occupation function $f(E_\text{t},z,\text{P},t)$ over trap energy $E_\text{t}$ and other trap parameter set $\text{P}$
\begin{equation*}
\rho_\text{ox,t}(z,t) = \iint g_\text{t}(E_\text{t},z,\text{P})f(E_\text{t},z,\text{P},t)\mathrm{d}E_\text{t}\mathrm{d}\text{P}
\end{equation*}
Often, $z$ and $\text{P}$ will be omitted for clarity in the following. Each trap is modeled as a two-state system: state 1 corresponds to an electron-filled defect, and state 2 to an empty defect. The electron occupation $f(E_t,t)$ function satisfies
\begin{equation}
\frac{\partial f(E_\text{t},t)}{\partial t} = -k_{12}(E_\text{t})f(E_\text{t},t)+[1-f(E_\text{t},t)]k_{21}(E_\text{t})
\end{equation}
where $k_{21}(E_\text{t})$ and $k_{12}(E_\text{t})$ are the capture and emission rate, respectively, and are dependent on temperature and the local potential  $\varphi(z)$ \cite{grasserStochasticChargeTrapping2012}. As illustrated in Fig. \ref{fig_chargecapture}, a charge in a trap in the oxide can exchange charge with three types of reserviors: (i) the CB and/or VB of a semiconductor/metal across the interface with the oxide \cite{shinModelingCarrierTrapping2025} (Fig. \ref{fig_chargecapture}a), (ii) the CB and/or VB of the oxide itself (Fig. \ref{fig_chargecapture}b) \cite{multipletrapping_Curtis},  and (iii) another trap in the oxide \cite{TATpartII_Schleich2022,schleichSingleMultiStepTrap2022}, which is not considered in the present study. 

All these charge-exchange processes are mediated by nonradiative multiphonon (NMP) processes \cite{grasserStochasticChargeTrapping2012,grasserBiasTemperatureInstability2014}. The potential energy surfaces (PES) for the two charge states of a defect are often  modeled as 1D parabolas as a function of a generalized reaction coordinate $q$, as shown in Fig. \ref{fig_chargecapture}a. For defects in a-SiO$_2$, the curvature ratio is assumed to be unity, consitent with linear electron-phonon coupling \cite{goesIdentificationOxideDefects2018}.  In the the high-temperature-strong-coupling limit, the state-to-state transition probabilities are, respectively,
\begin{subequations}
\begin{align}
k_{12,0}(E_\text{t}) &= \frac{c_0 D(E_\text{t},z_t)}{\sqrt{\beta E_\textsc{r}}}\exp\left(-\beta\varepsilon_{12}\right),\\
k_{21,0}(E_\text{t}) &= \frac{c_0 D(E_\text{t},z_t)}{\sqrt{\beta E_\textsc{r}}}\exp\left(-\beta\varepsilon_{21}\right)
\end{align}
\end{subequations}
in which $D(E_\text{t})$ is the tunneling factor, $E_\textsc{r}$ the relaxation energy, and $c_0$ an energy and temperature independent constant \cite{huangTheoryLightAbsorption1950, kuboApplicationMethodGenerating1955,goesIdentificationOxideDefects2018}. In addition, $\varepsilon_{12}$ and $\varepsilon_{12}$ are the activation energies for the $1\rightarrow2$ and $2\rightarrow1$ transitions, respectively, and for linear electron-phonon coupling they are
\begin{equation}
\varepsilon_{12} = \frac{\left(E_{21}+E_\textsc{r}\right)^2}{4E_\textsc{r}},
\varepsilon_{21} =\varepsilon_{12} -E_{21}
\label{eqn_barrierNMP}
\end{equation}
where $E_{21}=V_2(q_2)-V_1(q_1)$ is the PES energy difference \cite{grasserStochasticChargeTrapping2012}. When a defect interacts with charge reservoirs across the interface, the total rates can be obtained according to the band-edge approximation \cite{Comphy_Rzepa2018}. For example, the emission and capture rates for interaction with the CB of a semiconductor are respectively
\begin{subequations}
\begin{align}
k_{12}(E_t) &= \exp\left[\beta \left(E_\textsc{cf}-\varphi_\text{s}\right)\right]n(T,\varphi_\text{s})v_\text{th}(T)\sigma_\text{s}k_{12,0}(E_\text{t})\\
k_{21}(E_\text{t}) &= n(T,\varphi_\text{s})v_\text{th}(T)\sigma_\text{s}k_{21,0}(E_\text{t})
\end{align}
\end{subequations}
in which $n(T,\varphi_\text{s})$ is the electron concentration at the interface, $v_\text{th}(T)$ the thermal velocity of electrons and $\sigma_\text{s}$ the effective cross-section \cite{rzepaEfficientPhysicalModeling2018}. 

When a trap exchange with the CB and/or VB of the oxide, it can take place locally and non-locally as illustrated in Fig. \ref{fig_chargecapture}b. A trap can exchange charge locally with the CB and VB at the same spatial location $z$ for which the tunneling factor $D(E_\text{t},z_\text{t})$ in $k_{12,0}$ and $k_{21,0}$ are both unity. Such a model is referred to as multiple trapping (MT) model, and has been able to explain the features of experimental hole mobility in a-SiO$_2$ \cite{multipletrapping_Curtis}. The charge exchange can also occur via tunneling through a finite potential barrier via an NMP transition, a process often termed phonon-assisted tunneling \cite{makram-ebeidQuantumModelPhononassisted1982}. Considering the charge exchange with the VB of a-SiO$_2$, we can write the total emission and capture rates as,
\begin{subequations}
\begin{align}
k_{12}(E_\text{t}) &= \frac{c_0\sigma_\text{ox,h}}{\sqrt{\beta E_\textsc{r}}}\Big[\exp(-\beta\varepsilon_{12})J_\text{p}(z)  \nonumber\\ 
& {  } +\sum_{j=-n}^{j=n}J_\text{p}(z_{\textsc{fn}})\frac{3}{2}\Theta_j \exp\left(-\beta\varepsilon_{12,j}-2\Theta_j\right)\Big] \\ 
k_{21}(E_\text{t}) & = \frac{c_0\sigma_\text{ox,h}\bar{N_\text{p}}}{\sqrt{\beta E_\textsc{r}}}\Big[\exp(-\beta\varepsilon_{21})  \nonumber\\ 
&{ } +\sum_{j=-n}^{j=n}\frac{3}{2}\Theta_j \exp\left(-\beta\varepsilon_{21,j}-2\Theta_j\right)\Big]
\end{align}
\label{eqn_k12k21MT}
\end{subequations} 
where the hole fluxes $J_\text{p}(z)$ and $J_\text{p}(z_\textsc{fn})$ indicate the local and non-local transitions, respectively, and $\bar{N_\text{p}}$ characterizes an averaged flux of electrons in the VB. The barriers and phase-factor $\Theta_j$ are as follows,
\begin{align*}
\varepsilon_{12,j} &= \left(E_\textsc{r}-j\hbar\omega\right)^2/4E_\textsc{r},\varepsilon_{21,j} = \varepsilon_{12,j}+j\hbar\omega, \\
\Theta_j &= \frac{1}{\hbar}\int_{z_{\textsc{fn}}}^{z_t}\sqrt{2m_{\text{ox},h}\big|E_{21}(z)+j\hbar\omega\big|}\mathrm{d}z,\\
 E_{21}(z) &= E_\text{v}(z)-E_t(z).\\
\end{align*}
Only one phonon-mode of single frequency $\omega$ is considered for the phonon-assisted tunneling processes, which are manifested in $\Theta_j\exp(-2\Theta_j)$ in Eq. \ref{eqn_k12k21MT}.

\subsection{Quantum Tunneling Models}
\label{sec:tunneling}

We restrict our attention to electron tunneling from the SiC conduction band into the conduction band of the gate dielectric (a-SiO$_2$), since this process dominates gate leakage in SiC MOSFETs; hole tunneling follows the same principles. In most conventional treatments, the tunneling current density is computed via Bardeen’s transfer-Hamiltonian formalism \cite{Tunneling_BardenPRL1961} combined with the Tsu–Esaki approach \cite{Tunneling_TsuEsaki}, and the transmission probability is taken from Fowler–Nordheim or Gundlach analytic expressions for triangular or trapezoidal barriers \cite{FowlerNordheim1928, trapezoidbarrier_Gundlach}, respectively. Two key approximations underlie these methods, yet both break down for reliability assessment of thick (tens of nanometers) oxides in power devices. First, semiconductor carriers in the inversion or accumulation layer are modeled as free-electron gases, neglecting the quantization of subbands that arises under strong surface fields. Second, the tunneling is assumed to be scattering-free such that inelastic processes, which are unavoidable for transport at high electric field,  are ignored completely.

To overcome these limitations, we introduce a two-stage ``tunneling + transport'' framework. In the first stage, we map the confined inversion/accumulation layer states onto an equivalent energy-resolved distribution of free electrons incident on the oxide, and calculate their transmission into a-SiO$_2$ using a modified transfer-Hamiltonian approach that includes image-force corrections and electric-field-induced band bending. In the second stage, the EDF of tunneled electrons is advanced through the oxide by a 1D BTE, which explicitly accounts for electron-phonon scattering and impact ionization (see Sec. \ref{sec:transport}).

\subsubsection{Tunneling of electrons in the space-charge region}
\begin{figure}
\centerline{\includegraphics[width=\linewidth]{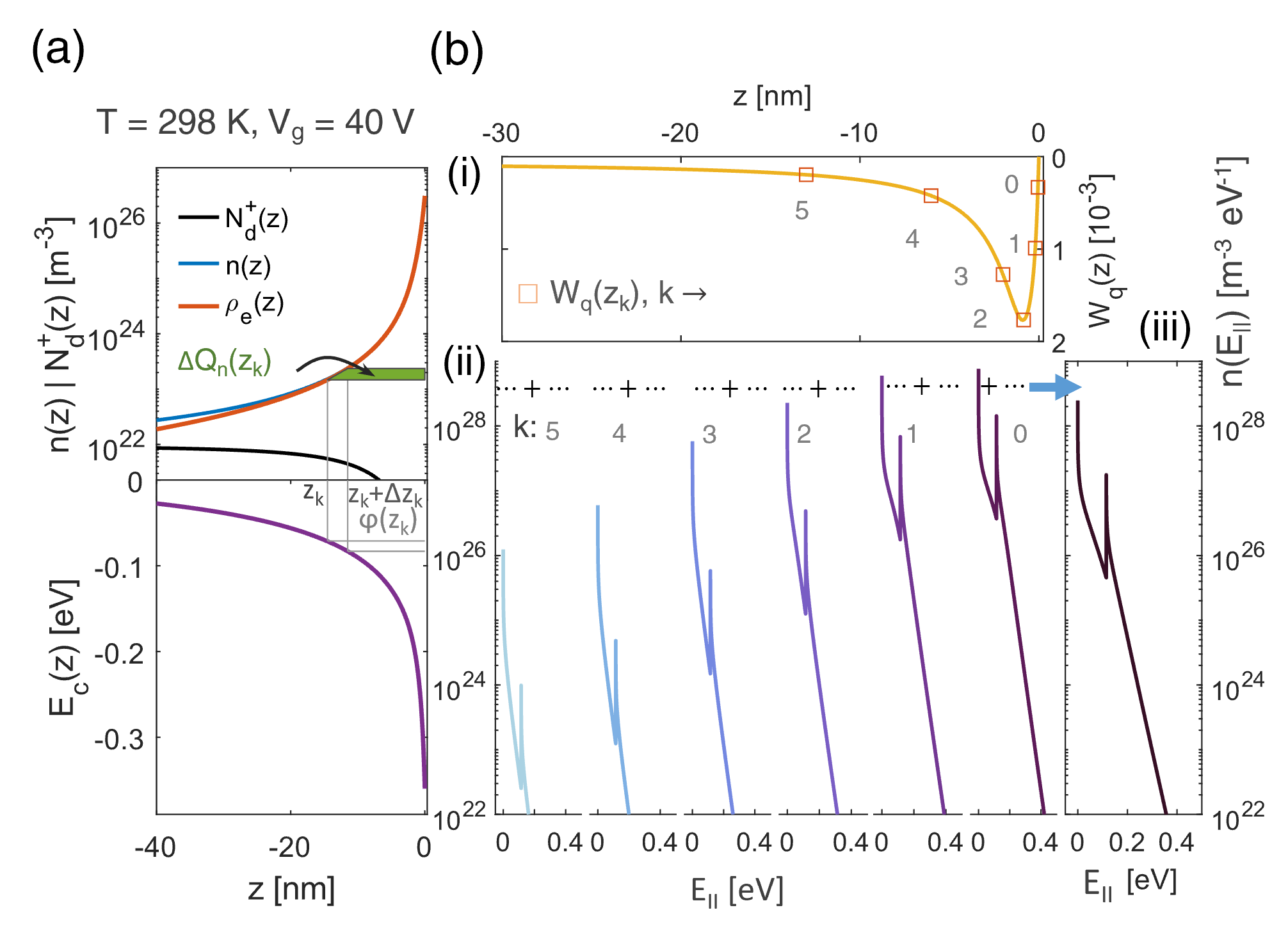}}
\caption{Methodology for modeling tunneling from the semiconductor space-charge region into the insulator. (a) Device electrostatics at $V_\text{g} = 40$ V, T = 298 K: carrier and ionized dopant profiles (top) under band bending (bottom), showing local carrier contributions $\Delta Q_n(z_k)$ to the surface charge. (b) Implementation of a weighting scheme for tunneling: (i) weight function $W_q(z)$ derived from surface charge contributions, (ii) energy-resolved electron distributions $n(E_\parallel,z)$ at selected depths, (iii) total weighted distribution for tunneling calculation. }
\label{fig_illustrationFN}
\end{figure}

In a fully quantum-mechanical treatment, the inversion or accumulation layer at the semiconductor/oxide interface is described by the solutions ($\Psi_i,E_{\parallel,i}$) of the coupled Poisson-Schrodinger equation \cite{invlayer_Stern1972,SPequation_nonuniform, pictureQBS_RegisterAPL1999, tunnelingmodel_Gehring}. Each eigenstate $\Psi_i$ contributes a two-dimensional sheet charge density $n_i(E_{\parallel,i})$ at energy $E_{\parallel,i}$ that can tunnel through the oxide barrier \cite{shinModelingCarrierTrapping2025}. In a classical treatment,  electrons are assumed to be plane waves that can tunnel through a barrie and the associated concentration is determined by the DOS. To marry these two viewpoints, we require an energy-resolved sheet density $n(E_\parallel, E_\perp)$ from the 3D electron concentration $n(E_\parallel, E_\perp,z)$ in the space charge region  where the influence of band-bending is retained via the local potential $\varphi(z)$.  Therefore, we implement this mapping by the transformation
\begin{subequations}
\begin{align}
n(E_\parallel,E_\perp) &= \int_{z(\varphi=0)}^{z(\varphi_s)} n(E_\parallel, E_\perp, \varphi) W_q[\varphi(z)] \mathrm{d}z,\\
n(E_\parallel) &= \int_0^{E_\perp}n(E_\parallel,E_\perp) \mathrm{d}E_\perp
\end{align}
\label{eqn_eleconc4tunnel}
\end{subequations}
where the weight function $W_\mathrm{q}[\varphi(z)]$ is a normalized weight that captures the fraction of the total surface charge arising from depth $z$. As illustrated in Fig. \ref{fig_illustrationFN}a,  $W_\mathrm{q}[\varphi(z)]$ is often built directly from the self-consistent Poisson solution under bias. The space-charge region is divided into thin slabs  $[z_k,z_k+\Delta z_k]$, and the incremental sheet-charge contributed by the slab at $z_k$ can be approximated as
\begin{equation}
\Delta Q_n[\varphi(z_k)]\approx \left\{\rho_e[\varphi(z_k+\Delta z_k)]-\rho_e[\varphi(z_k)] \right\}z_k.
\end{equation}
in which the net negative space-charge density at $z$ is defined as
\begin{equation}
\rho_e[\varphi(z)] = n[\varphi(z)]-\sum_{j=0}^MN_{\text{d},j}^{+}[\varphi(z),j].
\end{equation}
where all $M$ distinctive dopants (sites) are considered. After normalization with respect to all $z_k$, the weight function is expressed as 
\begin{equation}
W_\mathrm{q}[\varphi(z_k)] = \frac{\Delta Q_n[\varphi(z_k)]}{\sum_{k} \Delta Q_n[\varphi(z_k)]}
\end{equation}
such that $\sum_{k}W_\mathrm{q}(z_k)=1$. The $W_\mathrm{q}$ for a MOSC at $V_\text{g} = 40 $V at 298 K is shown in Fig. \ref{fig_illustrationFN}b, which suggests that only electrons that are 1-3 nm from the interface contribute significantly to the sheet charge. 

After this mapping, the electrons incident on the oxide can be treated as a continuum of free-electrons  parameterized by their longitudinal energy $E_\parallel$ (along $z$) and transverse energy $E_\perp$ (in-plane), i.e., $n(E_\parallel,E_\perp)$.  The concentration of electrons that actually tunnel into the CB of a-SiO$_2$ is
\begin{equation}
 n_\textsc{bte}(E_\parallel,E_\perp) = n(E_\parallel,E_\perp)D(E_\parallel,E_\perp)
\end{equation}
where $D(E_\parallel,E_\perp)$ is the tunneling coefficient.  $n_\textsc{bte}(E_\parallel,E_\perp)$ serves as the initial condition for the subsequent transport and impact ionization in the CB of a-SiO$_2$. 

\subsubsection{Tunneling coefficient}
\begin{figure}
\centerline{\includegraphics[width=\linewidth]{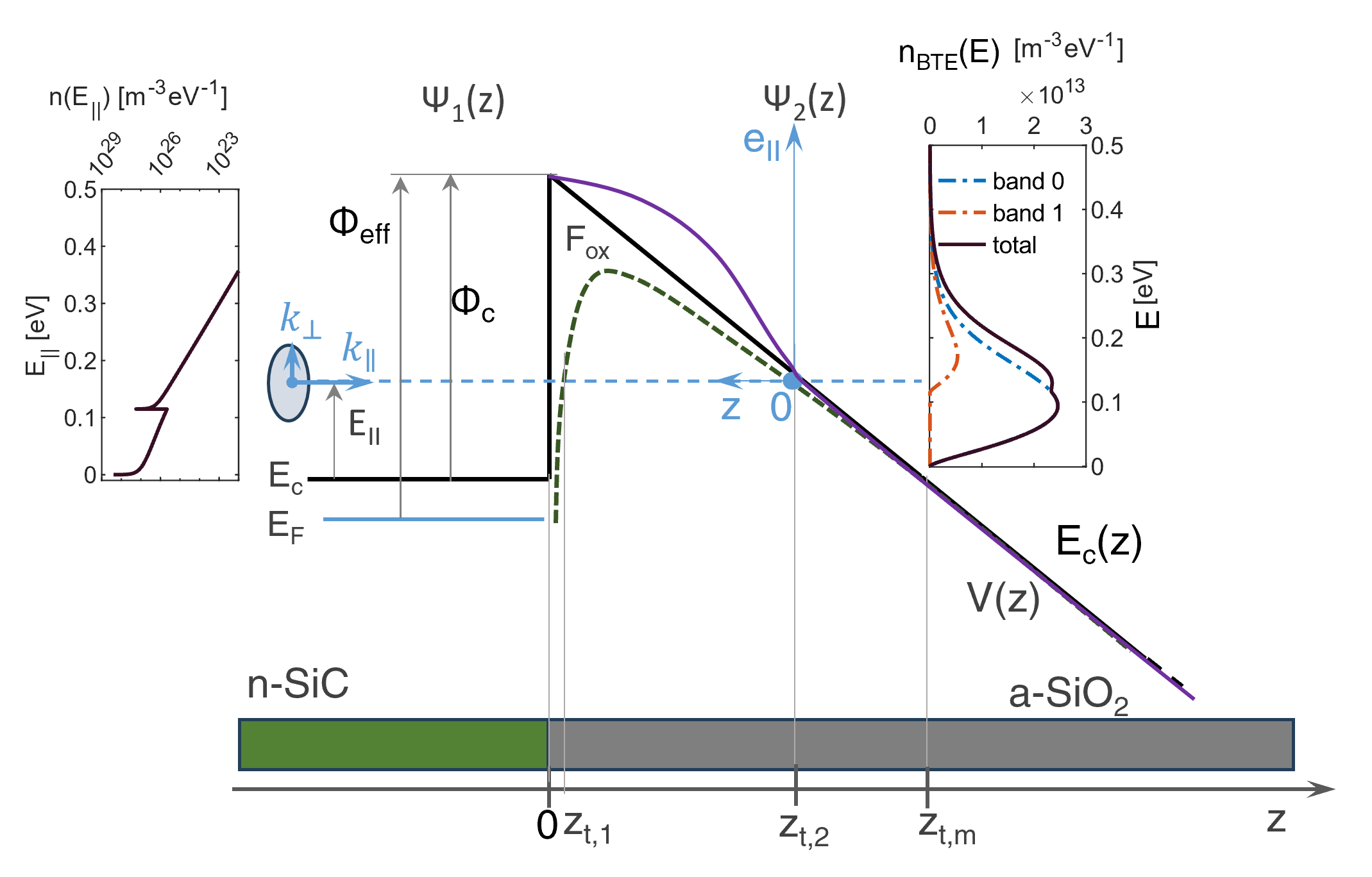}}
\caption{Methodology for calculating the tunneling coefficient. An electron is described as a plane wave $\Psi_1(\text{z})$ and can tunnel into the CB at the classical turning-point $z_\text{t,2}$, near which its wave-function is denoted as $\Psi_2(\text{z})$.  The variable $\text{z}$ is defined as The potential $V(\text{z})$ is referenced with respect to the CB of n-SiC. Note that $\text{z} = z_\text{t,2}-z$. The dashed olive (solid purple) curve depicts the $V(\text{z})$ due to image-force correction (charge trapping in the oxide) alone. The largest tunneling distance is depicted as $z_\text{t,m}$.}
\label{fig_TunnelingCoeff}
\end{figure}

The methodology of calculating the tunneling coefficient $D(E_\parallel,E_\perp)$ is illustrated in Fig. \ref{fig_TunnelingCoeff}, in which the variable z is defined as $\text{z} = z_\text{t,2}-z$.  According to the mapping Eq. \ref{eqn_eleconc4tunnel}, electrons in the space-charge region can be modeled as plane-waves  characterized by sheet density $n(E_\parallel,E_\perp)$ and momentum $(k_\parallel,k_\perp)$. The wave-function of these incident electrons penetrate into a-SiO$_2$'s forbidden band gap, whose band-structure for electrons is modeled by a spherical parabolic band with effective mass $m_\text{ox}$, and eventually leak into the CB of a-SiO$_2$ at the classical turning point $z_{\text{t},2}$.  The transmitted wave-function $\Psi_2(\text{z})$ follow the  Airy function with out-going characteristics for at least several nanometers.  Denoting the complex magnitudes of the incident and transmitted wave by $A$ and $C_\textsc{a}$ respectively, we write the tunneling coefficient \cite{tunnelingCoeff_FobesPRSA2011}
\begin{equation}
D(E_\parallel,E_\perp) = \frac{|C_\textsc{a}/\sqrt{\pi}|^2}{|A|^2},
\label{eqn_DExdef}
\end{equation} 
which denote the probability ratio of these two waves. Here, the scalar $\sqrt{\pi}$  accounts for z-dependence of the modulus $|\Psi_2(\text{z})|$. By matching the wave-functions at the SiC/a-SiO$_2$ interface ($\text{z}=z_\text{t,2}$) according to the  Ben Daniel-Duke boundary conditions \cite{BenDanielDukecondition_PhyRev1966}, we obtain the general tunneling coefficient as
\begin{subequations}
\begin{align}
&D(E_\parallel,E_\perp) =\nonumber\\
& \frac{4\pi^{-1}\omega^{-1}}{\frac{1}{\omega}\left[\text{Ai}(\xi)^2+\text{Bi}(\xi)^2 \right] + \omega \left[\text{Ai}^\prime(\xi)^2+\text{Bi}^\prime(\xi)^2 \right] + \frac{2}{\pi}},  \\
&\omega = \frac{m_{\parallel}\xi(\text{z})^{\prime}\bigr\rvert_{z_\text{t,2}}}{m_{\text{ox}}k_1}, k_1 = \frac{\sqrt{2m_{\parallel}}}{\hbar}\sqrt{E_{\parallel}\left(1+\alpha_{\parallel}E_{\parallel}\right)}
\end{align}
\label{eqn_TEx_full}
\end{subequations}
in which the variable $\xi(E_\parallel,E_\perp)$ can be calculated as \cite{bookAiryfunc_Olivier}:
\begin{subequations}
\begin{align}
&\xi(E_\parallel,E_\perp) = \text{sgn}(\text{z})\left[\frac{3}{2}\Theta(E_\parallel,E_\perp)\right]^{2/3},\\
&\Theta(E_\parallel,E_\perp) = \int_{0}^{z_\text{t,2}}k_{\parallel,\text{ox}}(E_\parallel,E_\perp,\text{z})\mathrm{d}\text{z}, \\
&k_{\parallel,\text{ox}}(E_\parallel,E_\perp,\text{z}) = \frac{\sqrt{2m_\text{ox}}}{\hbar}\times\nonumber\\
&\sqrt{\left|V(\text{z})-E_{\parallel}-E_{\perp}\left[1-\frac{m_{\perp}}{m_\text{ox}}\left(1+\alpha_{\perp}E_{\perp}\right)\right]\right|},\\
&V(\text{z}) = E_\text{c}(\text{z})+V_\text{img}(\text{z})
\end{align}
\label{eqn_xi_kparallel_ox}
\end{subequations}
Here, $E_\text{c}(\text{z})$ is the CB edge of a-SiO$_2$ and $V_\text{img}(z)$  the classical image-force potential; both are reference to the CB edge of SiC. The $E_{\perp}$ term in $k_{\parallel,\text{ox}}$ emerges from the conservation of total energy and the conservation of $k_\perp$ but without considering the mismatch of band structure at the SiC/a-SiO$_2$ interface \cite{tunnelingQBSPrefact_KriegerJAP1981, bandmismatch_WeinbergJAP1983}. 

For tunneling farther away from below the barrier maximum, the tunneling coefficient reads (Appendix \ref{sec:APP_DExdeeptunnel})
\begin{subequations}
\begin{align}
D(E_\parallel,E_\perp)&= \frac{\omega^{-1}\times P_\xi\exp[-2\Theta(z_\text{t,2})]}{1 + \frac{1}{2}P_\xi\exp[-2\Theta(z_\text{t,2})] + \frac{1}{4}\exp[-4\Theta(z_\text{t,2})]},  \\
P_\xi &= \frac{4\left(k_1/m_{\parallel}\right)\left(k_{\parallel,\text{ox}}/m_{\text{ox}}\right)}{\left(k_1/m_{\parallel}\right)^2 + \left(k_{\parallel,\text{ox}}/m_{\text{ox}}\right)^2}
\end{align}
\label{eqn_DExdeeptunnel}
\end{subequations}
which is strongly influenced by the phase-shift $\Theta(z_\text{t,2})$. The formula from the Wentzel-Kramers-Brillouin (WKB) approximation is
\begin{equation}
D(E_\parallel,E_\perp)_\textsc{wkb} = \omega^{-1}\exp[-2\Theta(z_\text{t,2})]
\label{eqn_DExWKB}
\end{equation}
which completely neglects the pre-factor $P_\xi$. 

Since the  $E_\perp$-dependence of  $D(E_\parallel,E_\perp)$ mainly arises from the mismatch of the effective masses at the SiC/a-SiO$_2$ interface, $D(E_\parallel,E_\perp)$ can be factorized as 
\begin{equation*}
D(E_\parallel,E_\perp)= D(E_\parallel,0)D_\perp(E_\parallel,E_\perp)
\end{equation*}
where $D(E_\parallel,0)$ is the zero-transverse-energy coefficient and $D_\perp(E_\parallel,E_\perp)$ contains the remaining variation with $E_\perp$.  Assuming electrons to follow a Maxwell-Boltzmann distribution with respect to $E_\perp$,  we obtain the averaged transverse-energy tunneling coefficient
\begin{equation*}
T_\perp(E_\parallel) = \beta^{-1}\int_0^{\infty} D_\perp(E_\parallel,E_\perp)\exp(-\beta  E_\perp)\mathrm{d}E_\perp.
\end{equation*}
Furthermore, the EDFs of electrons can be expressed with respect to $E_\parallel$ as follows
\begin{equation}
n_\textsc{bte}(E_\parallel) =  n(E_\parallel)D(E_\parallel, 0)T_\perp(E_\parallel)
\label{eqn_nEiifortunnel}
\end{equation}
from which the initial EDF for the 1D BTE $n_\textsc{bte}(E)$ can be obtained by properly converting $E_\parallel$ to the total energy $E$.

\subsubsection{Influence of charge-trapping and classical image-force correction}
Classical image-force interaction and  trapped charges in a-SiO$_2$ change the tunneling potential $V(\text{z})$ and thus alter the tunneling probability via $\Theta(E_\parallel,E_\perp)$, as illustrated in Fig. \ref{fig_TunnelingCoeff}.  Trapped charges in a-SiO$_2$ change the CB edge $E_\text{c}(\text{z})$ according to Eq \ref{eqn_dVox}. The potential for classical image-force interaction $V_\text{img}(z)$ for thick oxide is approximated as
\begin{equation}
V_\text{img}(\text{z}) = -\frac{\gamma_1}{4(z_\text{t,2}-\text{z})},\gamma_1  = \frac{1}{4\pi \epsilon_{\text{ox}}}\frac{\epsilon_{\text{s}}-\epsilon_{\text{ox}}}{\epsilon_{\text{s}}+\epsilon_{\text{ox}}}
\label{eqn_Vimg}
\end{equation}
which deviates only slightly from the full solution \cite{imageforceMOS_KleefstraJAP1980}. Quantum mechanical corrections for the divergence at $\text{z}\rightarrow z_\text{t,2}$ are not considered for $V_\text{img}(\text{z})$ \cite{Fischetti_shortcutpaperJAP1995}. For thin oxide, Schenk and Heiser proposed a pseudopotential method for image-force correction with trapezoidal barriers \cite{tunnelingthinoxide_Schenk},  but is less relevant here for thick oxide.  

The charge-trapping and image-force correction are decoupled in terms of their influence on the total potential $V(\text{z})$. It is therefore feasible to simplify the calculation of the phase shift $\Theta$ by approximating $E_\text{c}(\text{z})$ properly.  For a given $E_\parallel$, the CB edge $E_\text{c}(\text{z})$ can be approximated by connecting selected points on  $E_\text{c}(\text{z})$ with straight lines, for which the field constant. The zero$^\text{th}$-order approximation to $E_\text{c}(\text{z})$ is a straight line that connects $E_\text{c}(\text{z}=z_\text{t,2})$ and $E_\text{c}(\text{z}=0)$. The first-order approximation to $E_\text{c}(\text{z})$ are two straight lines, with the first one connecting $E_\text{c}(\text{z}=z_\text{t,2})$ and $E_\text{c}(\text{z}=\text{z}_\text{c})$, and the second one connecting $E_\text{c}(\text{z}=\text{z}_\text{c})$ and $E_\text{c}(\text{z}=0)$ where $\text{z}_\text{c}\in(0,z_\text{t,2})$. Hence, the total phase shift $\Theta$ that incorporates the influence of charge-trapping and image-force correction can be calculated with the help of elliptical integrals. More details are given in Appendix \ref{sec:APP_totalTheta}.

\subsection{Carrier transport in a-SiO$_2$}
\label{sec:transport}
Carrier transport in amorphous a-SiO$_2$ under high stress voltages involves coupled electron-hole dynamics with charge trapping and emission. To make the problem tractable, we assume quasi–steady-state conditions over short intervals: long enough for carrier scattering to equilibrate, yet short enough that trap occupation remains quasi-stationary. This decoupling is justified because electron flux is dominated by tunneling injection, which depends strongly on the electric field rather than small variations in capture or emission rates. The position-dependent EDFs of electrons are obtained by solving  the BTE as the their high-energy tail govern impact ionization. In contrast, the hole current is calulated from a drift equation because their mobility is low and hol-initiated impact ionization is negligible. The following sections detail this framework, the treatment of impact ionization, and relaxation processes.

\subsubsection{Boltzmann transport equation}
In the 1D MOSC model, the external electric field is one-dimensional along $z$. The 3D transport of electrons is then projected onto the transport direction, making this transport problem 1D. Following \cite{thermaltail_LacaitaAPL1991}, we define the force $\vec{F_\text{ox}}(z)=-F_\text{ox} \hat{z}$ where  $F_\text{ox}$ is the potential gradient. If the CB of a-SiO$_2$ follows a parabolic band ($E=\frac{\hbar^2k^2}{2m_\text{ox}}$), the simplified 1D BTE describes the evolution of the normalized EDF $n_\textsc{bte}(E,z)$
\begin{align}
\left[F_\text{ox}(z)\frac{\partial}{\partial E} + \frac{\partial}{\partial z}\right]n_\textsc{bte}(E,z)=& -\frac{n_\textsc{bte}(E,z)-n_\text{eq}(E,z)}{\lambda(E)}\nonumber\\
&+ \frac{S(E,z)}{{\lambda(E)}}
\label{eqn_BTE1d}
\end{align}
where $\lambda(E) = v_\text{g}(E)\tau_\text{m}(E)$ represents the momentum relaxation length (MRL), $\tau_\text{m}(E)$ the momentum relaxation time projected onto transport direction, and $v_\text{g}(E)$ the energy-dependent group velocity. The term $S(E,z)/{\lambda(E)}$ represents the generation of carriers due to impact ionization. Only the relaxation rate of electron-phonon scattering and impact ionization are considered for the solution of the 1D BTE. Electron-electron scattering is discarded because of the small electron concentration, while Coulomb scattering with charged impurities can be also ignored according to Ref. \cite{offstoichSiO_DiMariaJAP1985,fischettiTheoryHighfieldElectron1985}. 

The 1D BTE is solved in time windows where $F_\text{ox}$ is frozen but updated after each interval to include trapped-charge effects. Within each window, the equation is integrated along the characteristics in the energy-coordinate plane, reducing it to an ordinary differential equation (ODE). The solution is advanced at discrete nodes using an operator-splitting approach: transport and relaxation are computed first without impact ionization, followed by carrier generation treated as ballistic with first-order kinetics.  The two contributions are combined to obtain the EDF at the current node, which seeds the next step. This process is repeated across the domain and time steps. The numerical details are provided in Appendix \ref{sec:App_1DBTE}.

Key transport parameters are extracted from the EDF  $n_\textsc{bte}(E,z)$. The average drift velocity along the transport direction is \cite{adlerIntroductionSemiconductorPhysics1964,highFieldStatistics_BtinguierPRB1994}:
\begin{equation}
v_\text{d}(z_i) = \frac{1}{4}\int_0^{\infty}n_\textsc{bte}(E,z_i)v_\text{g}(E)\mathrm{d}E
\label{eqn_Jevg}
\end{equation}
where the factor $1/4$ accounts for flux projection  \cite{adlerIntroductionSemiconductorPhysics1964}. Impact ionization is quantified by the impact ionization coefficient (II coefficient )
\begin{equation}
\gamma_\text{ox}(z_i) = \ln[1+\delta n_\textsc{bte}(E,z_i)]/\delta z_{i}
\end{equation}
in which $\delta n_\textsc{bte}(E,z_i)$ is the electron concentration generated in  $[z_{i-1},z_{i}]$ . The II coefficient determines the amplification of electron current and the associated hole current in the oxide. The cumulative effect  is expressed by the quantum yield,
\begin{equation}
\text{QY}(z) = \exp\left(\int_{z_{\text{t,m}}}^z\gamma_\text{ox}(x)\mathrm{d}x\right) -1
\end{equation}
which represents the total number of electron-hole pairs generated between $z_{\text{t,m}}$ and $z$.

\subsubsection{Impact ionization rate}
\begin{figure}
\centerline{\includegraphics[width=0.94\linewidth]{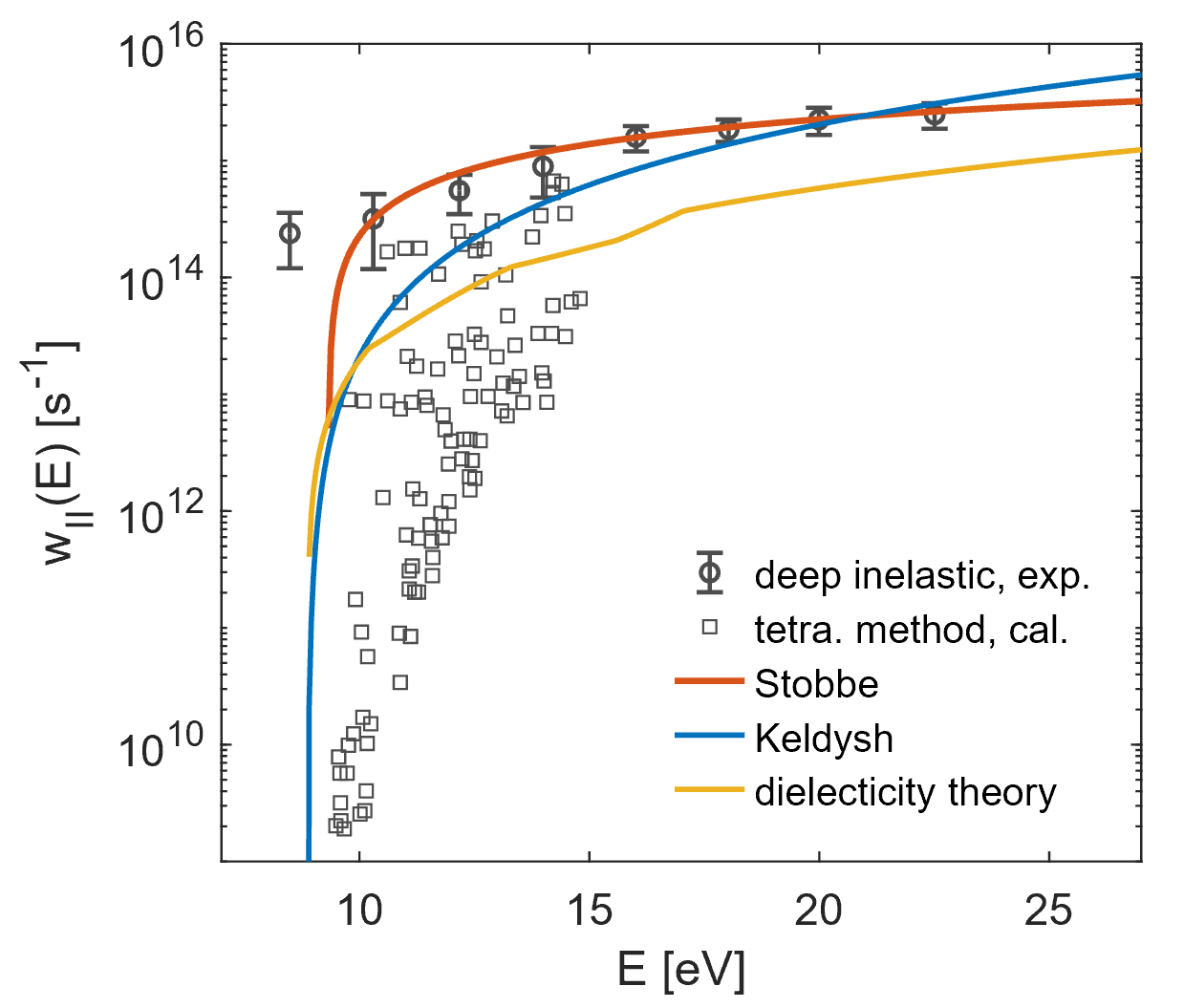}}
\caption{Comparison of the bulk impact ionization rate $w_\textsc{ii}$ in a-SiO$_2$ predicted by different models (Keldysh \cite{transportMC_ArnoldPRB1994}, modified Stobbe \cite{5bandsaSiO2MC_Fitting},  dielectricity theory \cite{IIrate_Murat}, and tetrahedral method calculation) against experimental deep inelastic scattering data \cite{corelevel_CartierPRB1991}. The modified Stobbe model, with parameters from Ref. \cite{5bandsaSiO2MC_Fitting} and adjusted $w_0$, shows excellent agreement with measurements and is adopted in this work.}
\label{fig_IIrate}
\end{figure}

The most widely used parameterization of the impact ionization rate is the Keldysh formula
\begin{equation}
w_\textsc{ii}(E) = w_0\left(\frac{E}{E_\text{th}}-1\right)^a, E>E_\text{th}
\end{equation}
where $w_0$ and $a$ are fitting parameters ($a=2$ for the classical form). However, this expression fails to reproduce the $1/\sqrt{E}$ scaling at high energies predicted by Beth's theory \cite{Bethe_meanfreepath,meanfreepath_Ziaja}. To address this limitation, Stobbe \textit{et al.}  proposed a modifed form \cite{STobbeIIrate_StobbePRB1991}, later extended by Schreiber and Fitting for a-SiO$_2$ \cite{5bandsaSiO2MC_Fitting}
\begin{equation}
w_\textsc{ii}(E) = w_0\left[\frac{E/E_\text{th}-1}{1 + D\left(E/E_\text{th}\right)^2}\ln\left(\frac{E}{E_\text{th}}\right)\right]^a, E>E_\text{th}
\label{eqn_wIIStobbe}
\end{equation}
where  $w_0$, $D$, and $a$ are fitted to experimental data.  Alternative approaches include calculations based on the complex dielectric function \cite{ashleyEnergyLossesMean1981,muratSpatialDistributionElectronhole2004}, and first-principle methods using tetrahedral integration and realistic band structures  \cite{IIrateDFTMC_MizunoJAP} .  

Fig. \ref{fig_IIrate} compares these models with deep inelastic scattering measurements  \cite{corelevel_CartierPRB1991}. In this work, we adopt the modified Stobbe model with  parameters from Ref. \cite{5bandsaSiO2MC_Fitting}, adjusting  $w_0$ to $2.45\times10^{15}\text{ s}^{-1}$. This choice ensures both physical consistency and excellent agreement with expererimental data, making it well-suited for modeling impact ionization in a-SiO$_2$.


\subsubsection{Relaxation rate for electron-phonon scattering}

\begin{figure*}
\centerline{\includegraphics[width=\textwidth]{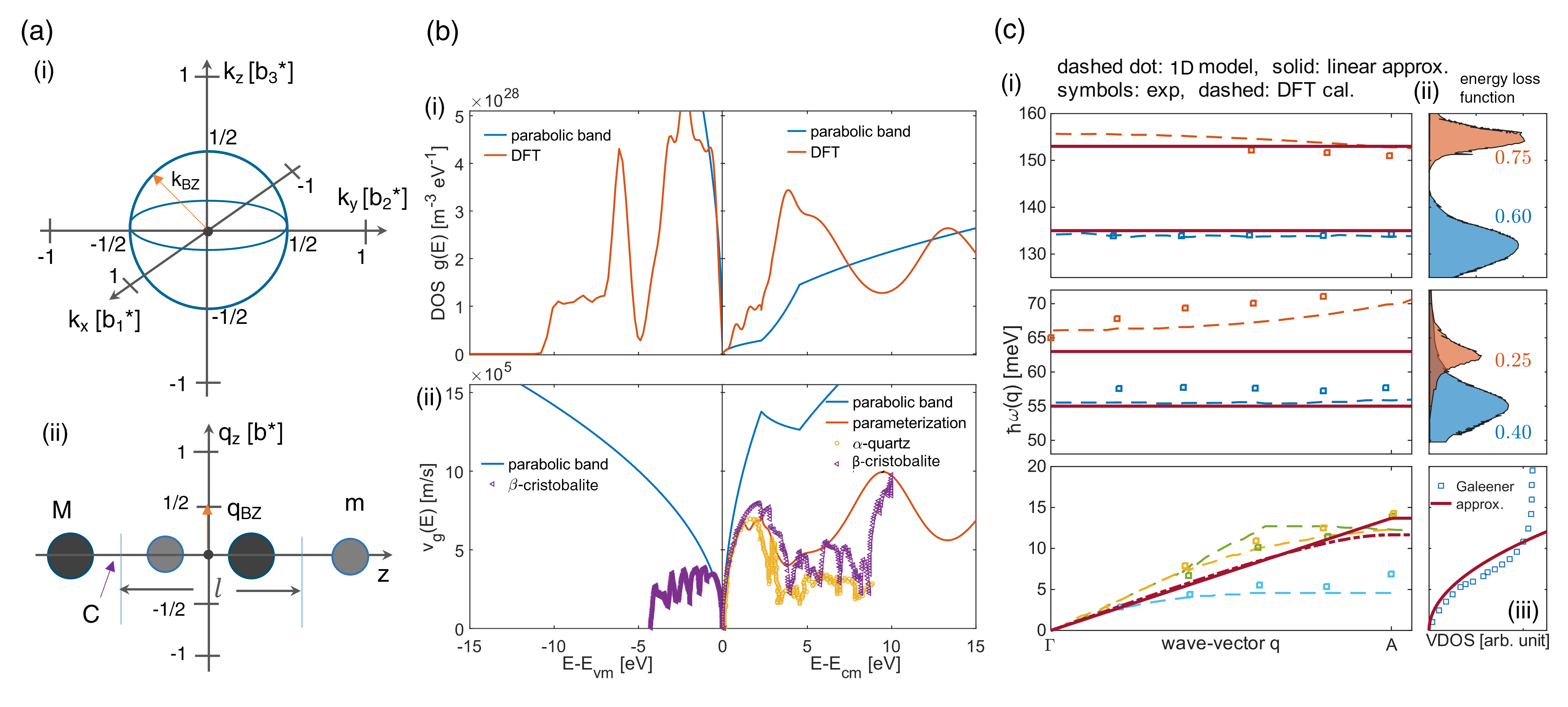}}
\caption{Structural and band models for amorphous SiO$_2$. (a) Reciprocal space structure: (i) Isotropic 3D $k$-space with Brillouin zone edge at $k_\textsc{bz}$, (ii) 1D phonon wave-vector space with diatomic model (periodicity $l$, masses $M$, $m$, spring constant $C$). (b) Electronic properties: (i) Density of states $g(E)$ and (ii) group velocity $v_\text{g}(E)$ for CB and VB, comparing simplified models with DFT calculations (a-SiO$_2$ DOS from \cite{TATpartII_Schleich2022}, $\beta$-cristobalite $v_\text{g}(E)$ from \cite{vgDOSbCrystobalite_RudanPhysicaB}, $\alpha$-quartz from \cite{groupvelocityscintillators_HuangPSSRRL2016}). (c) Phonon properties: (i) Dispersion model for acoustic and optical modes compared with $\alpha$-quartz data along $\Gamma$A (symbol: experiment \cite{Dorner_1980_alphaQuartzPhononDispersion,StrauchBorner_phonondispersion1993}, dashed-line: DFT calculation \cite{Bosak_NewInsightslatticeDynamics}, dash-dotted-line: full 1D model, solid-line: linear approximation); (ii) Energy-loss functions showing relative weights \cite{imqfuncaSiO2_GundePhysicaB2000} of two modes for TL and LO phonons, (iii) Acoustic phonon VDOS versus experimental data \cite{aSiO2vDOS_GaleenerPRB271052}.}
\label{fig_structure}
\end{figure*}

To initiate the scattering analysis, we first construct a physically representative model of amorphous silicon dioxide (a-SiO$_2$). Despite its lack of long-range order, a-SiO$_2$ is approximated here as a periodic yet isotropic crystal, with a primitive cell containing a single SiO$_2$ unit. This abstraction enables the use of a highly isotropic reciprocal space, where the Brillouin zone edge is defined by the Wigner-Seitz radius, denoted as $k_\textsc{bz}$ (Fig. \ref{fig_structure}a(i)). Using the bulk density of a-SiO$_2$ (2.2 g$/\text{cm}^3$) \cite{semiprocbookElKareh}, we obtain $k_\textsc{bz}=10.93$ nm$^{-1}$ following the methodology of Ashlley \textit{et al.} \cite{Acratecal_Ashley1987}. This value aligns with prior estimates from Bradford \textit{et al.} (8.07 nm$^{-1}$) \cite{Acinvlambda_BradfordJAP}, DiMaria \textit{et al.} \cite{VEballistic_DiMariaJAP1988} or Arnold \textit{et al.} \cite{transportMC_ArnoldPRB1994} (12.08 nm$^{-1}$). For consistency, we adopt a one-dimensional $q$-space representation  with $q_\textsc{bz}=k_\textsc{bz}$ (Fig.  \ref{fig_structure}a(i)).

Both electron and phonon dispersions are treated using the extended zone scheme. The CB of a-SiO$_2$ is modeled as a spherical parabolic model with energy-dependent mass $m(E)$ \cite{avelancheaSparks_PRB1981,Acratecal_Ashley1987,Acinvlambda_BradfordJAP,VEballistic_DiMariaJAP1988,transportMC_ArnoldPRB1994}, expressed as:
\begin{equation}
\frac{m(E)}{m_0} = \left\{\begin{array}{cc}
m_\text{eff,0}, & 0\leqslant E<\frac{E_\textsc{bz}}{2} \\
 \left(2m_\text{eff,0}-1\right) + \frac{\left(1-m_\text{eff,0}\right)E}{0.5\times E_\textsc{bz}}, & \frac{E_\textsc{bz}}{2}\leqslant E < E_\textsc{bz} \\
1.0, & E\geqslant E_\textsc{bz}
\end{array} \right.
\label{eqn_meff_E}
\end{equation}
Here, $m_\text{eff,0}$ represents the band-edge effective mass (0.42 derived from tunneling mass \cite{meffSiO2_MaserjianJVST1974,tunnelingthinoxide_Schenk}), and Brillouin edge energy is given by $E_\textsc{bz}= (\hbar k_\textsc{bz})^2/2m_0 = 4.536$ eV, consistent with prior reports (5.52 eV \cite{VEballistic_DiMariaJAP1988}).  The VB is modeled as three parabolic bands with identical energy maxima, and their effective masses are summarized in Table \ref{tab:conduction_band}. The resulting band structure captures key features of the group velocity $v_\text{g}(E)$ and the DOS from DFT calculations \cite{TATpartII_Schleich2022}, as illustrated in Fig. \ref{fig_structure}b. For energies below $0.5E_\textsc{bz}$, the  parabolic behavior of $v_\text{g}(E)$ is consistent with observations in $\beta$-cristobalite \cite{vgDOSbCrystobalite_RudanPhysicaB} and $\alpha$-quartz \cite{groupvelocityscintillators_HuangPSSRRL2016}. To extend the description to higher energies,  we propose an empirical formula
\begin{equation}
\frac{2\times 10^6 \text{ m/s}}{v_\text{g}(E)} = \frac{g_{\text{c},\textsc{dft}}(E)}{10^{28} \text{m}^{-3}\text{eV}^{-1}}+\frac{10^{28}\text{m}^{-3}\text{eV}^{-1}}{g_{\text{c},\text{parab}}(E)}
\label{eqn_vg_parameterization}
\end{equation}
in which $g_{\text{c},\textsc{dft}}(E)$ is the DOS from DFT and $g_{\text{c},\text{parab}}(E)$ is that of a parabolic band with energy-independent effective mass. The first term reflects the inverse relationship between $v_\text{g}$ and DOS at high energy, a behavior which is also observed in Si \cite{MCsimIIvgDOS_Fischetti1991}. Despite its simplicity, this empirical parameterization reproduces the $v_\text{g}(E)$ obtained from DFT with good accuracy and therefore provides a practical means to extrapolate $v_\text{g}(E)$ into the high-energy regime where neither DFT data nor experimental measurements are available.

The phonon dispersion in a-SiO$_2$ is described by a 1D diatomic chain model (Fig. \ref{fig_structure}c).  Acoustic phonons are approximated by a linear dispersion
\begin{equation}
\omega_\textsc{ac}(q) = \left\{\begin{array}{cc}
\frac{\sqrt{2}}{\pi}v_\text{s} q, \quad 0\leq q<q_\textsc{bz} \\
\frac{\sqrt{2}}{\pi}v_\text{s}q_\textsc{bz}, \quad q\geq q_\textsc{bz}
\end{array} \right.
\label{eq_acousticphonon}
\end{equation}
in which $v_\text{s}$ denotes the average sound velocity  \cite{avelancheaSparks_PRB1981,Acratecal_Ashley1987}. This simplified form reproduces the results of the full 1D model,  experimental data, and the DFT results  \cite{phononaSiOdft_Bosak,alphaQuartzphonon_StrauchJPCondMatter}. Optical phonons exhibit flat dispersions
\begin{equation}
\omega_\textsc{lo}(q) = \omega_\textsc{lo}, \omega_\textsc{to}(q) = \omega_\textsc{to}
\end{equation}
for both longitudinal and transverse polarization. There are two dominant modes in each polarization, and their relative weights were determined from energy-loss functions \cite{imqfuncaSiO2_GundePhysicaB2000} (Fig. \ref{fig_structure}c(ii)). For LO phonons, large polarons will modify the mass of CB electrons \cite{frohlichElectronsLatticeFields1954} and the modified band-edge mass 
\begin{equation}
\frac{m_p}{m_0} \approx \frac{m_\text{meff,0}}{1-\gamma_\textsc{f}/6}
\label{eqn_meff_polaron}
\end{equation}
in which $\gamma_\textsc{f}$ is the Fr\"olich coupling constant.

Within the Born approximation, the scattering of an electron state $|k\rangle$ into state $|k'\rangle$ proceeds via absorption (``+'') or emission (``$-$'') of a phonon with wave-number $q$ and polarization $p$ from (or to) a quasi-stationary phonon bath. For an interaction Hamiltonian $H_\text{ep}$, time-dependent perturbation theory yields the transition rate \cite{transporttheory_Lundstrom_2000,elphscatter_Ridley}:
\begin{equation}
w(k,k')_{\pm} = \frac{2\pi}{\hbar}|V_p|^2I_{k,k'}^2|A_q|^2\delta[E_{k'}-E_{k}\mp\hbar\omega(q)]\delta_{k\pm q-k',G} \label{eqn_rate_wif}
\end{equation}
where $I_{k,k'}=\langle k'|k\rangle$ is the overlap integral, $|V_p|^2 = \langle k'|H_\text{ep}(q)|k\rangle$  the squared matrix element for the electron-coordinate-dependent part of $H_\text{ep}$, and $A_q$  the phonon normal coordinate. The Dirac delta $\delta(E_\text{f}-E_\text{i})$ enforces energy conservation, while the Kronecker delta $\delta_{k\pm q-k',G}$ ensure momentum conservation with $G$ a reciprocal lattice vector.  Here, $G=0$ suffices since Umklapp processes are implicitly included in the extended zone scheme. The phonon normal coordinate $A_q$ is given by \cite{transporttheory_Lundstrom_2000}
\begin{equation}
|A_q(\pm)|^2 =\frac{\hbar\left\{n_p\left[\omega_p(q)\right]+\frac{1}{2}\mp \frac{1}{2}\right\}}{2\rho_\text{c}V_\text{c}\omega_p(q)}, 
\end{equation}
where $V_\text{c}$ is the unit cell volume, $\omega_p(q)$ the phonon angular frequency, and  
\begin{equation}
n_p[\omega_p(q)]= \left[\exp(\beta \hbar\omega_p(q))-1\right]^{-1}
\label{eqn_phonnumb}
\end{equation}
the corresponding phonon number.

We assume isotropic scattering, i.e., no explicit angular dependence. The momentum relaxation rate along the direction $\hat{\textsc{f}}$ of an external field $\mathbf{F}$ is then given by \cite{transporttheory_Lundstrom_2000,efffactor_WuPRB2015}:
\begin{equation}
w_m(k,k')_\pm = w(k,k')_\pm\times\left(1-\frac{k'}{k}\cos\theta\right)
\label{eqn_wm_kkp}
\end{equation}
where $\theta$ is the scattering angle. The inverse momentum relaxation length (MRL) is
\begin{equation}
\lambda^{-1}(k,k') = w_m(k,k')/v_k(k)
\end{equation}
with $v_k(k)=\hbar k/m(k)$ the  velocity. The total momentum relaxation rate of an initial state $|k\rangle$ due to one mode $p$ of electron-phonon scattering follows from integrating $w_m(k,k')_\pm$ over $q$ \cite{Llacer_LOintmethodJAP1969}
\begin{equation}
w_m(k)_{\pm} = \frac{V_\text{c}}{2\pi\hbar}\int_{q_\text{min}^{\pm}}^{q_\text{max}^{\pm}}|V_p|^2|A_q|^2\frac{q^2}{v_k k}\left(\frac{q}{2k}\mp\frac{\omega_p(q)}{v_kq}\right)\mathrm{d}q 
\label{eq_totwm_atk}
\end{equation}
with integration limits determined by energy and momentum conservation
\begin{subequations}
\begin{align}
q_\text{max}^{\pm} &= k\left[1 + \sqrt{1\pm\hbar\omega_p(q)/E(k)}\right]\\
q_\text{min}^{\pm} &= k\left[\mp 1 \pm \sqrt{1\pm\hbar\omega_p(q)/E(k)}\right]
\end{align}
\end{subequations}
Evaluation of Eq. \ref{eq_totwm_atk} requires explicit models for the electron band structure $E(k)$, the phonon dispersion $\omega_p(q)$, and the interaction Hamiltonian $H_\text{ep}(q)$, which will be introduced shortly.

\begin{table}[t]
\centering
\begin{threeparttable}
\caption{Parameters for the models of electron-phonon scattering and impact ionization}
\begin{ruledtabular}
\begin{tabular}{llcccccc}
\textsc{lo} & & $\hbar\omega$ & $1/\bar{k}$ & $m_p$ & weight & \multirow{2}{*}{Ref.} \\
phonon  & $|V_p|^2$ & [meV] & [1] & [$m_0$] & [1] &  \\
 \cmidrule(lr){3-7}
 & \multirow{2}{*}{$\frac{\omega(q)^2\rho_\text{c}}{\epsilon_0\bar{\kappa}q^2}$} & 63 &0.0814 &0.4823\tnote{a} & 0.25\tnote{b} & \multirow{2}{*}{\cite{Acratecal_Ashley1987}}\\
 &  &153 &0.1124 &0.4743\tnote{a} & 0.75\tnote{b} &  \\
\colrule
\textsc{to} & & $\hbar\omega$ & $D_\textsc{to}$ & $q_\alpha$ & weight & \multirow{2}{*}{} \\
phonon  &$|V_p|^2$ & [meV] & [$\frac{\text{eV}}{\text{nm}}$] & [$\frac{1}{\text{nm}}$] & [1] &  \\
 \cmidrule(lr){3-7}
 & \multirow{2}{*}{$\frac{D_\textsc{to}^2f_\text{m}(q)}{\left[1+(q/q_\alpha)^2\right]^2}$} &55.3 &200 &14.15\tnote{c} & 0.40\tnote{b} & \multirow{2}{*}{}\\
 &  &135 &200 &14.15\tnote{c} & 0.60\tnote{b} &  \\
\colrule
\textsc{ac} & & $v_\text{s}$ & $D_\textsc{ac}$ & $q_\alpha$ & weight & \multirow{2}{*}{Ref.} \\
phonon  &$|V_p|^2$ & [$\frac{\text{m}}{\text{s}}$] & [eV] & [$\frac{1}{\text{nm}}$] & [1] &  \\
 \cmidrule(lr){3-7}
 & $\frac{q^2D_\textsc{ac}^2f_\text{m}(q)}{\left[1+(q/q_\alpha)^2\right]^2}$ & 4233 &3.0 &14.15\tnote{c} & 1.0 & \cite{statsTranspaSiO2_Ac3eV}
\end{tabular}
\end{ruledtabular}
\label{tab:scattering_param}
\begin{tablenotes}
\item[a] calculated according to Eq. \ref{eqn_meff_polaron}
\item[b] Ref. \cite{imqfuncaSiO2_GundePhysicaB2000}
\item[c] calculated according to Eq. \ref{eqn_qalpha_Ac}
\end{tablenotes}
\end{threeparttable}
\end{table}


Within deformation potential theory, the squared matrix element for scattering by acoustic (\textsc{ac}) phonons is \cite{avelancheaSparks_PRB1981}
\begin{equation*}
|V_p|^2 = D_\textsc{ac}^2q^2
\end{equation*}
where $D_\textsc{ac}$ is the deformation potential of a-SiO$_2$ and is reckoned energy-independent. Following Bradford and Woolf \cite{Acinvlambda_BradfordJAP} and Ashley \textit{et al.} \cite{Acratecal_Ashley1987}, we adopt a modified squared matrix element
\begin{subequations}
\begin{align}
|V_p|^2 &= \frac{q^2D_\textsc{ac}^2}{\left[1 + (q/q_\alpha)^2\right]^2}f_\text{m}(q)
\label{eqn_Vasqrd_Ac}\\
f_\text{m}(q) &=
\left\{\begin{array}{cc}
1+ r_\text{m}(q/{k_\textsc{bz}})^2, \quad q<k_\textsc{bz} \\
1 + r_\text{m}, \quad q\geq k_\textsc{bz}
\end{array} \right.
\label{eq_fm_q}
\end{align}
\end{subequations}
Here, the screened Coulomb potential term $1/[1+(q/q_\alpha)^2]$ dampens high-energy scattering beyond $E_\text{bz}$, while $f_\text{m}(q)$ accounts for smooth variation of scattering mass with energy. The constant $r_\text{m}$ represents the mass ratio between light and heavy constituents in the unit cell. The characteristic wave-vector $q_\alpha$ is given by 
\begin{equation}
q_\alpha = \sqrt{\frac{Z}{\kappa_\infty\epsilon_0 V_\text{c} D_\textsc{ac}}}
\label{eqn_qalpha_Ac}
\end{equation}
where $Z=8$ corresponds to the dominant oxygen scatterer \cite{Acinvlambda_BradfordJAP}, and $\kappa_\infty$ is the high-frequency relative dielectric constant. For non-polar (transverse) optical phonons, the deformation potential theory \cite{avelancheaSparks_PRB1981} yields an analogous form
\begin{equation}
|V_p|^2 = \frac{D_\textsc{to}^2}{\left[1 + (q/q_\alpha)^2\right]^2}f_\text{m}(q)
\end{equation}
with $D_\textsc{to}$ the deformation potential ($\approx$ 200 eV/nm \cite{statsTranspaSiO2_Ac3eV}). Although some studies neglected \textsc{to} phonon scattering \cite{Unmklapp_ZakharovSSComm1988, transportMC_ArnoldPRB1994}, we include it due to its demonstrated importance in high-field transport \cite{statsTranspaSiO2_Ac3eV,VEballistic_DiMariaJAP1988}. For longitudinal optical (\textsc{lo}) phonons, the Fr\"ohlich formalism \cite{transporttheory_Lundstrom_2000,elphscatter_Ridley} applies
\begin{equation}
|V_p|^2=\frac{\omega(q)^2\rho_c}{\epsilon_0q^2}\frac{1}{\bar{\kappa}}, \quad \frac{1}{\bar{\kappa}} =\frac{1}{\kappa_\infty}-\frac{1}{\kappa_0}
\end{equation}
where $\kappa_0$ is static dielectric constant. The quantity $1/\bar{\kappa}$ can be extracted from the imaginary part of the dielectric function \cite{Acratecal_Ashley1987}. 

The numerical values of the relevant parameters for \textsc{ac}, \textsc{to}, and \textsc{lo} phonons used in this work are summarized in Table~\ref{tab:scattering_param}, together with their respective weights and references.

\section{Results and Discussions}

\subsection{Electrostatics: comprehensive vs simplified approaches}
The electrostatics for n-SiC and n$^+$-Si will be studied via a comprehensive model and a simplified model. The electrostatics of a-SiO$_2$ will be discussed in sections focused on charge trapping. The comprehensive model uses Fermi-Dirac statistics for carriers, incomplete ionization models for dopants, and multiple Kane-type nonparabolic bands for CB (and VB) if applicable.  The simplified model utilizes Boltzmann statistics for carriers, complete ionization model for dopants, and only one parabolic band for CB (and VB).

\begin{figure}
\centerline{\includegraphics[width=\linewidth]{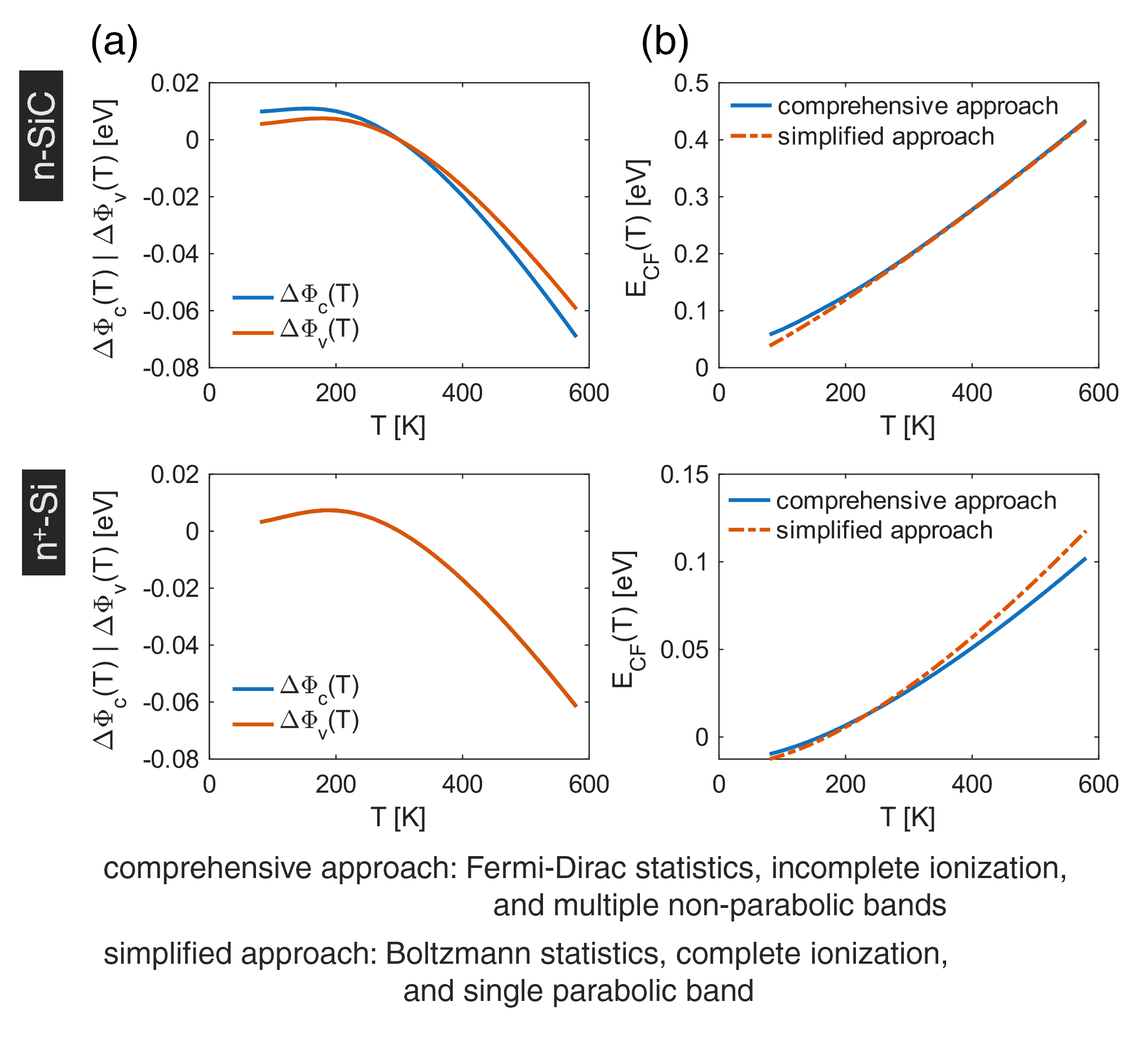}}
\caption{Temperature-dependent band offset shifts and Fermi-level offset at 80-573 K. Top panels: n-SiC; bottom panels: n$^+$-Si. (a) Evolution of band offsets $\Delta\Phi_\text{c}(T)$ and $\Delta\Phi_\text{v}(T)$ relative to room temperature (298 K), showing asymmetric shifts in n-SiC and symmetric behavior in n$^+$-Si. (b) Comparison of the Fermi level offset $E_\textsc{cf}(T)$ calculated using the comprehensive approach and the simplified approach.}
\label{fig_Tdepbarrier}
\end{figure}

\begin{figure*}
\centerline{\includegraphics[width=\textwidth]{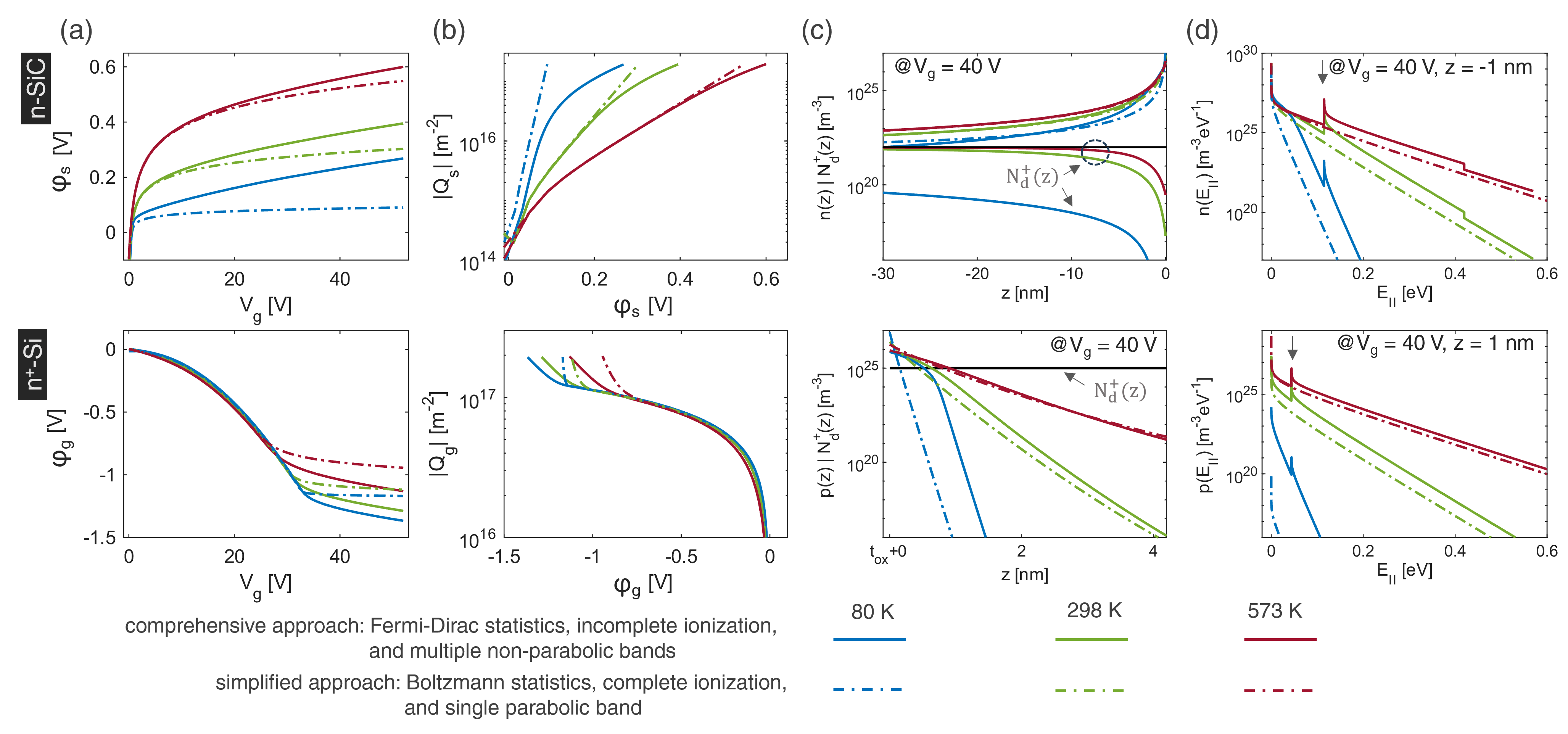}}
\caption{Electrostatic analysis of  an n-SiC/SiO$_2$/n$^+$-Si MOSC (54 nm oxide) comparing comprehensive and simplified approaches across reliability-relevant temperatures. Top panels: n-SiC; bottom panels: n$^+$-Si. (a) Surface potential evolution with gate bias, showing significant model differences at high $V_\text{g}$ and low temperatures. (b) Surface charge density characteristics revealing accumulation in n-SiC and depletion-to-inversion transition in n$^+$-Si gate. (c) Spatial distribution of carriers ($n(z)$ for n-SiC, $p(z)$ for n$^+$-Si) and ionized dopants ($N_\text{d}^+(z)$) at $V_\text{g}=40$ V, demonstrating the importance of incomplete ionization near interfaces. (d) Energy-resolved carrier distributions at 1 nm from respective interfaces ($V_\text{g}=40$ V), with vertical arrows indicating contributions from higher bands, crucial for tunneling current calculations.}
\label{fig_electrostatics}
\end{figure*}

The temperature-dependent Fermi-level offsets ($E_\textsc{cf}$ and $E_\textsc{fv}$) and band offset shifts [$\Delta\Phi_\text{c}(T)$ and $\Delta\Phi_\text{v}(T)$)] are calculated according to the temperature dependent band-gap, and CB and VB edges. For example, band offset shifts for the SiC/a-SiO$_2$ interface can be calculated as
\begin{subequations}
\begin{align}
\Phi_\text{c}(T) &= E_\text{c}(T,\text{SiO}_2)-E_\text{c}(T,\text{SiC}),\\
\Phi_\text{v}(T) &= E_\text{v}(T,\text{SiC}) - E_\text{v}(T,\text{SiO}_2),\\
\Delta\Phi_\text{c}(T) &=\Phi_\text{c}(T)-\Phi_\text{c}(298\text{ K}),\\
\Delta\Phi_\text{v}(T) &= \Phi_\text{v}(T)-\Phi_\text{c}(298\text{ K}).
\end{align}
\end{subequations}
The results are shown in Fig. \ref{fig_Tdepbarrier} for an n-SiC/SiO$_2$/n$^+$-Si MOSC in the temperature range from 80 K to 573 K.  The first striking observation is that Fermi-level offset $E_\textsc{cf}(T)$ and band offset shift $\Delta\Phi_\text{c}(T)$ show the opposite temperature dependency, above room temperature. This fact will help to resolve the controversy regarding the potential barrier for electron tunneling, i.e., whether it is the band offset $\Phi_\text{c}(T)$ or the effective barrier defined as 
\begin{equation}
\Phi_\text{eff}(T) = \Phi_\text{c}(T) + E_\textsc{cf}(T)
\end{equation}
as illustrated in Fig. \ref{fig_TunnelingCoeff}. The barrier $\Phi_\text{eff}(T)$ [$\Phi_\text{c}(T)$] increases (decreases) with increasing temperature, and hence result in smaller (largers) gate leakage current. A detailed discussion will be given in later sections. The second observation is that the comprehensive approach predicts a smaller range of $E_\textsc{cf}(T)$ than the simplified approach does. Since the Fermi-level offset will significantly influence the concentration of free carriers (eq. \ref{eqn_elhconc}), it is thus necessary to use the comprehensive approach for electrostatics.


We analyzed the electrostatics of a 1D MOSC under gate bias $V_\text{g}$ from 0 to 52 V at 80, 298, and 573 K using both the comprehensive and the simplified approaches, as shown in Fig. \ref{fig_electrostatics}. At a given $V_\text{g}$, the comprehensive approach consistently predicts higher surface potentials ($\psi_\text{s}$ and $\psi_\text{g}$) and lower charge densities ($|Q_\text{s}|$, $|Q_\text{g}|$) than the simplified approach does, as shown in Fig. \ref{fig_electrostatics}a-b. The difference between both approaches become particularly significant when the Fermi level starts crossing the band edges, which happens at $V_\text{g}\approx10$ V (27 V) for SiC (n$^+$-Si) at 298 K. The spatial distributions of carriers and ionized dopants are shown in Fig. \ref{fig_electrostatics}c. The comprehensive approach shows incomplete ionization effects, in the n-SiC near the SiC/SiO$_2$ interface, which becomes more pronounced at lower temperatures, and negligible incomplete ionization for the degenerate n$^+$-Si gate. The simplified approach predicts that concentration of free carrier is more concentrated near the interface. The carrier concentrations as a function of $E_\parallel$ at 1 nm from the respective interfaces are shown in Fig. \ref{fig_electrostatics}d for $V_\text{g}=40$ V. The comprehensive model shows larger concentration at higher energies and the contribution of sub-bands of CB and VB. These carrier concentration will enhance the gate leakage current due to smaller tunneling barriers. A more detailed analysis will be given in later sections.

Interface charges, whether pre-existing or generated by hot carriers under gate leakage stress, can alter global electrostatics (Eq. \ref{eqns_surfpot}) and locally perturb the field near the interfaces. Their impact is more pronounced on the lightly doped SiC side, where modest interface charge can significantly modify the anode-side potential and influence tunneling and trapping. In contrast, the heavily doped gate maintains large surface charge (Fig. \ref{fig_electrostatics}a) and thus screens electrostatic shifts. Without explicit modeling or experimental input, the quantitative influence of these traps remains uncertain.

\subsection{Gate leakage current due to tunneling} 
In this section, the gate leakage current due to tunneling will be simulated without considering the effect of impact ionization and later compared with experimental observation. According to the ``tunneling + transport" procedure, the gate leakage current without impact ionization can be calculated as
\begin{equation}
J_\text{g} = v_\text{d}(L)n_\textsc{fn}(z_\text{t,m})
\label{eqn_JgFNalone}
\end{equation}
in which  $n_\textsc{fn}(z_\text{t,m})=\int n_\textsc{fn}(E)\mathrm{d}E$ is the total electron concentration that has tunneled into the CB of a-SiO$_2$ and $v_\text{d}(L)$ is the averaged drift velocity at the n$^+$-Si/a-SiO$_2$ interface. We adopt $v_d = 1.52\times10^5$ m/s for all temperatures and field strengths, which is supported by experimental results \cite{driftvelocity_HughesPRL1973}. The influence of the classical image-force and charge trapping in a-SiO$_2$ directly change the current $J_\text{g}$ via changing the electron concentration $n_\textsc{fn}(z_\text{t,m})$. 

According to the conventional Tsu-Esaki formalism together with the WKB formula for tunneling coefficient \cite{Tunneling_TsuEsaki,WKBTdeptunnel_Fromhold,tunnelingmodel_Gehring}, the gate leakage current $J_\text{g}$ can be calculated according to the Lenzlinger-Snow formula \cite{tunneling_LenzlingerSnowJAP1969}
\begin{subequations}
\begin{align}
J_\text{g}\left(F_\text{ox}\right) =& \frac{F_\text{ox}^2\exp\left[-2\Theta_0(F_\text{ox})v(y)\right]m_\text{d}}{(4\pi)^2\hbar\Phi_\text{eff}(T)m_{\text{ox}}}\nonumber\\
&\times\left\{t(y)^2\text{sinc}\left[2\Theta_0(F_\text{ox})t(y)\frac{3k_\textsc{b}T/2}{\Phi_\text{eff}(T)}\right]\right\}^{-1}\\
\Theta_0(F_\text{ox})=&\frac{2}{3}\frac{\sqrt{2m_{\text{ox}}\Phi_\text{eff}(T)^3}}{\hbar F_\text{ox}}
\end{align}
\label{eqn_JgLS}
\end{subequations}
where $m_\text{d}$ denotes the DOS effective mass \cite{tunnelingmodel_Gehring} and $\Phi_\text{eff}(T)$ is the barrier height. Here, the classical image-force effects enter through the auxilliary functions $v(y)$ and $t(y)$ \cite{approxvy_ForbesAPL2006}, respectively
\begin{align*}
&v(y) \approx 1-y^2[1 - \ln(y)/3], \\
&t(y)\approx 1+\left(\frac{y}{3}\right)^2\left[1-\ln(y)\right]
\end{align*}
in which $y = \frac{\sqrt{|\gamma_1F_\text{ox}|}}{\Phi_\text{eff}(T)-E_{\parallel}}$.  The barrier height $\Phi_\text{eff}(T)$ increases and thus the dominant part of the tunneling coefficient $\exp[-2\Theta_0(F_\text{ox})v(y)]$ decreases with increasing temperature. This predicts a negative correlation between temperature and $J_\text{g}$ at a given gate bias $V_\text{g}$. 


\begin{figure}
\centerline{\includegraphics[width=\linewidth]{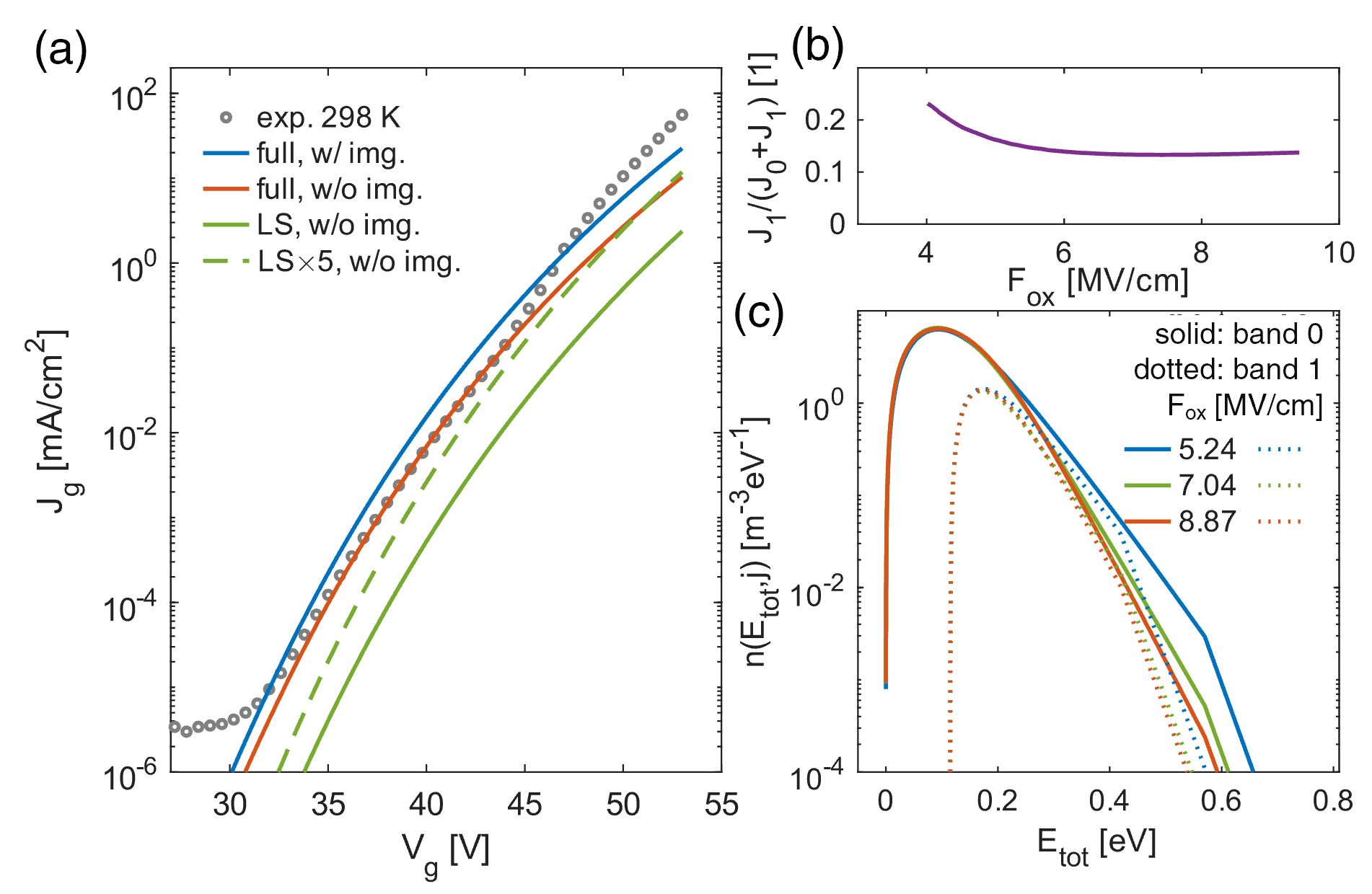}}
\caption{Gate leakage current analysis at $T=298$ K. (a) Experimental $J_\text{g}\text{-}V_\text{g}$ (circles) comparing with the simulation using the full model (Eq. \ref{eqn_JgFNalone}) and LS model (Eq. \ref{eqn_JgLS}) combined with the influence image-force correction. (b) Relative contribution of the second sub-band of CB to the total current versus oxide field. (c) Energy distribution functions of tunneled electrons at different oxide fields $F_\text{ox}$ for the first (dotted) and second (solid)  sub-bands of CB. }
\label{fig_JgVgT298K}
\end{figure}

We calculate the gate leakage current $J_\text{g}$ at 298 K using both Eq. \ref{eqn_JgFNalone} (full model) and Eq. \ref{eqn_JgLS} (LS model), without including electron/hole traps nor impact ionization. The effect of classical image force correction has also been examined. The features of the gate leakage current are summarized in Fig. \ref{fig_JgVgT298K}. In Fig. \ref{fig_JgVgT298K}a, the full model without image-force correction yields an excellent agreement with the experimental gate leakage current $J_\text{g}$, while the image-force correction enhances $J_\text{g}$ by a factor of about 2. The LS model without image-force correction predicts $J_\text{g}$ that is about 5 times smaller than the one from the full model. Hence, the full model not only predicts better $J_\text{g}$ but also provides $n_\textsc{fn}(E)$ as an initial condition for the 1D BTE, for which the LS model cannot be used. As shown in \ref{fig_JgVgT298K}b-c, the second sub-band of the CB of n-SiC contributes around 15\% of the total $J_\text{g}$, and the EDFs of tunneled elctrons  associated with this sub-band show highly energetic electrons ($E>0.115$ eV) regardless of the field $F_\text{ox}$. The tail of these EDFs reflects a Boltzmann distribution because of the termination of the sub-bands of the CB of SiC ([001]). Hence it is evident that the full model should be used to calculate gate leakage currents.


\begin{figure}
\centerline{\includegraphics[width=\linewidth]{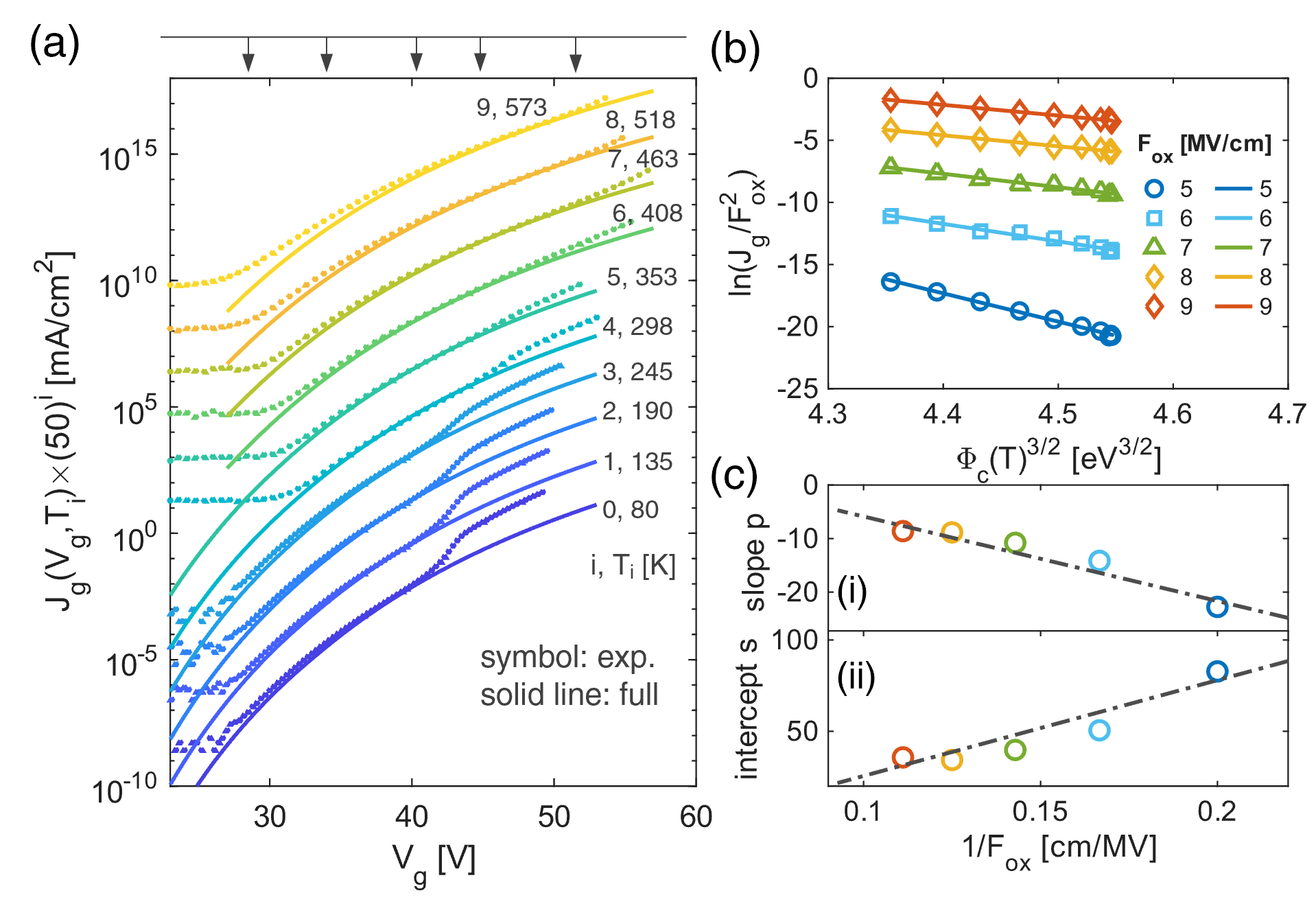}}
\caption{Temperature-dependent gate leakage characteristics. (a) Experimental (triangles) and calculated (lines) $J_\text{g}(V_\text{g})$ curves for temperatures $T_i=80,135,\cdots,573$ K ($i=0,1,\cdots,9$), each scaled by $50^i$ for clarity. Arrows indicate gate voltages corresponding to fixed fields $F_\text{ox}=5.0\text{-}9.0$ MV/cm. (b) Linear relationship between $\ln(J_\text{g}/F_\text{ox}^2)$ and $\Phi_\text{c}(T)^{3/2}$ at constant fields. (c) Field dependence of linear regression parameters: (i) slope $p$ and (ii) intercept $s$ versus $1/F_\text{ox}$.}
\label{fig_JgVgTdepAll}
\end{figure}

Fig. \ref{fig_JgVgTdepAll} summarizes the temperature dependence of gate leakage current $J_\text{g}$ calculated by using the full model without image-force correction. Good agreement was achieved between the experimental observation and the simulated value for temperatures from 80 K to 573 K, as shown in Fig. \ref{fig_JgVgTdepAll}a where the $J_\text{g}$-$V_\text{g}$ curves were scaled for clarity. The abrupt current amplification of $J_\text{g}$ at larger $V_\text{g}$ stems from impact ionization and hole capture in a-SiO$_2$, which will be covered in the following sections.

As shown in Fig. \ref{fig_JgVgTdepAll}b, the $J_\text{g}$ values at fixed fields $F_\text{ox}=5.0,6.0,7.0,8.0,9.0$ MV/cm (corresponding gate voltages marked by arrows in Fig. \ref{fig_JgVgTdepAll}b) reveal a linear relationship between $\ln(J_\text{g}/F_\text{ox}^2)$ and $\Phi_\text{c}(T)^{3/2}$, with its slope $p$ and intercept $s$ shown in Fig. \ref{fig_JgVgTdepAll}c. Both experiment and simulation reveal that $J_\text{g}(F_\text{ox})$ increases with increasing temperature because $\Phi_\text{c}(T)$ decreases with increasing temperature. This indicates $\Phi_\text{c}(T)$ is the barrier height for tunneling rather than $\Phi_\text{eff}(T)$ as conventionally assumed in the Lenzlinger-Snow formalism. The linear dependency of the slope $s$ (intercept) on $1/F_\text{ox}$ can be readily understood from the phase-shift Eq. \ref{eqn_JgLS}. 

The fact that $\Phi_\text{c}(T)$ is the barrier height for tunneling is crucial for higher sub-bands of the CB of SiC. The barrier height will be overestimated if $\Phi_\text{eff}(T)$ is used as the barrier height because the Fermi-level offset $E_\textsc{cf}$ will be added by the min-gap of the two sub-bands.

\subsection{Electron transport in a-SiO$_2$}
The features of electron transport in the CB of a-SiO$_2$ and the electron-initiated impact ionization will be discussed in this section. At first, the inverse mean relaxation lengths (MRLs) $\lambda^{-1}(E)$ due to electron-phonon scattering will be calculated using methods introduced in previous sections. Then, the 1d BTE will be solved for the 1D MOSC, to determine the EDFs of electrons, the average drift velocity, averaged energy, II coefficient and QY for electron-hole pair generation. The dependency of these quantities on temperature and/or electric field yield well-known laws for impact ionization.

\subsubsection{The inverse mean relaxation length}
\begin{figure*}
\centerline{\includegraphics[width=0.96\linewidth]{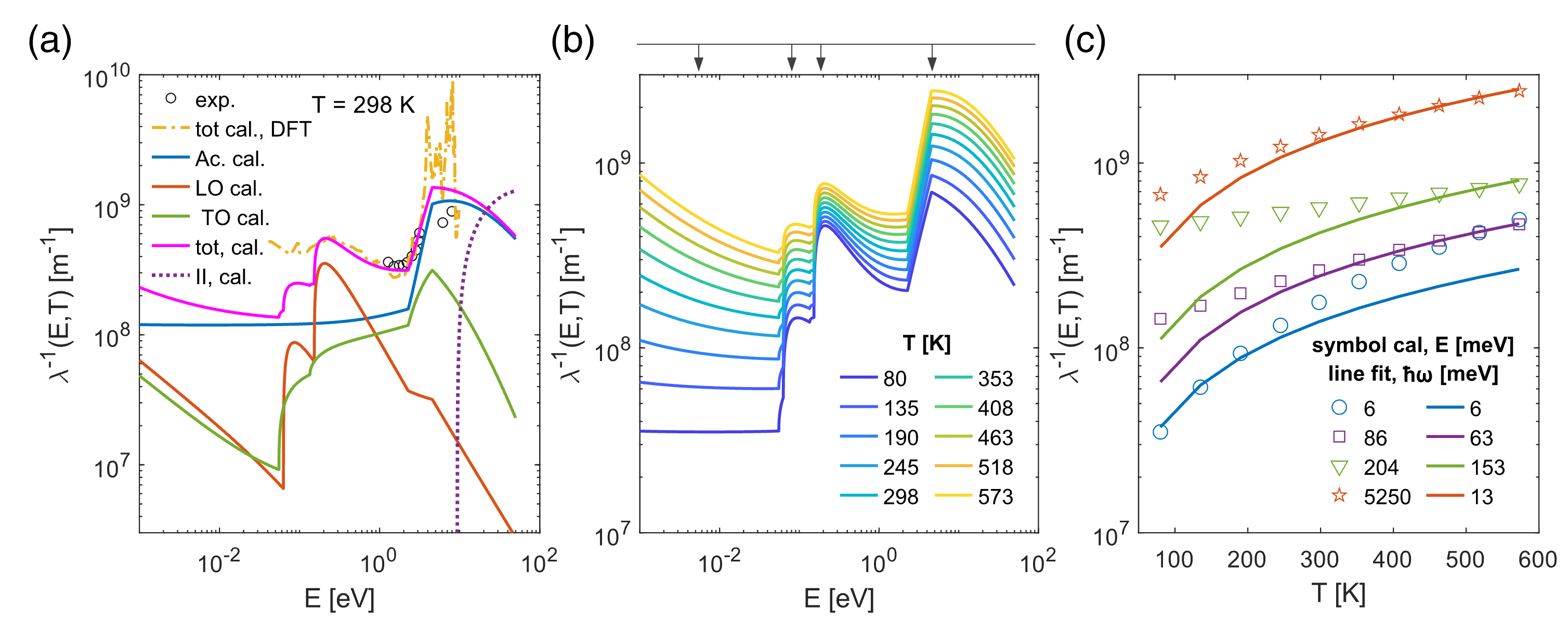}}
\caption{Electron-phonon scattering in a-SiO$_2$ expressed via the inverse mean relaxation length $\lambda^{-1}(E,T)$. (a) Contributions from acoustic (\textsc{ac}), longitudinal optical (\textsc{lo}), and transverse optical (\textsc{to}) phonons at 298 K, compared with DFT predictions for $\beta$-cristobalite \cite{vgDOSbCrystobalite_RudanPhysicaB} and experimental measurements \cite{Expinvlambda_CartierPhysScriptaT23}. The $\lambda^{-1}(E,T)$ for II is displayed for comparison.  (b) Temperature evolution of $\lambda^{-1}(E,T)$ from 80 K to 573 K, with arrows marking energies selected for detailed temperature analysis. (c) Temperature dependence at selected energies fitted to asymptotic formulae with characteristic phonon energies $\hbar\omega$.}
\label{fig_elphIIrate}
\end{figure*}
The inverse mean relaxation lengths (MRLs) $\lambda^{-1}(E)$ were calculated using parameters from Table \ref{tab:scattering_param} for all types of electron-phonon scattering mechanisms. The results at 298 K are shown in Fig. \ref{fig_elphIIrate}a and compared with DFT results for $\beta$-cristobalite \cite{vgDOSbCrystobalite_RudanPhysicaB} which were converted from scattering rates $w(E)$  and group velocity $v_\text{g}(E)$ according to
\begin{equation*}
\lambda^{-1}(E)=\frac{0.42}{0.68}\frac{w(E)}{v_\text{g}(E)}
\end{equation*}
where the factor 0.42/0.68 accounts for the difference of the effective mass between $\beta$-cristobalite and a-SiO$_2$. The calculated total $\lambda^{-1}(E)$ shows excellent agreement with both DFT results \cite{vgDOSbCrystobalite_RudanPhysicaB} and experimental measurements \cite{Expinvlambda_CartierPhysScriptaT23}. Notably, \textsc{lo} phonon contributions become negligible above $\sim 1.5$ eV, challenging previous interpretations that experimental rates can be decomposed solely into \textsc{ac} and  \textsc{lo} phonon contributions \cite{Expinvlambda_CartierPhysScriptaT23,corelevel_CartierPRB1991}. In the energy range $(E_\textsc{bz}/2,E_\textsc{bz})$, both \textsc{ac} and  \textsc{to} phonons contribute to a sharp increase in total $\lambda^{-1}$, primarily due to effective mass enhancement (Eq. \ref{eqn_meff_E}). Beyond $E_\textsc{bz}$, the total $\lambda^{-1}$ decays as $E^{-1}$, mainly due to \textsc{ac} phonons, and yield $E^{-1/2}$ law of   set by Arnold \textit{et al.} \cite{transportMC_ArnoldPRB1994}.

The inverse MRL results in Fig. \ref{fig_elphIIrate}a reveal significant implications for high-field transport. Electron-phonon scattering alone can stabilize the electron distribution function (EDF), with maxima at $\sim153$ meV and $\sim5.0$ eV from \textsc{lo} and \textsc{ac} phonons respectively preventing ``run-away" phenomena \cite{driftvelocity_FerryJAP1979}. Furthermore, additional relaxation mechanisms, particularly impact ionization, can provide extra stabilization channels beyond the impact ionization threshold $E_\text{th}$, potentially enabling higher breakdown fields.

Fig. \ref{fig_elphIIrate}b illustrates the temperature dependence of total $\lambda^{-1}(E)$ from 80 to 573 K. Higher temperatures yield larger $\lambda^{-1}(E)$ values, with increased curvature below $\sim 30$ meV. The temperature dependent $\lambda^{-1}(E)$ were extracted for several chosen energies (as indicated by arrows in Fig. \ref{fig_elphIIrate}b) and the results are shown in Fig. \ref{fig_elphIIrate}c. The temperature dependence exhibits distinct characteristics at different energies. At 5250 meV (> $E_\textsc{bz}$), the behavior follows
\begin{equation*}
\lambda^{-1}(T,E)= 0.5\lambda^{-1}_0(E)\text{coth}\left(0.5\beta\hbar\omega\right)
\label{eqn_invlambdaTdepAc}
\end{equation*}
with $\hbar\omega=13$ meV, indicating a dominance of \textsc{ac} phonons. Near 86 meV, the dependence follows \begin{equation*}
\lambda^{-1}(E,T)=  \lambda^{-1}_0(E) n_p[\omega_p(q)]
\label{eqn_invlambdaTdepOpt}
\end{equation*}
with $\hbar\omega=63$ meV above 200 K, revealing \textsc{lo} phonon dominance with possible contributions of \textsc{ac} phonons at lower temperatures. The behavior at 6 meV shows a transition from \textsc{ac} characteristics ($\hbar\omega=6$ meV) at low temperatures to \textsc{lo} behavior ($\hbar\omega=63$ meV) at high temperatures. Interestingly, at 204 meV, $\lambda^{-1}$ varies almost linearly with temperature, reflecting the combined influence of \textsc{ac},  \textsc{lo}, and  \textsc{to} contributions. These varied temperature dependencies demonstrate that $\lambda^{-1}(E)$ cannot be simply scaled from room temperature values but must be calculated directly for each temperature.

\subsubsection{Electron transport}
\begin{figure*}
\centering
\includegraphics[width=0.96\linewidth]{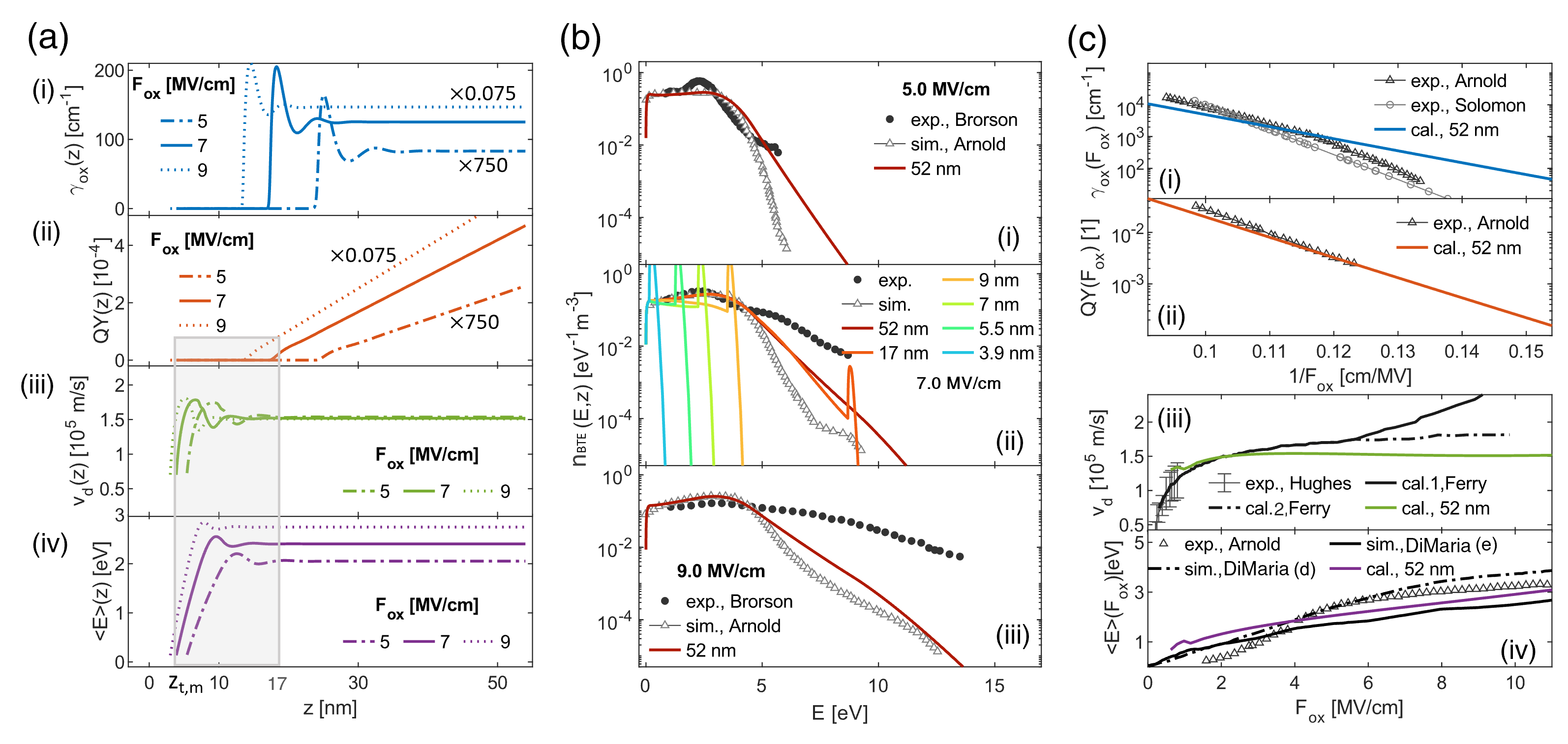}
\caption{Electron transport characteristics in a-SiO$_2$ under a constant field $F_\text{ox}$ at 298 K. (a) Spatial evolution of transport parameters versus distance $z$ from the SiC/a-SiO$_2$ interface: (i) II coefficient $\gamma_\text{ox}(z)$, (ii) quantum yield $\text{QY}(z)$, (iii) average drift velocity $v_d(z)$, and (iv) average electron energy $\langle E\rangle(z)$ for different field strengths. Note: $\gamma_\text{ox}$ and $\text{QY}(z)$ values for 9.0 MV/cm and 7.0 MV/cm are scaled by 0.075 and 750 respectively for clarity. The shaded area indicates the dark space region. (b) Normalized EDFs $n_\textsc{bte}(E,z)$ for fields of (i) 5.0 MV/cm, (ii) 7.0 MV/cm, and (iii) 9.0 MV/cm at selected distance $z$, compared with Monte Carlo results \cite{transportMC_ArnoldPRB1994} and experimental data from DiMaria \textit{et al.} \cite{trapcreatation_DiMariaJAP1989} and Brorson \textit{et al.} \cite{EDFsdirectmeas_BrorsonJAP1985}. (c) Field dependence of transport parameters at $z=52$ nm: (i) II coefficient $\gamma_\text{ox}$ and (ii) quantum yield $\text{QY}$ compared with experimental data \cite{transportMC_ArnoldPRB1994,IIcoefaSiO2_Solomon}, showing agreement with the Chynoweth model \cite{IIcoef_Chynoweth1958}; (iii) average drift velocity $v_\text{d}$ compared with experiments \cite{driftvelocity_HughesPRL1973} and Ferry's calculations \cite{driftvelocity_FerryJAP1979}; (iv) average energy $\langle E\rangle$ compared with experimental measurements \cite{transportMC_ArnoldPRB1994} and DiMaria's Monte Carlo simulations with and without quantum effects \cite{EavgthrdMCsim_DiMaria}.}
\label{fig_IIFoxdep}
\end{figure*}

We begin our discussion on electron transport in a-SiO$_2$ by solving the 1D BTE under uniform electric fields. The solution provides the electron's EDF $n_\textsc{bte}(E,z)$ and enables extraction of the average drift velocity $v_\text{d}(z)$, average electron energy $\langle E\rangle(z)$, II coefficient $\gamma_\text{ox}(z)$, and  quantum yield $\text{QY}(z)$ as a function of  oxide field $F_\text{ox}$. 

Fig. \ref{fig_IIFoxdep} presents results at 298 K for fields ranging from 1.0 to 11.0 MV/cm.  The spatial evolution of transport characteristics at field strengths of 5.0, 7.0, and 9.0 MV/cm is given in Fig. \ref{fig_IIFoxdep}a. Taking $F_\text{ox}=7.0\text{ MV/cm}$ as an example, electron transport starts at $z_\text{t,m}$, but $\gamma_\text{ox}(z)$ remains negligible upto a field-dependent distance $l_\text{d}$, beyond which it rises sharply and then stabilizes with minor variations. This behavior demonstrates the ``dark space" concept \cite{transportMC_ArnoldPRB1994}, where electrons must traverse a minimum distance to accumulate energy from the field before being able to initiate impact ionization. Hence, the quantum yield $\text{QY}(z)$ increases significantly only beyond this ``dark space". Both average drift velocity $v_\text{d}(z)$ and average energy $\langle E\rangle(z)$ saturate approximately 5 nm beyond $z_\text{t,m}$. Interestingly,  $\langle E\rangle(z)$ shows a linear relationship with $z$ within $3\sim4$ nm after $z_\text{t,m}$ and thus exhibits near-ballistic behavior, which is consistent with experimental observations \cite{ballistictransportaSiO2_DiMariaPRL1986}.

The EDFs at different field strengths (Fig. \ref{fig_IIFoxdep}b) reveal crucial details about the transport mechanisms. For $F_\text{ox}=7.0\text{ MV/cm}$ ($V_\text{g}\approx40$ V), the initial EDF at $z_\text{t,m}\approx 3.9$ nm peaks around $E\approx0.1$ eV. After 13 nm transport ($z\approx 17$ nm), the EDF develops a significant tail above the impact ionization threshold $E_\text{th}$ ($\sim8.7$ eV at 298 K). Hence, the ``dark space" phenomenon originate from a simple fact that carriers must transverse sufficient distance to develop high-energy tails to trigger substantial impact ionization. Experimental EDFs from DiMaria et al. \cite{trapcreatation_DiMariaJAP1989} and Brorson \textit{et al.} \cite{EDFsdirectmeas_BrorsonJAP1985} show a secondary high-energy tail at energy between 5.0 and 8.0 eV. While Fischetti attributed this to the DOS of a-SiO$_2$ and quantum mechanical effects \cite{fischettiQuantumMC_PRL552475}, Bradford and Woolf suggested it results from mixing of EDFs caused by the pinholes in the aluminum contact of the MOSC studied \cite{gateFilter_Bradford01091991}. Beyond 9 eV, the EDFs decrease with energy because of relaxation due to impact ionization and the EDFs can thus stabilize. The presence of ballistic transport is evidenced by the spikes in these EDFs.


Fig. \ref{fig_IIFoxdep}c presents the field dependence of the transport parameters at $z=52$ nm, which is consistent with that from Anorld \text{et al.} \cite{transportMC_ArnoldPRB1994} and DiMaria \text{et al.} \cite{EavgthrdMCsim_DiMaria} for comparison. The calculated II coefficient $\gamma_\text{ox}(F_\text{ox})$ under homogeneous fields shows good agreement with the results from previous studies \cite{transportMC_ArnoldPRB1994,IIcoefaSiO2_Solomon}, and it follows the Chynoweth formula \cite{IIcoef_Chynoweth1958}:
\begin{equation}
\gamma_\text{ox}(F_\text{ox})= a\exp\left( -b/F_\text{ox}\right)
\label{eqn_Chynoweth}
\end{equation}
where the constants $a$ and $b$ vary with temperature. The simulated quantum yield $\text{QY}(F_\text{ox})$, which is typically the primary experimental observable from which $\gamma_\text{ox}$ is derived, matches well with carrier-separation measurements from Si FETs of comparable oxide thickness \cite{transportMC_ArnoldPRB1994}. Contrary to the assumption employed by Arnold \textit{et al.} \cite{transportMC_ArnoldPRB1994}, we believe that the $\text{QY}(F_\text{ox})$ measured at lower field is more reliable because other channels of injecting holes into the VB of a-SiO$_2$, such as anode hole injection \cite{DiMariaAnodeholeinjection_JAP80304}, can be active under high fields. The average drift velocity $v_\text{d}(F_\text{ox})$ initially exhibits a field-dependent growth  and then saturates  at approximately $1.52\times 10^{5}$ m/s. This value is consistent with experimental measurements at fields of about 1.0 MV/cm \cite{driftvelocity_HughesPRL1973}. For higher fields, where experimental data is unavailable, our results align more closely with Ferry's uncorrected Monte Carlo simulations (`cal. 2, Ferry' in Fig. \ref{fig_IIFoxdep}c-iii) than with his field-modified scattering calculations (`cal. 1, Ferry' in Fig. \ref{fig_IIFoxdep}c-iii) \cite{driftvelocity_FerryJAP1979}. The calculated average energy $\langle E\rangle(F_\text{ox})$ demonstrates a good agreement with DiMaria's quantum-corrected calculations (`sim., DiMaria (e)' in Fig. \ref{fig_IIFoxdep}c-iv) \cite{EavgthrdMCsim_DiMaria} and experimental data from 100 nm oxide measurements \cite{transportMC_ArnoldPRB1994}. Intriguingly, DiMaria's uncorrected calculations (`sim., DiMaria (d)' in Fig. \ref{fig_IIFoxdep}c-iv) show better agreement with experiments, underscoring the complexity of high-field electron transport phenomena.

\begin{figure*}
\centering
\includegraphics[width=0.96\linewidth]{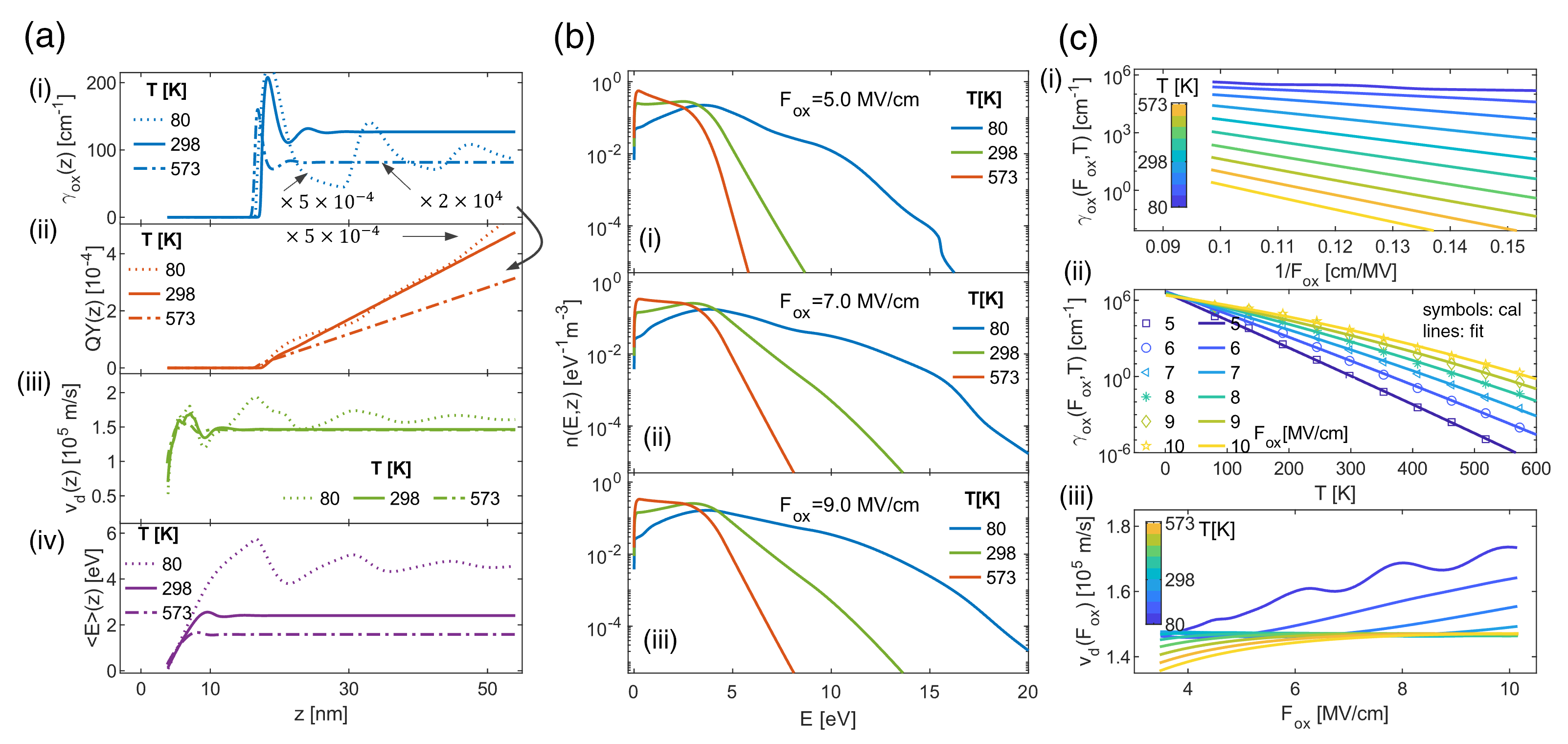}
\caption{Temperature-dependent electron transport in a-SiO$_2$ under uniform fields. (a) Spatial evolution at $F_\text{ox}=7.0$ MV/cm showing temperature dependence of: (i) II coefficient $\gamma_\text{ox}(z)$, (ii) quantum yield $\text{QY}(z)$, (iii) average drift velocity $v_\text{d}(z)$, and (iv) average electron energy $\langle E\rangle(z)$. Note: the $\gamma_\text{ox}$ and $\text{QY}(z)$ values at 80 K and 573 K are scaled by 5$\times 10^{-4}$ and 2$\times 10^{4}$ respectively. (b) Normalized EDFs at $z=52$ nm for temperatures 80, 298, and 573 K at fields: (i) 5.0 MV/cm, (ii) 7.0 MV/cm, and (iii) 9.0 MV/cm. (c) Temperature effects on: (i) field-dependent II coefficient across 80-573 K following Chynoweth's law, (ii) temperature dependence of $\gamma_\text{ox}$ at various fields following the model of Okuto and Crowell  \cite{okutoThresholdEnergyEffect1975}, and (iii) field-dependent drift velocity across the temperature range.}
\label{fig_IITdep}
\end{figure*}

The temperature dependence of electron transport, illustrated in Fig. \ref{fig_IITdep}, reveals significant variations across multiple parameters. At $F_\text{ox}=$ 7.0 MV/cm, both impact ionization coefficient $\gamma_\text{ox}(z,F_\text{ox})$ and quantum yield $\text{QY}(z,F_\text{ox})$ exhibit a strong temperature sensitivity, spanning eight orders of magnitude between 80 K and 573 K. The drift velocity $v_\text{d}(z)$ and average energy $\langle E\rangle(z)$ show pronounced spatial oscillations with reduced damping at lower temperatures, likely due to increased mean relaxation lengths. The EDFs at $z=52$ nm (Fig. \ref{fig_IITdep}b) become broader at lower temperatures for all field strengths (5.0, 7.0, and 9.0 MV/cm). This broadening reflects reduced scattering at lower temperatures, allowing electrons to reach higher energies due to smaller inverse MRLs $\lambda^{-1}(E,T)$. The field-dependent II coefficient $\gamma_\text{ox}(F_\text{ox},T)$ follows the Chynoweth's law \cite{IIcoef_Chynoweth1958} (Eq. \ref{eqn_Chynoweth}) across all temperatures, with temperature-dependent parameter $b$. The temperature dependence of $\gamma_\text{ox}$ follows the model of Okuto and Crowell \cite{okutoThresholdEnergyEffect1975}:
\begin{align}
\gamma_\text{ox}(F_\text{ox},T) &= a_\text{ref}\left[1+c(T-T_\text{ref})\right]F_\text{ox} \nonumber \\
&\times\exp\left[-\left(\frac{b_\text{ref}\left\{1+d(T-T_\text{ref})\right\}}{F_\text{ox}}\right)^2\right]
\label{eqn_gammaoxTdep}
\end{align}
where $a_\text{ref}$ and $b_\text{ref}$ are  parameters at a reference temperature (typically $T_\text{ref}=300$ K), and $c$, $d$ are  fitting parameters independent of temeprature. For semiconductor-like behavior ($c,d\sim10^{-4}$), this simplifies to:
\begin{equation}
\ln(\gamma_\text{ox})\propto -\left(\frac{b_\text{ref}}{F_\text{ox}}\right)^2\left[1+2d(T-T_\text{ref})\right]
\label{eqn_gammaoxTdepsimple}
\end{equation}
producing linear $\ln(\gamma_\text{ox})$-$T$ relationships with slopes proportional to $1/F_\text{ox}^2$, as confirmed in Fig. \ref{fig_IITdep}c-(ii). Quadratic corrections become significant only at high fields and low temperatures. The drift velocity $v_\text{d}(F_\text{ox},T)$ shows substantial temperature dependence, exceeding or falling below the room-temperature value of 1.52$\times 10^5$ m/s at low and high temperatures, respectively.

This analysis demonstrates that our simplified 1D BTE model, which incorporates electron-phonon scattering and impact ionization, successfully captures the essential features of temperature-dependent electron transport in a-SiO$_2$.

\subsection{Hole trapping in a-SiO$_2$ }
\label{sec:charge_capture}
The generation of electron-hole pairs due to impact ionization in a-SiO$_2$ will lead to hole current in the VB and subsequent hole capture into traps in the entire insulator. The hole capture can be described by the MT model which incorporates the phonon-assisted tunneling as well. The parameters for the NMP model and MT model are the following: $c_0=9.83$, $\bar{N_\text{p}}=10^{25}$ s$^{-1}$ m$^{-2}$, and $\hbar\omega=63$ meV (in previous sections). One acceptor-like trap-band and two donor-like trap-bands were used and their parameters were listed in Table \ref{tab:holetrap_parameters}. The trap levels for the donor-like traps are consistent with the experimental ones \cite{holetrapband_CaiJAP83_851}, and are likely due to oxygen vacancies (cite dft result).  and the cross-sections for hole capture are close  to those in Ref. \cite{holemobiaSiO2_HughesAPL264362,wagerLowfieldTransportSiO22017}. The parameters of the acceptor-like traps are close to those in Ref. \cite{SchleichIEDM19} where these traps were proven to be responsible for the BTI in SiC MOSFETs.

\begin{figure}
\centering
\includegraphics[width=1.0\linewidth]{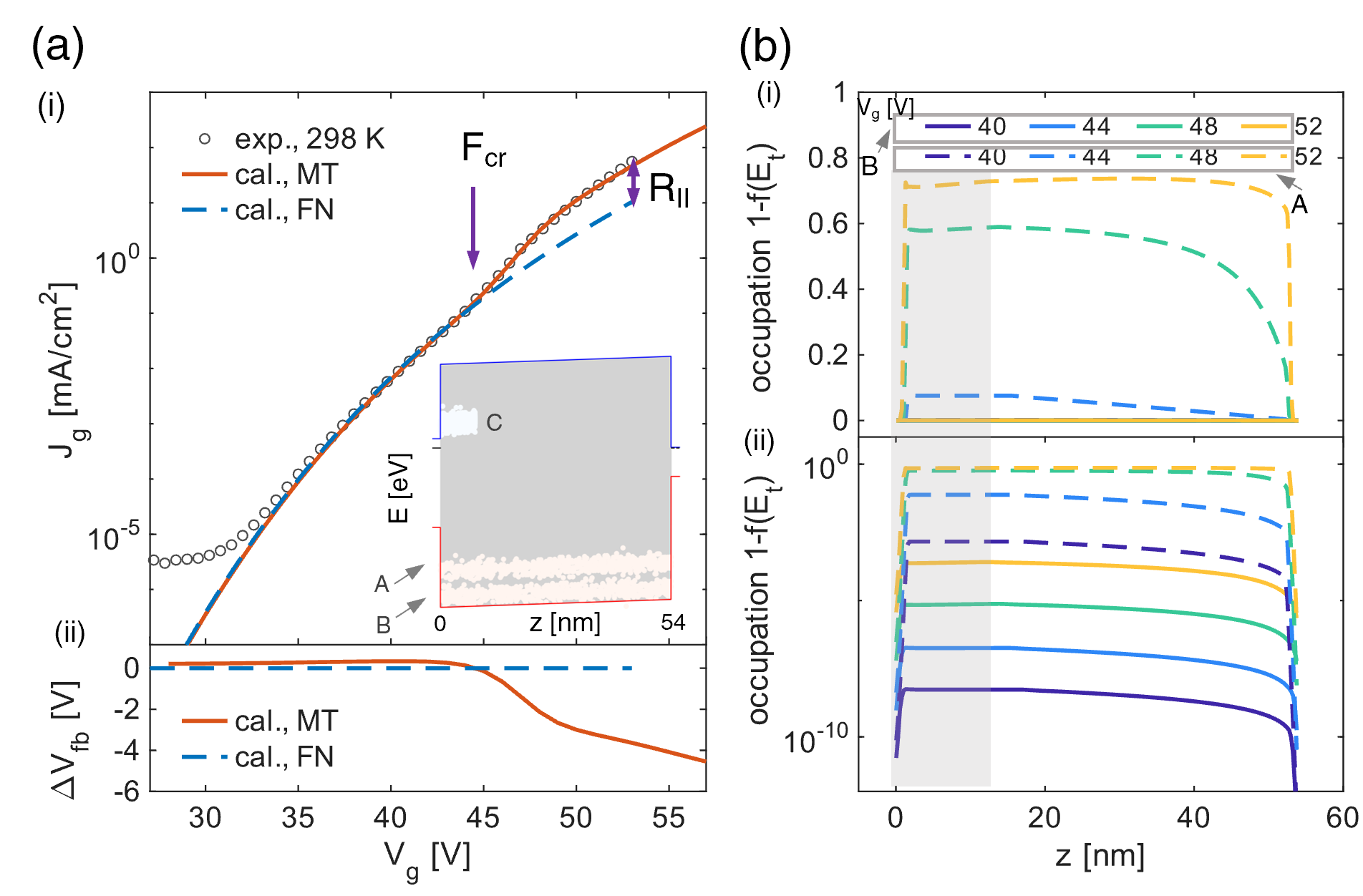}
\caption{Impact of hole capture in the a-SiO$_2$ bulk on gate leakage current  at 298 K. (a) Comparison of experimental and simulated results showing: ($i$) gate leakage current $J_\text{g}(V_\text{g})$ due to FN tunneling alone versus that due to  MT in a-SiO$_2$, with critical field $F_\text{cr}$ and amplification ratio $R_\textsc{ii}$ illustrated; ($ii$) the corresponding flat-band voltage shift $\Delta V_\text{fb}$. Inset: the band diagram illustrating donor-like trap-bands A and B, and acceptor-like trap-band C. (b) Spatial distribution of hole occupation $1-f(E_\text{t})$ for trap-bands A and B at their mean trap levels $E_\text{t}$, shown on a (i) linear and (ii) logarithmic scale, for gate bias $V_\text{g}=$ 40, 44, 48, and 52 V.}
\label{fig_T298Khump}
\end{figure}

\begin{table}
\begin{threeparttable}
\caption{Trap parameters for simulating charge capture in a-SiO$_2$ using the NMP model with linear electron-phonon coupling ( $R=1$). E$_\text{c}$ and E$_\text{v}$ are the CB and VB edges of a-SiO$_2$, respectively.}
\label{tab:holetrap_parameters}
\begin{ruledtabular}
\begin{tabular}{ccccccccc}
trap & type & $E_\text{t}$ & $\sigma_{\text{E}_t}$ & $E_\textsc{r}$ & $\sigma_{\text{E}_\textsc{r}}$ &  $\sigma_\text{ox,h}$ & $z$ & $N_t$ \\
 & &[eV] & [eV] & [eV] & [eV]  &[$\frac{\text{m}^{2}}{10^{18}}]$ & [nm] & [$\frac{10^{24}}{\text{m}^{3}}$]\\
\hline
A & donor & $E_\text{v}+1.45$  & 0.16 & 1.0 & 0.1  & 0.6&(0,54) & 0.5 \\
B & donor& $E_\text{v}+0.45$  & 0.16 & 1.0 & 0.1  & 0.6&(0,54) & 0.5 \\
C & acceptor & $E_\text{c}-2.29$  & 0.16 & 1.0 & 0.1  &$\frac{0.2}{10^{3}}$ \tnote{a} &(0,8.1) & 0.5 \\
\end{tabular}
\end{ruledtabular}
\begin{tablenotes}
\item[a]: cross-section for electrons
\end{tablenotes}
\end{threeparttable}
\end{table}

The gate leakage current $J_\text{g}$ during a $V_\text{g}$ sweep at 298 K was simulated with the MT model, and compared with the $J_\text{g}$ simulated by considering the FN tunneling alone. The results are presented in  Fig. \ref{fig_T298Khump}. $J_\text{g}$ from the FN tunneling mechanism alone provides a reference which is free from the influence of trapped charges in the oxide.  As shown in Fig. \ref{fig_T298Khump}a(i), the MT model quantitatively predicts the abrupt enhancement of $J_\text{g}$ at $V_\text{g}\approx 47$ V, which is absent in the the FN tunneling mechanism. When $V_\text{g}$ exceeds around 47 V, $J_\text{g}(V_\text{g})$ predicted by the MT model differs only by a scalar factor from that obtained from the FN tunneling model. This indicates that capture of holes into a-SiO$_2$ dominantly took place when $V_\text{g}$ was swept from 45 to 47 V and that  this capture is almost saturated when $V_\text{g}>47$ V. Such an evolution of trapped holes can be also understood from the shift of flat-band voltage $\Delta V_\text{fb}$ from the simulation shown in the (ii) panel of Fig. \ref{fig_T298Khump}a. The magnitude of $\Delta V_\text{fb}$ is consistent with experimental observations for SiC MOSFETs \cite{avramenkoThresholdVoltageDrift2024}. The electric field near the SiC/a-SiO$_2$ interface was thus enhanced in  a narrow $V_\text{g}$ window, leading to an abrupt enhancement (or ``hump") of $J_\text{g}$. 

We define a critical field $F_\text{cr}$ characterizing the onset of  hole capture and thus the onset of the  ``hump". A further study of the temperature dependence of  $F_\text{cr}$ will be valuable from the point of view of reliability tests.  To quantify the current enhancement within the MT model, we define an amplification ratio:
\begin{equation}
R_\textsc{ii}=\frac{J_\text{g}(V_\text{g},\textsc{mt})}{J_\text{g}(V_\text{g},\textsc{fn})}\big{|}_{V_\text{g}=V_\text{m}}
\end{equation}
where $V_\text{m}$ is the maximum $V_\text{g}$ after the $V_\text{g}$ for which the hump has fully developed.  Analogous to quantum yield $\text{QY}$, the yield of the amplification ratio can be defined as
\begin{equation}
\text{RY} = {R}_\textsc{ii}-1
\end{equation}
which is the net gain of $J_\text{g}$ due to impact ionization and hole capture. $\text{RY}$ will be zero when both effects are absent.

Fig. \ref{fig_T298Khump}b shows  the spatial variation of occupations of holes, i.e., $1-f(E_\text{t})$, at the mean trap levels $E_\text{t}$ for trap-band A and trap-band B at variable $V_\text{g}$, in both linear (i) and logarithmic (ii) scales. Hole occupations increase with increasing $V_\text{g}$ and show distinct spatial characteristics. The hole occupation decreases with decreasing distance to the n$^+$-Si/a-SiO$_2$ interface, which follows the distribution of hole current (Eq. \ref{eqn_holecurrent}) and thus the capture coefficient (Eq. \ref{eqn_k12k21MT}). The hole occupation remain almost constant within around 10 nm from the SiC/a-SiO$_2$ interface, because a ``dark-space" forms there and constant hole capture is expected. Trap-band B exhibits lower hole occupation due to smaller barriers for hole emission (Eq. \ref{eqn_barrierNMP}).

\begin{figure}
\centering
\includegraphics[width=1.0\linewidth]{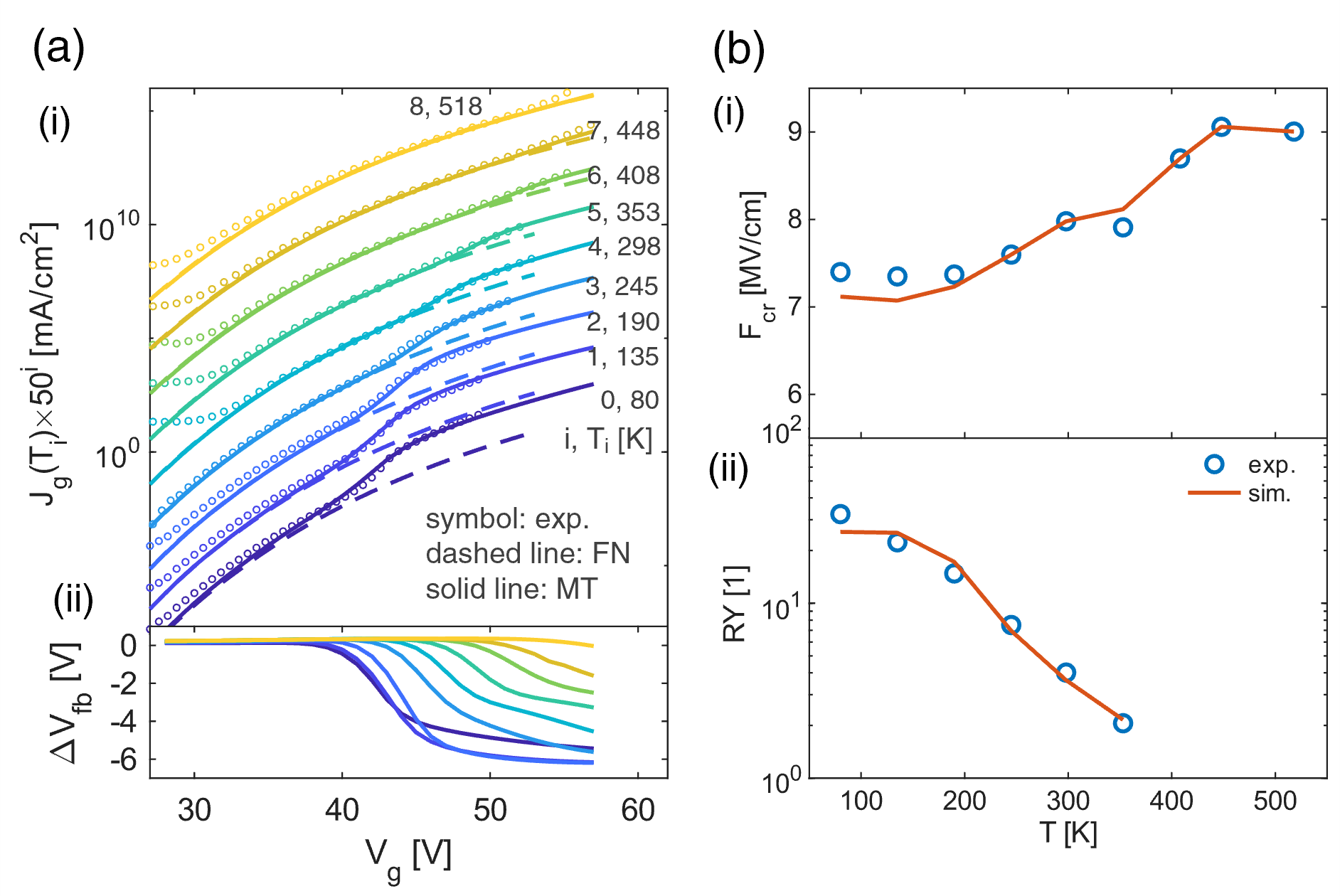}
\caption{Temperature-dependent gate current characteristics incorporating multiple trapping effects. (a) Gate current characteristics showing: (i) comparison of experimental data with simulations including and excluding multiple trapping model across different temperatures, and (ii) corresponding flat-band voltage shifts $\Delta V_\textsc{fb}$ from MT model. (b) Temperature dependence of (i) the critical field $F_\text{cr}$ marking ``hump" onset and (ii) the yield of amplification ratio $RY$, comparing experimental and simulated results. }
\label{fig_Tdephump}
\end{figure}
 
The $J_\text{g}(V_\text{g})$ were also simulated with the MT model for temperatures from 80 to 518 K, as shown in Fig. \ref{fig_Tdephump}. The simulated $J_\text{g}(V_\text{g})$ accurately reproduce the experimental humps across all temperatures considered (Fig. \ref{fig_Tdephump}a-(i)). The critical field $F_\text{cr}$ from the simulation agrees quite well with that from experiment, as shown in Fig. \ref{fig_Tdephump}b-i. In addition, the yield of amplication ratio RY shows a good agreement with the experimental ones [Fig. \ref{fig_Tdephump}b(ii)]. Interestingly, RY decreases almost exponentially with increasing temperature. The temperature dependency of RY is dominated by the II coefficient $\gamma_\text{ox}$. This is because RY is roughly proportional to the concentration of captured holes, which is proportional to the hole current in the VB of a-SiO$_2$. For example,  the hole current in the dark space $J_\text{p}(l_\text{d})$ can be approximated as 
\begin{equation*}
J_\text{p}(l_\text{d})= n_\textsc{fn}(z_\text{t,m})v_\text{d}(L)\text{QY}(L)\propto \bar{\gamma_\text{ox}}
\end{equation*}
which is proportional to the spatial average II coefficient $\bar{\gamma_\text{ox}}$. The exponential decay behavior can be readily understood from the temperature dependency of $\gamma_\text{ox}$ as shown in Eq. \ref{eqn_gammaoxTdepsimple}. 

These results demonstrate that gate leakage current under high-fields is dominated by the impact ionization and hole capture in a-SiO$_2$. The quantitative agreement between theory and experiment, achieved without artificial fitting parameters, represents progress toward predictive modeling of oxide charge capture-related reliability issues.

\section{Conclusions}
\label{sec:conclusion}
We have developed and validated a self-consistent modeling framework for gate leakage and hole trapping in the  gate oxide (a-SiO$_2$) of SiC MOSFETs. By coupling electrostatics, quantum tunneling, carrier transport, impact ionization, and charge trapping, the model reproduces measured leakage characteristics over a wide temperature range of 80-573 K and a wide bias range without adjustable parameters. The simulations reveal that electron-initiated impact ionization and subsequent hole capture in a-SiO$_2$ bulk significantly amplify leakage current by causing a shift of the flat-band voltage. This predictive capability closes a long-standing gap in reliability-oriented simulation for wide-bandgap devices, enabling more accurate lifetime projections and better-targeted stress tests. The methodology extends naturally to advanced Si technology with thick gate dielectric as is used in power devices.

\appendix
\section{Band structures}
\label{sec:App_bandstructure}
\begin{table*}[!ht]
\centering
\begin{threeparttable}
\caption{Band-structure parameterization for 4H-SiC, Si, and a-SiO$_2$ at 298 K with $k_\parallel=$ [001]. Electron masses are in units of $m_0$, the mass of an electron at rest. The  absence (NA) of non-parabolicity parameter $\alpha$  indicates the use of a parabolic band. The edge of the CB (VB) subbands $E_\text{c}$ ($E_\text{v}$) is referenced as zero. Spin degeneracy is not included in the valley degeneracy $g_{v}$. The `band 0' for the CB is a separate 3d spherical parabolic or Kane-type non-parabolic band model as a reference. The electron affinity $\chi$ is given at 298 K and includes the shift of $E_\text{c}$ due to band-gap narrowing.}
\begin{ruledtabular}
\begin{tabular}{l*{20}{c}}
\multirow{4}{*}{CB} 
& \multicolumn{6}{c}{band 1} & \multicolumn{6}{c}{band 2} & \multicolumn{5}{c}{'band 0'} &  &  &\\
 \cmidrule(lr){2-7} \cmidrule(lr){8-13} \cmidrule(lr){14-18} \cmidrule(lr){19-19}
& $m_\parallel$ & $\alpha_\parallel$ & $m_\perp$ & $\alpha_\perp$  & $E_\textsc{c}$ & $g_v$ &$m_\parallel$ & $\alpha_\parallel$ & $m_\perp$ & $\alpha_\perp$ & $E_\textsc{c}$ & $g_v$ & $m_\parallel$ &$m_\perp$& $\alpha$  & $E_\textsc{c}$  & $g_v$ & $\chi$ &  \multirow{2}{*}{Ref.}\\
& [$m_0$] & [$\frac{1}{\text{eV}}$] & [$m_0$] & [$\frac{1}{\text{eV}}$]  & [meV] & [1] &[$m_0$] & [$\frac{1}{\text{eV}}$] & [$m_0$] & [$\frac{1}{\text{eV}}$]  & [meV] & [1] & [$m_0$] & [$m_0$]  & [$\frac{1}{\text{eV}}$]  & [meV]  & [1] &[eV] & &\\
\cmidrule(lr){2-7} \cmidrule(lr){8-13} \cmidrule(lr){14-18} \cmidrule(lr){19-19}
4H-SiC & 0.31 & 1.013 & 0.441 & 0.23 &0 &3 &0.73 & 1.5 & 0.33 & 0.023 &115 & 3 & 0.28 &0.42 & 0.323 &0 & 3& 3.48\tnote{a} & \cite{massDFTplusexp_KaczerPRB1998,MCeltransp_MickevciusJAP1998}\\
\multirow{2}{*}{Si} &0.916 &0.28 &0.19 &0.25 &0 &2 & & & & & & &\multirow{2}{*}{0.32} & &\multirow{2}{*}{0.5}  &\multirow{2}{*}{0} &\multirow{2}{*}{6} &\multirow{2}{*}{4.08\tnote{a}} &  \multirow{2}{*}{\cite{bandstructSilicon_CohenPRB1974}} \\
 &0.191\tnote{d} &0.25 &0.42 &0.26 &0 &4 & & & & & & & &\\
a-SiO$_2$ & & & & & & & & & & &  & &0.42 &0.42&NA &0 &1 &0.75\tnote{a} &\cite{meffSiO2_MaserjianJVST1974,tunnelingthinoxide_Schenk} \\
\cmidrule{1-20}
\multirow{4}{*}{VB} 
& \multicolumn{6}{c}{heavy hole} & \multicolumn{6}{c}{light hole} & \multicolumn{6}{c}{split-off hole} & \\
 \cmidrule(lr){2-7} \cmidrule(lr){8-13} \cmidrule(lr){14-19}
& $m_\parallel$ & $\alpha_\parallel$ & $m_\perp$ & $\alpha_\perp$  & $E_\textsc{v}$ & $g_v$ &$m_\parallel$ & $\alpha_\parallel$ & $m_\perp$ & $\alpha_\perp$ & $E_\textsc{v}$ & $g_v$ &$m_\parallel$ & $\alpha_\parallel$ & $m_\perp$ & $\alpha_\perp$ & $E_\textsc{v}$ & $g_v$ &  \multirow{2}{*}{Ref.} \\
& [$m_0$] & [$\frac{1}{\text{eV}}$] & [$m_0$] & [$\frac{1}{\text{eV}}$]  & [meV] & [1] &[$m_0$] & [$\frac{1}{\text{eV}}$] & [$m_0$] & [$\frac{1}{\text{eV}}$]  & [meV] & [1] & [$m_0$] & [$\frac{1}{\text{eV}}$]  & [$m_0$]  & [$\frac{1}{\text{eV}}$] & [meV] &[1] & &\\
\cmidrule(lr){2-7} \cmidrule(lr){8-13} \cmidrule(lr){14-19}
4H-SiC & 1.60 & NA & 3.28 & NA &0 &1 &1.60 & NA & 0.325 & NA &-8.6 & 1 & 0.22 & NA  &1.49 & NA&-73 &1  &\cite{VBofSiC_PersonLindefeltJAP1997}\\
Si &0.254 &NA &0.254 &NA &0 &1 &0.188 &NA &0.188 &NA &0 &1 &0.222 &NA &0.222 &NA &-44  &1 & \cite{massDFT_JanssenPRB2016}\\
a-SiO$_2$ &2.0 &NA &2.0 &NA &0 &1 &1.0 &NA &1.0 &NA &0  &1  &1.75 &NA &1.75 &NA &0  &1 & \tnote{b}\\
\end{tabular}
\end{ruledtabular}
\label{tab:conduction_band}
\begin{tablenotes}
\item[a] Ref. \cite{elXiSiC_Fujimura_2016}  
\item[b] see section \ref{sec:transport}
\end{tablenotes}
\end{threeparttable}
\end{table*}

The CB and VB are parameterized for computation of the free-carrier concentration (Eq. \ref{eqn_elhconc}) and the sheet charge for tunneling into the oxide (Eq. \ref{eqn_eleconc4tunnel}). Hence, a cylindrical coordinate system in $k$-space is used, where $k_\parallel$ is along the tunneling direction $\vec{z}$, and rotational symmetry is assumed in the $k_x\text{-}k_y$ plane. The CB can be modeled by one (multiple) Kane-type nonparabolic band (subbands) for a wave-vector parallel to $k_z$ and a wave-vector in  the $k_x\text{-}k_y$ plane, respectively
\begin{subequations}
\begin{align}
&\frac{(\hbar k_\parallel)^2}{2m_\parallel} = E_\parallel(k_\parallel)[1+\alpha_\parallel E(k_\parallel)],\\
&\frac{\left(\hbar k_{\perp,1}\right)^2}{2m_{\perp,1}} + \frac{\left(\hbar k_{\perp,2}\right)^2}{2m_{\perp,2}} =\frac{(\hbar k_\perp)^2}{2m_\perp}= E_\perp(k_\perp)[1+\alpha_\perp E(k_\perp)]
\end{align}
\label{eqn_Ekparal}
\end{subequations}
in which $m_\parallel$ ($m_\perp$) and $\alpha_\parallel$ ($\alpha_\perp$) are the effective mass at the respective band-edge and the nonparabolicity parameter alone $k_\parallel$ ($k_\perp$ in the $k_x\text{-}k_y$ plane). When $\alpha_\parallel=0$, the band will be parabolic.  


The VB can often be modeled by three distinct bands, namely heavy-hole, light-hole, split-off hole bands. The VB of SiC from DFT and $\mathbf{k}\cdot\mathbf{p}$ calculations \cite{VBofSiC_PersonLindefeltJAP1997} show spin-orbit and crystal-field splittings of 8.6 meV and 73 meV, respectively. The split-off hole band of Si is spherical, while heavy-hole and light-hole bands show non-parabolicty \cite{VBofSilicon_Kane} but can be approximated as parabolic bands with appropriate averaged effective masses. The VB of SiO$_2$ is modeled by three parabolic bands with effective masses in agreement with reports in literature and no splitting was considered. The effective mass and non-parabolicity parameters for 4H-SiC, Si, and a-SiO$_2$ at 298 K with $k_\parallel=$ [001] are tabulated in Table \ref{tab:conduction_band}. 

\begin{table}
\centering
\begin{threeparttable}
\caption{Band gap and band gap narrowing (BGN) parameterization for 4H-SiC, Si, and a-SiO$_2$. The presence (absence) of $p$ implies the use of the Passler Eq. \ref{eqn_EgPassler} (O'Donnel and Chen Eq. \ref{eqn_EgODonnel}).}
\begin{ruledtabular}
\begin{tabular}{lcccccc}
\multirow{2}{*}{band gap} 
& $E_\text{g}(0\text{K})$ & $S_\infty$ & $\Theta_p$ & $p$ & $r_\text{g}$ & \multirow{2}{*}{Ref.} \\
& [eV] & [$k_\textsc{b}$] & [K] & [1] & [1] &  \\
\colrule
4H-SiC &3.297 &3.361 &414.3 &3.192 &0.435\tnote{a} & \cite{EgSiC_GrivicksJAP2007} \\
Si &1.170 &3.69 &406 &2.33 &0.5 & \cite{BandgapFit_PasslerPSSb} \\
a-SiO$_2$ &8.946 &12.21 &905.2 & / &0.5 & \cite{TdepEgaSiO2_SaitoPRB}\\
\colrule
\multirow{2}{*}{BGN} & $A_\text{nc}$ & $B_\text{nc}$ & $A_\text{nv}$ & $B_\text{nv}$ & $N_\text{ref}$  & \multirow{2}{*}{Ref.}\\
  & [meV] & [meV] & [meV] & [meV] & [$\frac{10^{24}}{\text{m}^{3}}$] &  \\
\colrule
n-4H-SiC &-17.91 &-2.2 &28.23 &6.24 &1.0 &\cite{BGN_PerssonJAP1999}\\
n-Si &-14.84 &0.78 &15.08 &0.74 &1.0 & \cite{BGN_PerssonJAP1999}\\
\end{tabular}
\end{ruledtabular}
\label{tab:bandgap}
\begin{tablenotes}
\item[a] Ref. \cite{bandshiftSiC_CannucciaPRMater}  
\end{tablenotes}
\end{threeparttable}
\end{table}

Crucial for evaluating the free-carrier concentration, the Fermi-level offset $E_\textsc{cf}$ ($E_\textsc{fv}$) depends on the shifts of CB edges $E_\textsc{c}$ (VB edge $E_\textsc{v}$) as a function of temperature and dopant concentration. The temperature-induced band-edge shifts are defined for the CB and VB as
\begin{subequations}
\begin{align}
&\Delta E_\text{g}(T) = E_\text{g}(T)-E_\text{g}(0),\\
&\Delta E_\textsc{c}(T) = E_\textsc{c}(T)-E_\textsc{c}(0) = r_\text{g}(T)\Delta E_\text{g}(T),\\
&\Delta E_\textsc{v}(T) = E_\textsc{v}(T)-E_\textsc{v}(0) = [r_\text{g}(T)-1]\Delta E_\text{g}(T), 
\end{align}
\end{subequations}
where $r_\text{g}(T)$ is the partition factor of $\Delta E_\text{g}(T)$ to $\Delta E_\textsc{c}(T)$. In practice, the temperature dependent band gap $E_\text{g}(T)$ and partition factor $r_\text{g}(T)$ are often parameterized to calculate the shifts of band-edges. Two parameterization schemes for $E_\text{g}(T)$ are popular among the many proposed band gap models \cite{bandgap_VarshiniPhysica1976, bandgap_ODonnelChenAPL91,BandgapFit_PasslerPSSb}. The one from Passler \cite{BandgapFit_PasslerPSSb} is
\begin{equation}
\Delta E_\text{g}(T) = -\frac{S_\infty\Theta_p}{2}\left(\left[1 + \left(2T/\Theta_p\right)^p\right]^{1/p}-1\right)
\label{eqn_EgPassler}
\end{equation}
where the parameters $S_\infty$, $\Theta_p$, and $p$ are associated with the entropy, average phonon temperature and anisotropy, respectively. The other one proposed by ODonnel and Chen \cite{bandgap_ODonnelChenAPL91} is,
\begin{equation}
\Delta E_\text{g}(T) = -\frac{S_\infty\Theta_p}{2}\left[\coth\left(\frac{\Theta_{p}}{2T}\right)-1\right].
\label{eqn_EgODonnel}
\end{equation}

Extrinsic dopants can cause significant band gap narrowing (BGN), which is specifically important for n$^+$-Si gate. BGN effects for n-type semiconductor can be modeled according to Ref. \cite{BGN_LindefeltJAP1998, BGN_PerssonJAP1999}:
\begin{subequations}
\begin{align}
\Delta E_\textsc{c} &= A_\text{nc}\left(\frac{N_{\text{d}}^+}{N_\text{ref}}\right)^{1/3} + B_\text{nc}\left(\frac{N_{\text{d}}^+}{N_\text{ref}}\right)^{1/2} \\
\Delta E_\textsc{v} &= A_\text{nv}\left(\frac{N_{\text{d}}^+}{N_\text{ref}}\right)^{1/4} + B_\text{nv}\left(\frac{N_{\text{d}}^+}{N_\text{ref}}\right)^{1/2}
\end{align}
\end{subequations}
where the ionized cocentration can be replaced by the nominal total dopant concentration without introducing much error \cite{BGN_SchenkJAP1998}. Similar models exist for p-type semiconductors as well \cite{BGN_LindefeltJAP1998, BGN_PerssonJAP1999}. The temperature dependence of BGN is negligible (a few meV) for temperature ranging from 1 to 1000 K, at least for Si according to Ref. \cite{BGN_SchenkJAP1998}. 

The parameterization of the band gap $E_\text{g}(T)$ and BGN for 4H-SiC, Si, and a-SiO$_2$ are tabulated in Table \ref{tab:bandgap}.
\begin{table}
\caption{Dopant energy level models $E_\text{dop}(N)$, $b$ model and the valley degeneracy model $g_v$ (without spin) for SiC and Si. }
\begin{ruledtabular}
\begin{tabular}{llccccccccc}
$E_\text{dop}$
& & \multicolumn{3}{c}{Altermat \textit{et al.}} & \multicolumn{2}{c}{Pearson and Bardeen }  & \multirow{4}{*}{Ref.}\\
\cmidrule(lr){3-5} \cmidrule(lr){6-7}
model
& dopant  & $E_\text{dop0}$ & $N_\text{ref}$ & $c$  & $E_\text{dop0}$ & $\alpha_d$ \\
&  & [meV] & [$\frac{10^{24}}{\text{m}^{3}}$]  &[1] & [meV] & [$\frac{\text{meV}}{10^8\text{m}^{-1}}$] & \\
\colrule
\multirow{2}{*}{4H-SiC} & N(h)  & & &  &70.9 &33.8& \cite{SiCdopantlevel_KagamiharaJAP2004}\\
& N(k)  & & &  &123.7 &46.5& \cite{SiCdopantlevel_KagamiharaJAP2004}\\
4H-SiC& Al &214.9 &81.2 &0.632 & & &\cite{incompleteionization_AlSiCJAP2019} \\
Si &P &45.5 &3 &2.0 & & &  \cite{IncompleteIonizationpaperII_AltermattJAP2006} \\
Si &B &44.4 &1.3 &1.4 & & &  \cite{IncompleteIonizationpaperII_AltermattJAP2006}\\
\colrule
$b$
& & \multicolumn{3}{c}{Altermat \textit{et al.}} & \multicolumn{2}{c}{Pearson and Bardeen}  & \multirow{4}{*}{Ref.}\\
\cmidrule(lr){3-5} \cmidrule(lr){6-7}
model
& dopant  & $N_\text{b}$ & $N_\text{ref}$ & $d$  & $b$ & & \\
&  & [$\frac{10^{24}}{\text{m}^{3}}$] & [$\frac{10^{24}}{\text{m}^{3}}$]  &[1] & [1] &  &\\
\colrule
\multirow{2}{*}{4H-SiC} & N(h) & & & &1.0 & &   \\
& N(k) & & & &1.0 & & & \\
4H-SiC & Al&450 &81.2 &2.9 & &  &\cite{incompleteionization_AlSiCJAP2019} \\
Si &P &6.0 &3.0 &2.3 & & & \cite{IncompleteIonizationpaperII_AltermattJAP2006} \\
Si &B &4.5 &1.3 &2.4 & & & \cite{IncompleteIonizationpaperII_AltermattJAP2006} \\
\colrule
$g_{v}$
& &\multicolumn{5}{c}{$g_{v}=g_{v,0} + \sum_{i=1}^{2}g_{v,i}\exp\left(-\frac{\Delta E_{\text{vo},i}}{k_\textsc{b}T}\right)$} & \multirow{4}{*}{Ref.}\\
\cmidrule(lr){3-7}
model
& dopant & $g_{v,0}$ &$g_{v,1}$ & $\Delta E_{\text{vo},1}$ &$g_{v,2}$  &$\Delta E_{\text{vo},2}$  & & \\
&  & [1] &[1] & [meV] &[1]  &[meV]   &  \\
\colrule
\multirow{2}{*}{4H-SiC} & N(h) &1 &2 &7.6 & & &\cite{Nitrogenhksite_GotzJAP1993}  \\
& N(k) &1 &2 &45.5 & & &\cite{muzafarovaElectronicStructureSpatial2016}\\
4H-SiC & Al&2 &/ &$\infty$ & &  \\
Si &P &1 &1 &11.7 &1 &13.0 & \cite{PinSiliconVOsplitting_JainPRB1976} \\
Si &B &2 &2 &39 & & &\cite{vosplitting_BoroninSilicon}  \\
\end{tabular}
\end{ruledtabular}
\label{tab:dopant_models}
\end{table}

\section{Incomplete ionization}
\label{sec:APP_incompleteionization}
The ionization energy $E_\text{d/a}$ of a dopant at a specific site depends on the total dopant concentration $N_\text{d/a}$. The Altermatt \textit{et al.} model \cite{IncompleteIonizationpaperI_AltermattJAP2006,IncompleteIonizationpaperII_AltermattJAP2006}, and Pearson and Bardeen model \cite{IncompleteIonization_PearsonBardenPhysRev1949} are described as, respectively
\begin{subequations}
\begin{align}
E_\text{d/a}(N_\text{d/a}) &= E_\text{d/a,0} - \alpha_\text{d/a} \times \left(N_\text{d/a}\right)^{\frac{1}{3}}, \\
E_\text{d/a}(N_\text{d/a}) &= E_\text{d/a,0}\left[1 + \left(\frac{N_\text{d/a}}{N_\text{ref}}\right)^c\right]^{-1}
\end{align}
\end{subequations}
where $E_\text{d/a,0}$ is the ionization energy at the dilute limit of dopants ($N_\text{d/a}\rightarrow0$).  The clustering parameter $b$ only comes into effect above a critical dopant concentration $N_\text{ref}$ and is modeled as \cite{IncompleteIonizationpaperI_AltermattJAP2006,IncompleteIonizationpaperII_AltermattJAP2006}
\begin{equation}
b(N_\text{d/a}) =  \left\{\begin{array}{cc}
1.0, & N_\text{d/a}\leq N_\text{ref} \\
\left[1+\left(\frac{N_\text{d/a}}{N_\text{b}}\right)^d\right]^{-1}, & N_\text{d/a}>N_\text{ref}
\end{array} \right.
\end{equation}
where $N_\text{b}$ is a characteristic dopant concentration. 

To account for temperature effects on valley-orbital splitting and excited states, we use the effective valley degeneracy
\begin{equation}
g_{v}=g_{v,0} + \sum_{i=1}^{N}g_{v,i}\exp\left(-\frac{\Delta E_{\text{vo},i}}{k_\textsc{b}T}\right), v=\text{d, a}
\end{equation}
where $g_{v,0}$ represents the ground state degeneracy, and the summation accounts for higher energy levels with splitting $\Delta E_{\text{vo},i}$ and degeneracy $g_{v,i}$. For computational efficiency, we typically limit $N$ to 2, capturing the most significant excited states. Table \ref{tab:dopant_models} provides the model parameters for 4H-SiC and Si.

\section{Tunneling cofficient Eq. \ref{eqn_DExdeeptunnel}}
\label{sec:APP_DExdeeptunnel}
The variablle $\xi>0$ for tunneling through beneath the barrier for which $\text{z}>0$.  Using the asymptotic formulae for Airy functions,  we expressed Airy functions as follows
\begin{subequations}
\begin{align}
\text{Ai}(\xi)&=\frac{\exp(-\Theta)}{2\sqrt{\pi}\xi^{1/4}}\text{eAi}(\xi), \\
 \text{Bi}(\xi)&= \frac{\exp(\Theta)}{\sqrt{\pi}\xi^{1/4}}\text{eBi}(\xi),\\
\left|\text{Ai}^\prime(\xi)\right|&=\frac{\exp(-\Theta)\xi^{1/4}}{2\sqrt{\pi}}\text{eAip}(\xi), \\
 \text{Bi}^\prime(\xi)&=\frac{\exp(\Theta)\xi^{1/4}}{\sqrt{\pi}}\text{eBip}(\xi),
\end{align}
\label{eqn_scaledAiry}
\end{subequations}
where the functions $\text{eAi}(\xi),\text{eBi}(\xi),\text{eAip}(\xi),\text{eBip}(\xi)$ aproach unity as $\xi\rightarrow\infty$.  The variable  $\omega$ can be expressed as follows
\begin{equation}
\omega = \text{sgn}(\text{z})\frac{r_\text{v}}{\sqrt{\left|\xi\right|}}.
\end{equation}
where the new variable $r_\text{v}$ is defined as below 
\begin{equation*}
r_\text{v} = \frac{k_{\parallel,\text{ox}}/m_\text{ox}}{k_1/m_{\parallel}}.
\end{equation*}

\begin{figure}
\centerline{\includegraphics[width=0.8\linewidth]{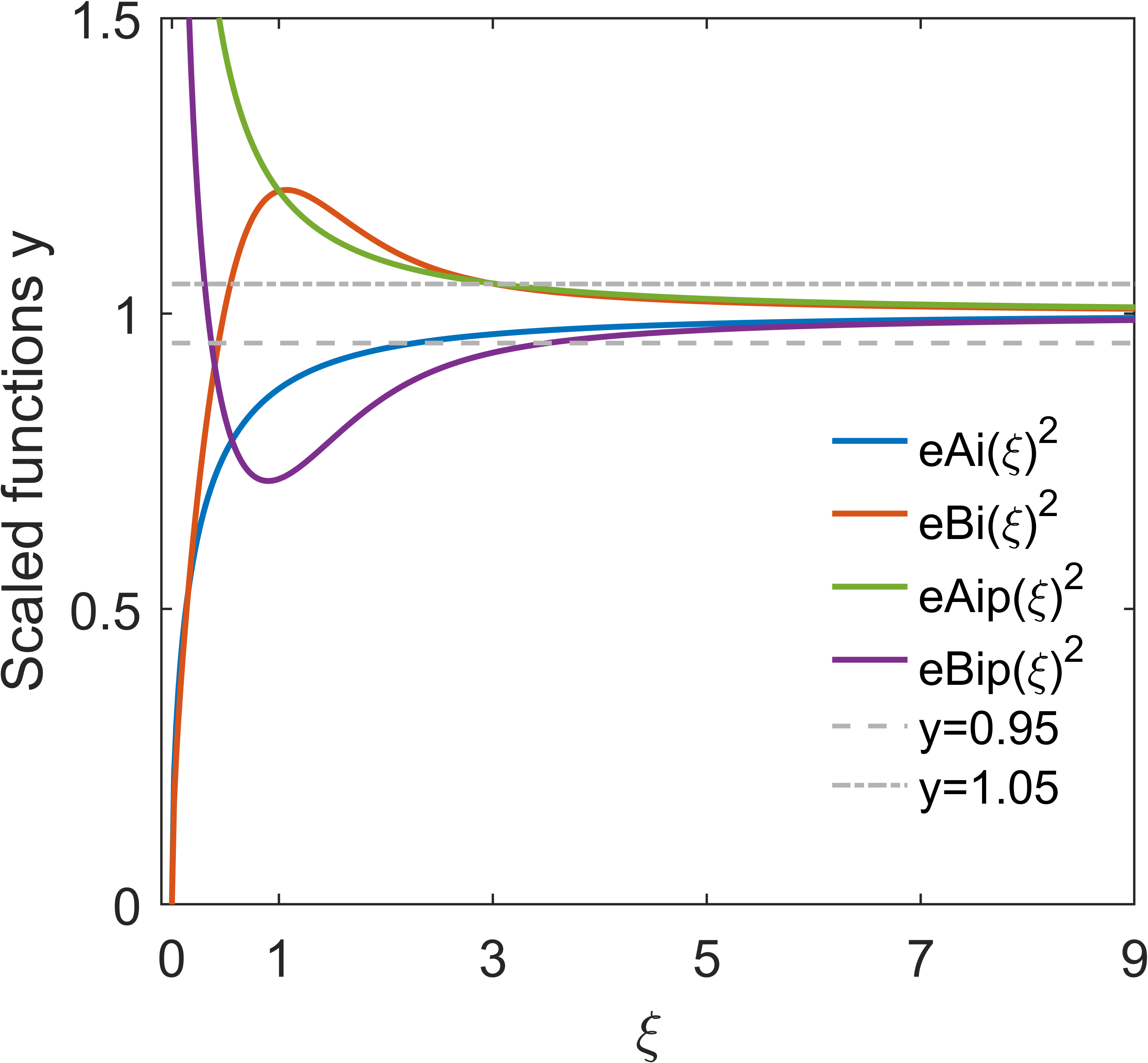}}
\caption{Functions $\text{eAi}(\xi)^2,\text{eBi}(\xi)^2,\text{eAip}(\xi)^2$, and $\text{eBip}(\xi)^2$, which deviate within 5\% from 1.0  when $\xi>\sim 3.5$. }
\label{fig_scaledAiry}
\end{figure}

Hence $D(\xi)$ (Eq. \ref{eqn_TEx_full}) can be expressed as
\begin{widetext}
\begin{subequations}
\begin{align}
D(\xi) &= \frac{4\pi^{-1}\omega^{-1}}{\frac{1}{r_\text{v}\pi} \left[\frac{1}{4}{\text{eAi}}(\xi)^2\exp(-2\Theta) + \text{eBi}(\xi)^2\exp(2\Theta)\right]+\frac{r_\text{v}}{\pi}\left[\frac{1}{4}\text{eAip}(\xi)^2\exp(-2\Theta)+\text{eBip}(\xi)^2\exp(2\Theta)\right]\exp(-2\Theta) + \frac{2}{\pi}} \nonumber \\
&=\frac{\omega^{-1}}{\frac{1}{4}\left[{r_\text{v}}^{-1}\text{eBi}(\xi)^2+r_\text{v}\text{eBip}(\xi)^2\right]\exp(2\Theta)+\frac{1}{16}\left[{r_\text{v}}^{-1}\text{eAi}(\xi)^2+r_\text{v}\text{eAip}(\xi)^2\right]\exp(-2\Theta)+1/2}\nonumber.
\end{align}
\end{subequations}
\end{widetext}
Fig. \ref{fig_scaledAiry} shows  $\text{eAi}(\xi)^2,\text{eBi}(\xi)^2,\text{eAip}(\xi)^2$, and $\text{eBip}(\xi)^2$, which deviate from 1.0 within 5\% when $\xi>\sim3.5$, and thus are approximated to be 1.0. Hence, the tunneling coefficient can be approximated as
\begin{align}
D(\xi)& \approx \frac{\omega^{-1}}{\frac{\exp(2\Theta)}{P_\xi}+\frac{\exp(-2\Theta)}{4P_\xi}+\frac{1}{2}} \nonumber\\
& = \frac{\omega^{-1}P_\xi \exp(-2\Theta)}{1+\frac{1}{2}P_\xi\exp(-2\Theta)+\frac{1}{4}\exp(-4\Theta)}
\label{eqn_Dxi_approx_derivation}
\end{align}
in which
\begin{equation}
P_\xi = \frac{4}{r_\text{v}+{r_\text{v}}^{-1}}=\frac{4\left(k_1/m_\parallel\right)\left(k_{\parallel,\text{ox}}/m_\text{ox}\right)}{\left(k_1/m_\parallel\right)^2+\left(k_{\parallel,\text{ox}}/m_\text{ox}\right)^2}.
\end{equation}
The variable $\xi$ is calculated to be 4.17 at the field of 10 MV/cm, assuming $\Phi=2.7$ eV and $m_\text{ox}=0.42$ $m_0$. Hence,  Eq. \ref{eqn_Dxi_approx_derivation} serves as an excellent approximation in practice.

\section{Total phase shift with band-bending and image-force correction}
\label{sec:APP_totalTheta}
\begin{figure}
\centerline{\includegraphics[width=0.92\linewidth]{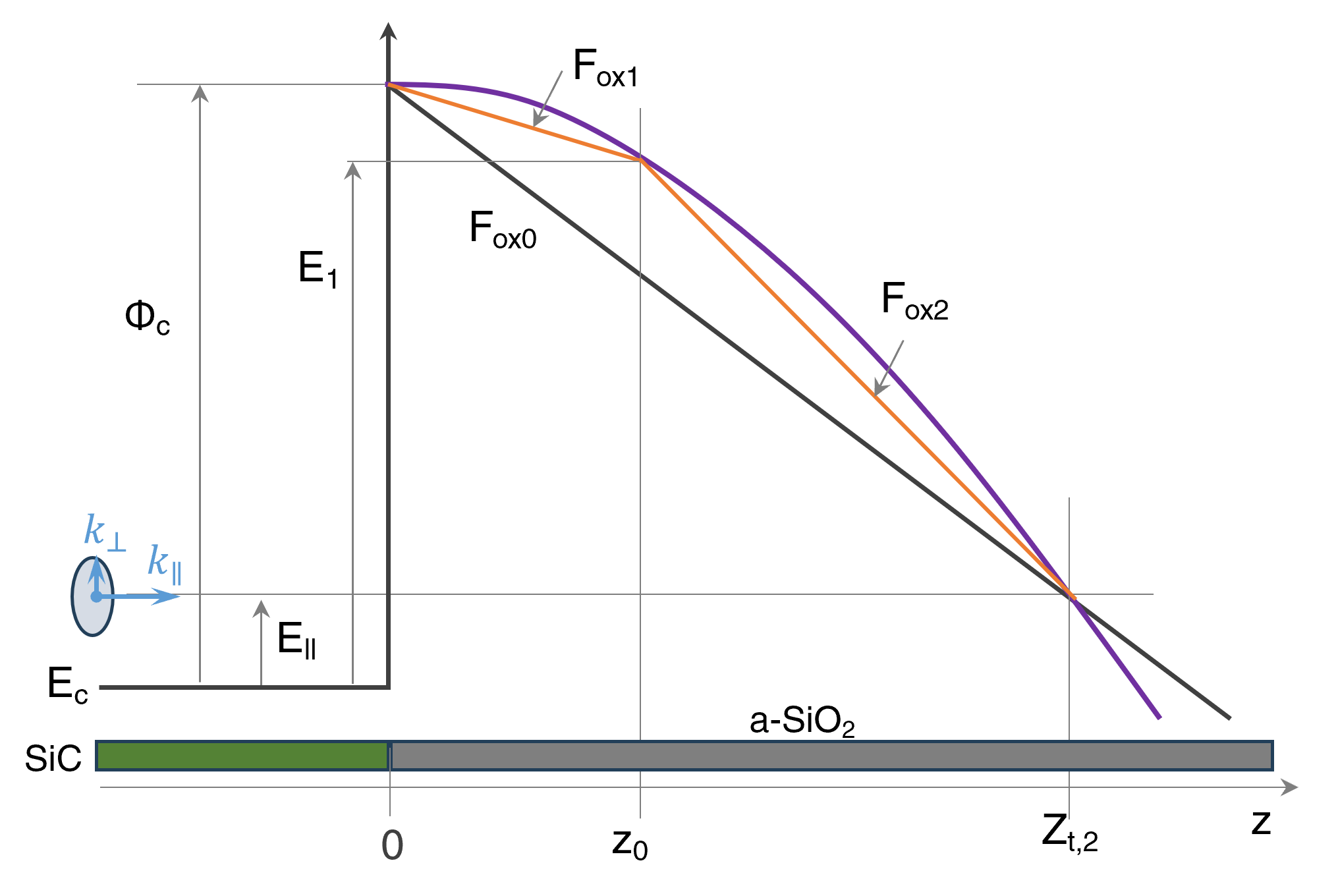}} 
\caption{Approximations to the CB edge $E_\text{c}(\text{z})$ of a-SiO$_2$. For each incident wave characterized by ($E_\parallel$, $E_\perp$), the CB edge $E_\text{c}(z)$ for $z\in(0,z_{\text{t},2})$ can be approximated by a straight line (zero$^\text{th}$-order approximation) or by two straight lines connected at $z_0$ (first-order approximation). The electric field strengths are denoted as $F_\text{ox0}$,$F_\text{ox1}$, and $F_\text{ox2}$, respectively. Note that $z=z_{\text{t},2}-\text{z}$ is used instead of z for clarity.} 
\label{figS_approx2tunnelpot}
\end{figure}

Fig. \ref{figS_approx2tunnelpot} illustrates the potential energy barrier for tunneling in the presence of charge trapping that changes the CB edge $E_\text{c}(z)$ of a-SiO$_2$ and the classical image-force correction, where  the variable $z=z_{\text{t},2}-\text{z}$ instead of $\text{z}$ is used for clarity. The effective mass of SiC is assumed to match with that of a-SiO$_2$, as a starting approximation. Hence only $E_\parallel$ matters for the tunneling coefficient. The first-order approximation of the tunneling potential profile yields $V_1(z)$ and $V_2(z)$:
\begin{subequations}
\begin{align}
V_1(z) &= V_{h1} - F_\text{ox1}z-\frac{\gamma_1}{4z},\quad V_{h1}=\Phi_\text{c}-E_\parallel\\
V_2(z) &= V_{h2} - F_\text{ox2}z-\frac{\gamma_1}{4z}, \quad V_{h2} = F_\text{ox2}z_{\text{t},2}
\end{align}
\end{subequations}
while the zero$^\text{th}$-order approximation yields
\begin{equation*}
V_0(z) = V_{h0} - F_\text{ox0}z-\frac{\gamma_1}{4z}, \quad V_{h0}=\Phi_\text{c}-E_\parallel.
\end{equation*}

The total phase shift comprises two components:
\begin{subequations}
\begin{align}
\Theta(E_\parallel) &= \Theta_1(E_\parallel) + \Theta_2(E_\parallel), \\
\Theta_1(E_\parallel) &= \frac{\sqrt{2m_\text{ox}}}{\hbar}\int_{z_{-}}^{z_0}\sqrt{|V_1(z)|}\mathrm{d}z ,\\
\Theta_2(E_\parallel) &= \frac{\sqrt{2m_\text{ox}}}{\hbar}\int_{z_0}^{z_{+}}\sqrt{|V_2(z)|}\mathrm{d}z .
\end{align}
\label{eqnS_phashiftdef}
\end{subequations}
in which $z_{-}$ and $z_{+}$ are the two classical turning point for the incident wave ($z_{+} > z_{-}$). These partial phase shifts are evaluated as, respectively
\begin{subequations}
\begin{align}
&\Theta_1(E_\parallel) = \frac{\sqrt{2m_{\text{ox}}}}{\hbar F_\text{ox1}}\left(\frac{V_{h1}}{2}\right)^{3/2}\int_{-a_1}^{t_{0-}}\sqrt{\frac{a_1^2-t^2}{t+1}}\mathrm{d}t \\
&\Theta_2(E_\parallel)= \frac{\sqrt{2m_{\text{ox}}}}{\hbar F_\text{ox2}}\left(\frac{V_{h2}}{2}\right)^{3/2}\int_{t_{0+}}^{a_2}\sqrt{\frac{a_2^2-t^2}{t+1}}\mathrm{d}t
\end{align}
\end{subequations}
with parameters:
\begin{subequations}
\begin{align}
&a_j = \sqrt{1-y_j^2},y_j = \frac{\sqrt{\gamma_1F_{\text{ox}j}}}{V_{hj}},\\
&t_j = \frac{2F_{\text{ox}j}}{V_{hj}}z-1,  j=0,1,2
\end{align}
\end{subequations}
with the limits $t_{0-} = t_1(z_0), t_{0+} = t_2(z_0)$. Through the variable transformation
\begin{equation}
t_j = a_j(1-2\sin(\theta)^2),\lambda_j = \sqrt{\frac{2a_j}{1+a_j}}, j=1,2
\end{equation}
we obtain the final expressions for $\Theta_1$ and $\Theta_2$
\begin{widetext}
\begin{subequations}
\begin{align}
\Theta_1(E_\parallel) &= \frac{2}{3}\frac{\sqrt{2m_{\text{ox}}}}{\hbar F_\text{ox1}}V_{h1}^{3/2}\left\{I(t_{0-},a_1)+\sqrt{\frac{1+a_1}{2}}\left[(a_1-1)\left[K(\lambda_1)-K_\text{inc}(\theta_{0-},\lambda_1)\right] + E(\lambda_1)-E_\text{inc}(\theta_{0-},\lambda_1)\right]\right\}\\
\Theta_2(E_\parallel) &= \frac{2}{3}\frac{\sqrt{2m_{\text{ox}}}}{\hbar F_\text{ox2}}V_{h2}^{3/2}\left\{-I(t_{0+},a_2)+\sqrt{\frac{1+a_2}{2}}\left[(a_2-1)K_\text{inc}(\theta_{0+},\lambda_2) + E_\text{inc}(\theta_{0+},\lambda_2)\right]\right\}
\end{align}
\end{subequations}
\end{widetext}
in which
\begin{subequations}
\begin{align}
I(t,a) &= \frac{\sqrt{(t+1){a^2-t^2}}}{2^{3/2}},\\
E_\text{inc}(\theta,\lambda) &= \int_0^{\theta}\sqrt{1-\lambda^2 \sin(t)^2}\mathrm{d}t, \\ 
K_\text{inc}(\theta,\lambda) &= \int_0^{\theta}\frac{1}{\sqrt{1-\lambda^2 \sin(t)^2}}\mathrm{d}t 
\end{align}
\end{subequations}
and the value of $\theta_{0-}$ and $\theta_{0-}$ are computed as
\begin{equation*}
\theta_{0-} = \sin^{-1}\left(\sqrt{\frac{1}{2}-\frac{t_{0-}}{2a_1}}\right),\theta_{0+} = \sin^{-1}\left(\sqrt{\frac{1}{2}-\frac{t_{0+}}{2a_2}}\right)
\end{equation*}

For the zero$^\text{th}$-order approximation, we have $t_{0-}=t_{0+}$ and $\theta_{0-}=\theta_{0+}$ such that the total phase-shift is obtained as
\begin{align}
&\Theta(E_\parallel) = \Theta_1(E_\parallel) +  \Theta_2(E_\parallel) \nonumber\\
&= \frac{2}{3}\frac{\sqrt{2m_{\text{ox}}}}{\hbar F_\text{ox0}}V_{h0}^{3/2}\left\{\sqrt{\frac{1+a_0}{2}}\left[(a_0-1)K(\lambda_0)+E(\lambda_1)\right]\right\}
\label{eqn_App_Theta0order}
\end{align}
in which 
\begin{equation*}
E(\lambda_0) = E_\text{inc}(\pi/2,\lambda_0), K(\lambda_0) = K_\text{inc}(\pi/2,\lambda_0).
\end{equation*}
are the complete elliptical integral with $\lambda_0=\lambda_1=\lambda_2$. The term in the curly brace of Eq \ref{eqn_App_Theta0order} has a compact approximate formula \cite{approxvy_ForbesAPL2006}
\begin{equation*}
v(y_0)\approx 1-(y_0)^2\left[1-\text{ln}(y_0)/3\right]
\end{equation*}

\section{The solution of 1D BTE}
\label{sec:App_1DBTE}
\begin{figure}
\centerline{\includegraphics[width=\linewidth]{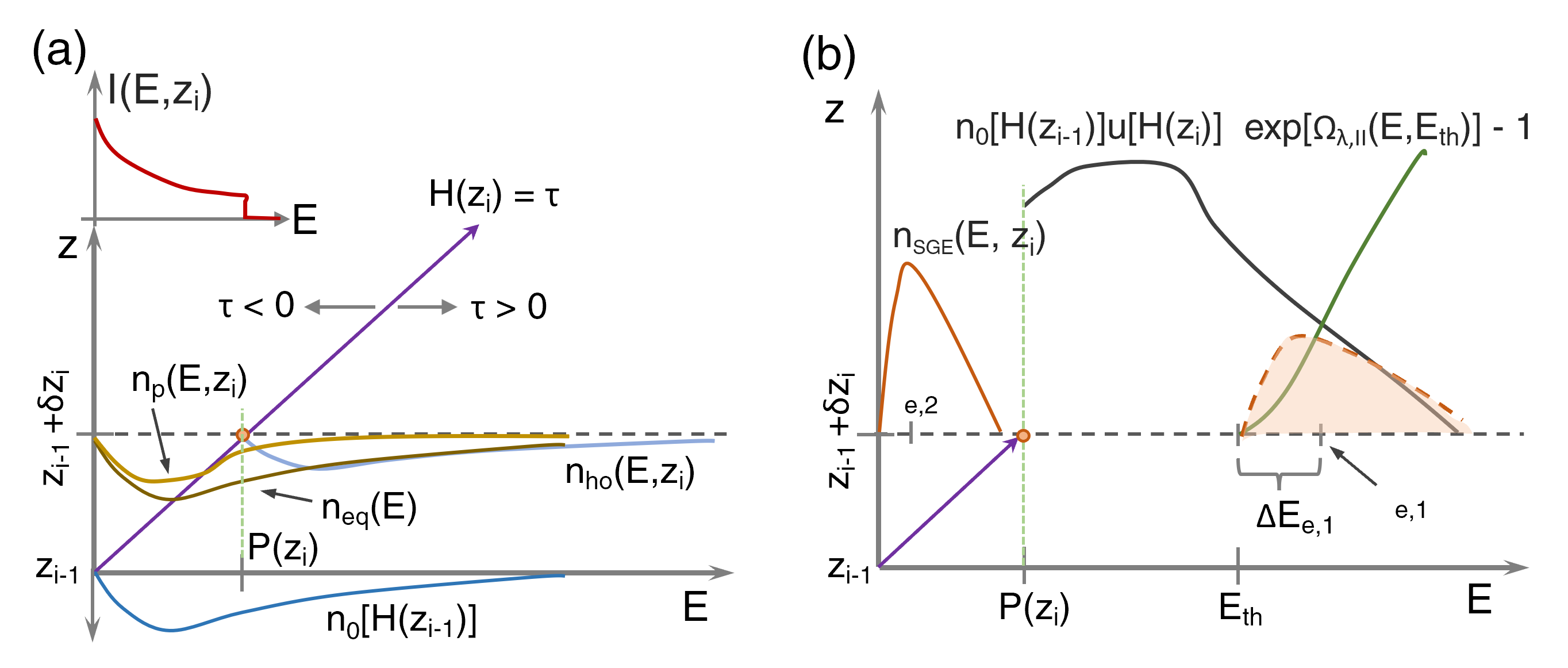}}
\caption{Two-step operator-splitting scheme for solving the 1D BTE. (a) Transport-relaxation substep: the EDF is propagated from $z_{i-1}$ to $z_i$  along the characteristics defined by $H=E- P(z)$, including attenuation and the thermal source term.  (b) Impact ionization substep: electrons with $E>E_\text{th}$ generate electron-hole pairs; the excess energy is partitioned, and secondary generated electrons are obtained according to the Maxwellian distribution.}
\label{fig_BTEII}
\end{figure}

The 1D BTE can only be solved iteratively at discrete nodes $(z_0,z_1,\cdots,z_{i-1}, z_i,\cdots, z_\textsc{n})$ under a  frozen potential profile in the chosen time window. The first node $z_{0}$ is approximated at the maximum tunneling distance $z_{\text{t,m}}$, and the last node  $z_\textsc{n}=L$ is at the a-SiO$_2$/gate interface.  The potential difference $\Delta V=\varphi(z_\textsc{n})-\varphi(z_0)$ are partitioned into $N$ intervals. For each interval $[z_{i-1},z_i]$, we define according to the spirit of the $H$-transform \cite{GnudiSHE_HTransform,ruppReviewRecentAdvances2016} 
\begin{equation}
P(z) = \int_{z_{i-1}}^{z}F_\text{ox}(x)\mathrm{d}x, \quad H(z) =  E - P(z)
\end{equation}
where $E$ is the kinetic energy of electrons and the local invariant $H$ characterizes ballistic transport.  After introducing the  relaxation energy
\begin{equation}
E_\lambda[E,P(z)]=F_\text{ox}\left[z(P)\right]\lambda(E)
\end{equation}
the 1D BTE Eq. \ref{eqn_BTE1d} reduces to an ordinary differential equation (ODE) along the characteristics $P$
\begin{equation}
\frac{dn_\textsc{bte}}{dP} = -\frac{n_\textsc{bte}-n_\text{eq}}{E_\lambda(E)}+\frac{S_\textsc{ii}(E)}{E_\lambda(E)}.
\end{equation}
We advance from $z_{i-1}$ to $z_i$ using the operator splitting (Lie-Trotter splitting) method along the characteristics into:  (i) the transport-relaxation substep without impact ionization and (ii) an impact-ionization substep where transport is ballistic. Both use the EDF from the previous node as initial condition. Fig. \ref{fig_BTEII} illustrates the construction of solutions for both substeps.

For the transport-relaxation substep, the ODE reads
\begin{equation}
\frac{dn_\textsc{bte}}{dP} = -\frac{n_\textsc{bte}-n_\text{eq}}{E_\lambda(E)}.
\end{equation}
Denoting the initial condition for the interval $[z_{i-1},z_i]$ as $n_0[H(z_{i-1})]$, we obtain the
the solution at $z_i$
\begin{equation}
n_{i}^\textsc{a}(E,z_i) = n^\text{ho}(E,z_i) + n^\text{p}(E,z_i)
\end{equation}
where the homogeneous solution $n^\text{ho}(E,z_i)$ and particular solution  $n^\text{ho}(E,z_i)$ are, respectively,
\begin{subequations}
\begin{align}
n^\text{ho}(E,z_i) =& n_0[H(z_{i-1})]\exp\left[-\Omega(E,z_i)\right]u[H(z_i)] \label{eqn_1dBTEsolhomo}  \\
n^\text{p}(E,z_i)=&  \int_0^\infty I(E-\mathcal{E},z_i)\frac{n_\text{eq}(\mathcal{E})}{E_\lambda(\mathcal{E})}\mathrm{d}\mathcal{E}
 \label{eqn_1dBTEsolparticular}
\end{align}
\label{eqn_1dBTEsol}
\end{subequations}
with auxiliary functions
\begin{subequations}
\begin{align}
\Omega_\lambda(E,z_i) &= \int_{H(z_i)}^{E} \frac{u[\mathcal{E}-P(z_i)]}{E_\lambda(\mathcal{E})}\mathrm{d}\mathcal{E}, \\ 
I(E,z_i) &= \exp[-\Omega(E,z_i)]\{u(z_i-z_{i-1})-u[H(z_i)]\}.
\end{align}
\end{subequations}

The homogeneous solution (Eq. \ref{eqn_1dBTEsolhomo}) describes ballistic transport with attenuation factor $\exp[-\Omega(E,z_i)]$, while the particular solution (Eq. \ref{eqn_1dBTEsolparticular}) shows generation through the convolution of equilibrium distribution $n_\text{eq}(E)$ with response function $I(E,z_i)$. It is noteworthy that $E_\lambda$ includes the contribution of relaxation of carriers energy via impact ionization.

The second substep describe the electron generation by impact ionization under the ballistic transport of solution at the previous node $n^\textsc{a}(E,z_{i-1})$. The increment of distribution $\delta n(E,z)$ follows first-order kinetics
\begin{equation}
\frac{\partial n^\textsc{a}(E,z)}{\partial P} = \frac{1}{E_{\lambda,\textsc{ii}}(E,E_\text{th})} n^\textsc{a}(E,z_{i-1}), E>E_\text{th}
\end{equation}
which yields the electrons generated with kinetic energy above $E_\text{th}$
\begin{align*}
\delta n(E,z_i) = & n[H(z_{i-1})] \left\{\exp\left[\Omega_{\lambda,\textsc{ii}}(E,E_\text{th})\right]-1\right\} \\
& \times u[H(z_i)]u(E-E_\text{th}),
\end{align*}
where $\Omega_\textsc{ii}(E,E_\text{th})$ is defined as
\begin{equation}
\Omega_{\lambda,\textsc{ii}}(E,E_\text{th}) = \int_{H(z_i)}^{E} \frac{1}{E_{\lambda,\textsc{ii}}(\mathcal{E},E_\text{th})}\mathrm{d}\mathcal{E}.
\end{equation}
The rate of generation is calculated as
\begin{equation}
G_\textsc{ii}(z_i) = \int_{E_\text{th}}^\infty\delta n(E,z_i)v_\text{g}(E)\mathrm{d}E
\end{equation}

The generated electrons undergo relaxation to near the conduction band edge within very short time scale ($\sim$ fs). These relaxed electrons are often refered as secondary generated electrons (SGE). It is challenging to calculate the spectrum of SGE.  Full-band Monte Carlo simulations showed that in a-SiO$_2$ the SGE follows a Maxwellian distribution with average energy $\bar{E}_{\text{e},2}$ \cite{IIrateDFTMC_MizunoJAP}. The excess of kinetic energy above threshold $\Delta E_{\text{e},1}=E-E_\text{th}$ will be partitioned according to 
\begin{equation}
\bar{E}_{\text{e},2} = \frac{2.5}{6}\Delta E_{\text{e},1},\bar{E}_{\text{h},2} = \frac{1.25}{6}\Delta E_{\text{e},1}.
\end{equation} 
where $\bar{E}_{\text{h},2}$ is the averaged energy of the secondary generated holes (SGH). Notably, the linear relationship between $\Delta E_{\text{e},1}$, $\bar{E}_{\text{e},2}$, and the Maxwellian SGE distribution appear in silicon and other semiconductors \cite{kunikiyoModelImpactIonization1996,harrisonImpactIonizationRate1999,impactionization_SiMC}, suggesting broader applicability of this approach. Enforcing the continuity of the generation rate, the concentration of the SGE can be obtained
\begin{equation}
n_\textsc{sge}(E,z_i) = \frac{{v}_{\text{d},1}(z_i)}{{v}_{\text{d},2}(z_i)}\frac{g_\text{c}(E)\exp\left(-\frac{E}{\bar{E}_{\text{e},2}}\right)\times \int_{E_\text{th}}^\infty \delta n(E,z_i)\mathrm{d}E}{\int_0^{\infty}g_\text{c}(E)\exp\left(-\frac{E}{\bar{E}_{\text{e},2}}\right)\mathrm{d}E}
\end{equation}
where the average drifted velocities are defined as 
\begin{subequations}
\begin{align}
{v}_{\text{d},1}&=\frac{G_\textsc{ii}(z_i)}{\int_{E_\text{th}}^\infty \delta n(E,z_i)\mathrm{d}E},\\
{v}_{\text{d},2}&=\frac{\int_0^{\infty}v_\text{g}(E)g_\text{c}(E)\exp\left(-\frac{E}{\bar{E}_{\text{e},2}}\right)\mathrm{d}E}{\int_0^{\infty}g_\text{c}(E)\exp\left(-\frac{E}{\bar{E}_{\text{e},2}}\right)\mathrm{d}E}.
\end{align}
\end{subequations}
Hence, the solution at node $z_i$ is now
\begin{equation}
n_\textsc{bte}(E,z_i) = n^\textsc{a}(E,z_i) + n_\textsc{sge}(E,z_i).
\end{equation}


\bibliography{SiCMOSbib.bib}


\end{document}